\documentclass[reprint,twocolumn,longbibliography,superscriptaddress]{revtex4-2}

\usepackage{graphicx}
\usepackage{amsmath,amssymb,latexsym}
\usepackage{natbib}

\begin{document}

\title{Study of dust-induced beam losses in the cryogenic arcs of the CERN Large Hadron Collider}
\author{A. Lechner}\email[]{Anton.Lechner@cern.ch}
\affiliation{European Organization for Nuclear Research (CERN), Espl.~des Particules 1, 1211 Geneva, Switzerland}
\author{P. B\'{e}langer}
\affiliation{European Organization for Nuclear Research (CERN), Espl.~des Particules 1, 1211 Geneva, Switzerland}
\affiliation{TRIUMF, 4004 Wesbrook Mall, Vancouver, BC, V6T 2A3, Canada}
\author{I. Efthymiopoulos}
\affiliation{European Organization for Nuclear Research (CERN), Espl.~des Particules 1, 1211 Geneva, Switzerland}
\author{L. Grob}\altaffiliation[Present address:]{Toptica Photonics AG, Lochhamer Schlag 19, 82166 Graefelfing, Germany.}
\affiliation{European Organization for Nuclear Research (CERN), Espl.~des Particules 1, 1211 Geneva, Switzerland}
\author{B.~Lindstrom}
\affiliation{European Organization for Nuclear Research (CERN), Espl.~des Particules 1, 1211 Geneva, Switzerland}
\affiliation{Uppsala University, Department of Physics and Astronomy, Box 516, 75120 Uppsala, Sweden}
\author{R. Schmidt}
\affiliation{European Organization for Nuclear Research (CERN), Espl.~des Particules 1, 1211 Geneva, Switzerland}
\author{D. Wollmann}
\affiliation{European Organization for Nuclear Research (CERN), Espl.~des Particules 1, 1211 Geneva, Switzerland}

\begin{abstract}
  The interaction of dust particles with the LHC proton beams accounts for a major fraction of irregular beam loss events observed in LHC physics operation. The events cease after a few beam revolutions because of the expulsion of dust particles from the beam once they become ionized in the transverse beam tails. Despite the transient nature of these events, the resulting beam losses can trigger beam aborts or provoke quenches of superconducting magnets. In this paper, we study the characteristics of beam-dust particle interactions in the cryogenic arcs by reconstructing key observables like nuclear collision rates, loss durations and integral losses per event. The study is based on events recorded during 6.5~TeV operation with stored beam intensities of up to $\sim 3\times 10{^{14}}$ protons per beam. We show that inelastic collision rates can reach almost $10^{12}$ collisions per second, resulting in a loss of up to $\sim 1.6\times 10^{8}$ protons per event. We demonstrate that the experimental distributions and their dependence on beam parameters can be described quantitatively by a previously developed simulation model if dust particles are assumed to be attracted by the beam. The latter finding is consistent with recent time profile studies and yields further evidence that dust particles carry a negative charge when entering the beam. We also develop different hypotheses regarding the absence of higher-loss events in the measurements, although such events are theoretically not excluded by the simulation model. The results provide grounds for predicting dust-induced beam losses in presence of higher-intensity beams in future runs of the High-Luminosity LHC. 
\end{abstract}

\maketitle

\section{Introduction}

The Large Hadron Collider (LHC) \cite{Bruning2004} at CERN is the first machine with positively charged hadron beams where interactions with micrometer-sized dust particles caused perceivable disruptions of beam operation \cite{Baer2011,Nebot2011,Nebot2012,Baer2012,Baer2013,Goddard2012,Auchmann2015a,Papotti2016,Lechner2016}. The trapping of ionized dust particles in the beam, accompanied by a drop of beam lifetime, is a well-known phenomenon in electron storage rings \cite{Hiroshi1991,Sagan1993,Zimmermann1995,Kelly1997,Kling2006,Tanimoto2009,Suetsugu2016}. Dust or macroparticle-related beam losses were, however, not expected to be perturbing for a proton collider since dust particles become rapidly ionized and repelled from the beams. Simulations and experimental observations in the LHC indeed support the hypothesis that dust grains are expelled before reaching the beam core \cite{Auchmann2014,Rowan2015,Rowan2016,Lindstrom2018,Lindstrom2020}. The events typically last less than a millisecond, i.e. less than twelve beam revolutions, with many events being as fast as one or two revolutions \cite{Baer2013}. Despite the short loss duration, it was already apparent in LHC run~I (2009-2013) that dust particles can still generate sufficient beam losses to provoke beam aborts by the Beam Loss Monitor (BLM) system \cite{Baer2011,Nebot2011,Baer2012,Goddard2012,Baer2013}. Beam aborts already occurred at stored intensities as low as 10$^{12}$ protons \cite{Baer2013}, which is two orders of magnitude below the LHC design intensity. After increasing the operation energy from 3.5~TeV and 4~TeV in run~I to 6.5~TeV in run~II (2015-2018), the first dust-induced quenches of superconducting magnets were observed \cite{Auchmann2015a,Papotti2016,Lechner2016}. All quenches concerned bending dipoles in the arcs or dispersion suppressors. The quench events typically required a recovery time between 8 and 12~hours before regular cryogenic conditions could be restored. In total, about a hundred BLM abort triggers and eight quenches attributed to dust particles were observed in the first two LHC runs, resulting in the loss of several hundred hours of beam time. No other beam loss mechanism caused more magnet quenches than the interaction of the beam with dust particles. 

Dust-induced loss events occur all around the LHC circumference, including the room temperature insertion regions and the cryogenic arcs and dispersion suppressors \cite{Baer2011,Nebot2011,Nebot2012,Baer2012,Goddard2012,Baer2013,Auchmann2015a,Papotti2016,Lechner2016}. Besides the events causing beam dumps or quenches, a copious amount of smaller events has been registered by the BLMs every operational year. These events appear as a localized transient loss spike on BLMs, but do not have any detrimental effect on operation and luminosity production. The beam intensity loss is smaller than the resolution of the LHC beam current monitors ($\sim10^9$ charges) and cannot be measured directly. The secondary particle showers can nonetheless be detected by the BLMs even for beam losses as small as 10$^{4}$-10$^{5}$ protons. Although these events are harmless, they are still carefully monitored as they provide insight about the correlation with beam parameters and the long-term evolution of event rates.

Dust particles were not the only cause of beam-induced aborts and quenches in run~II. Performance limitations arose also from a macroscopic obstacle in one of the dispersion suppressor dipoles \cite{Mirarchi2015,Mirarchi2019} and from solid micrometer-sized aggregates of residual air molecules \cite{Mether2017,Jimenez2018,Lechner2018,Mirarchi2019}. The latter were leftovers from an accidental air inflow in a certain arc cell. While the occurrence of these localized loss events was limited in time, dust-induced quenches remain a concern for future runs of the LHC, in particular in the High-Luminosity (HL)-LHC era \cite{HL2020}. A large increase of event rates can potentially be expected after long shutdowns, as was the case when restarting the LHC for its second run in 2015 \cite{Mirarchi2019}. Such an increase can have a detrimental impact on the operational performance in the first year after a shutdown. Increasing the operation energy from 6.5~TeV in run~II to 6.8~TeV in run~III (2022-2024) and further to 7~TeV in the HL-LHC era, will in addition increase the susceptibility for magnet quenches since the temperature rise, which superconducting coils can tolerate, will decrease. Besides the reduced quench margin, the beam parameters will become more challenging in future runs, with the stored beam intensity increasing from $3.2\times10^{14}$ protons in run~II to possibly $5\times10^{14}$ protons in run~III and further to $6.1\times10^{14}$ in the HL-LHC era (see Table~\ref{tab:beamparam}).

\begin{table}[!t]
\caption{\label{tab:beamparam} Proton beam parameters (beam energy $E$, bunch spacing $\Delta t_b$, number of bunches per beam $N_b$, bunch intensity $I_b$, and normalized transverse emittance $\varepsilon_n$) in previous and future runs of the LHC, and in the HL-LHC era. The values correspond to the start of physics production, referred to as the stable beams period. Besides the standard beam production scheme, an alternative low-emittance scheme is used, called Batch Compression, Merging and Splitting (BCMS) scheme. For run~I and run~II, maximum performance values are given, corresponding to standard 50~ns beams in 2012 \cite{Lamont2013} and to 25~ns BCMS beams in 2017/18 \cite{Steerenberg2019}, respectively. For run~III, optimistic and pessimistic values are given for the transverse emittance (BCMS beams), whereas the bunch intensity is the maximum intensity expected in run~III \cite{Karastathis2019}. The last column shows nominal HL-LHC beam parameters for standard 25~ns beams \cite{HL2020}.}
\begin{ruledtabular}
\begin{tabular}{lcccc}
  &run~I&run~II&run~III&HL-LHC\\
&(2009-2013)&(2015-2018)&(2022-2024)&(2027-)\\
\hline
$E$ (TeV)& 4 & 6.5 & 6.8 & 7\\
$\Delta t_b$ (ns) & 50 & 25 & 25 & 25 \\
$N_b$& 1380 & 2556 & 2748 & 2760\\
$I_b$ ($10^{11}~\text{p}$)& 1.55-1.65 & 1.2 & 1.8 & 2.2\\
$\varepsilon_n$ ($\mu$m\,rad) & $\approx$2.5 & $\approx$2 & 1.8-2.5 & 2.5\\
\end{tabular}
\end{ruledtabular}
\end{table}

Dust-induced loss events have been put under scrutiny since their first occurrence in run~I, by studying event rates and empirical correlations with beam parameters \cite{Baer2011,Nebot2011,Nebot2012,Baer2012,Baer2013,Auchmann2015a,Papotti2016}, by analysing dust samples from the vacuum chamber of magnets \cite{Goddard2012,Grob2019}, and by modeling the motion of dust particles in the beam \cite{Zimmermann2010,Fuster2011,Auchmann2014,Rowan2015,Rowan2016}. In combination with bunch-by-bunch beam diagnostics, individual dust particle trajectories could be reconstructed \cite{Lindstrom2018,Lindstrom2020}. The simulation studies and measurements also indicated that the radius of dust particles is smaller than 100~$\mu$m \cite{Rowan2015,Rowan2016} and that the dust grains likely carry a negative charge when entering the beam \cite{Lindstrom2020,Belanger2021}. Despite these findings, a better understanding of these events is needed for quantifying the performance impact on the LHC, the HL-LHC and other future high-energy proton colliders like the FCC-hh \cite{Abada2019}. In particular, the physical mechanism governing the release of dust particles into the beam still lacks a theoretical explanation, which is fundamental for predicting the likelihood of events. It is also essential to reliably quantify dust-induced beam losses as a function of beam parameters and dust properties. The induced beam losses have been estimated previously by means of simulations \cite{Zimmermann2010,Fuster2011}, but the predictions still lack a systematic experimental verification.

In this paper, we study the characteristics of beam-dust particle interactions through a comprehensive experimental analysis of inelastic nuclear collisions between beam protons and dust grains. Inelastic collisions are the main mechanism for dust-induced beam losses in superconducting magnets. The characteristic features of loss profiles, like peak collision rates and integral losses, are of practical importance since they directly relate to the risk of magnet quenches. They also reveal more about the nature of these events and the properties of dust particles. Owing to a better coverage of arc dipoles with beam loss monitors in run~II, observables related to beam-dust collisions could be systematically reconstructed for a large ensemble of events. Based on these data, we probe the ability of the previously developed simulation model \cite{Zimmermann2010,Fuster2011,Auchmann2014,Rowan2015,Rowan2016,Lindstrom2020} to reproduce different experimental distributions by constraining the properties of dust particles. We further probe the ability of the model to reproduce the dependence of proton losses on beam parameters. This enables more accurate predictions for future runs with higher-intensity beams.

\begin{figure}[!t]
\centering
\includegraphics[width=0.48\textwidth]{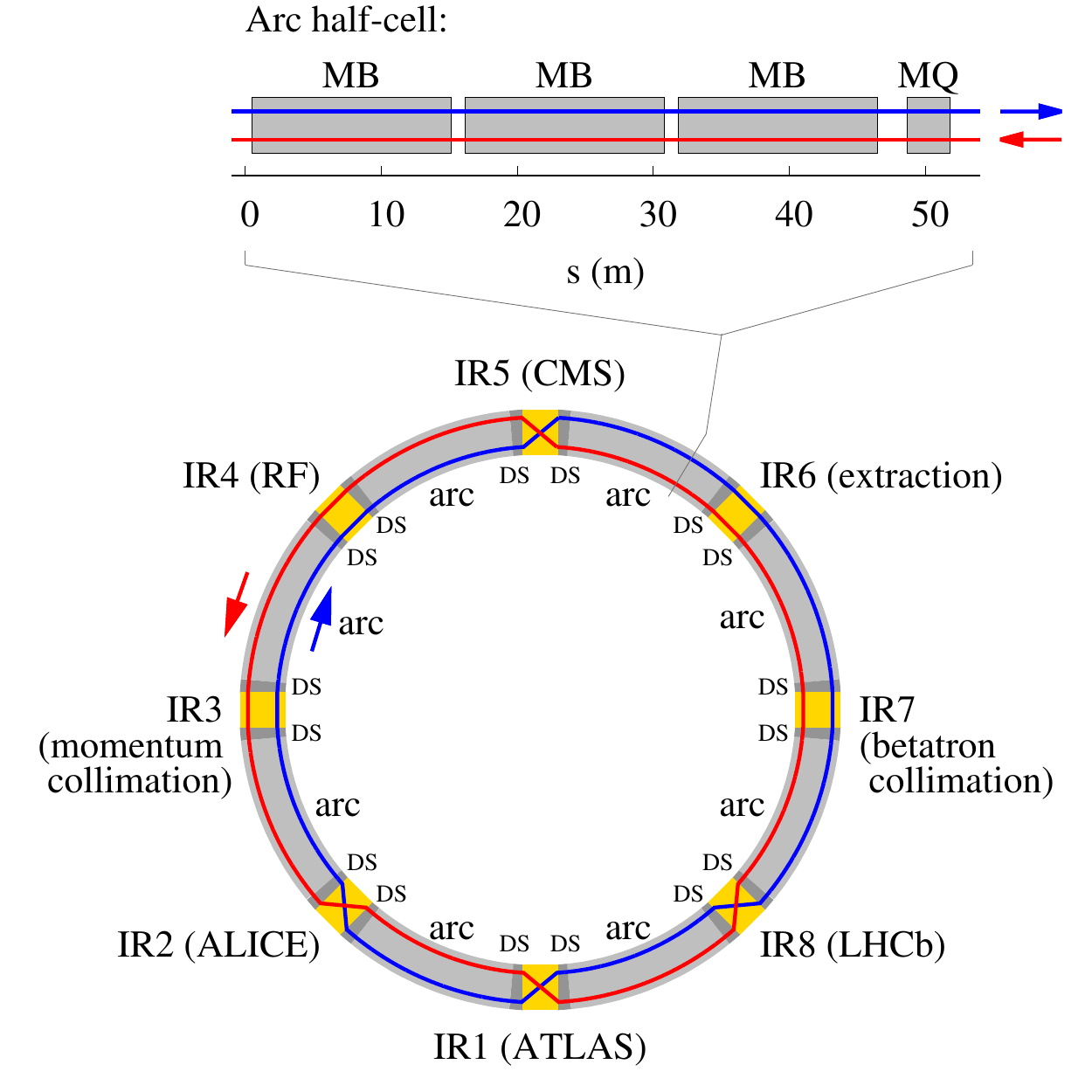}
\caption[]{Illustration of the LHC layout, showing the different insertion regions (IRs - yellow), dispersion suppressors (DS - dark gray) and arcs (gray). Each of the eight arcs consists of 23 cells. A cell is made up of two 53.45~m long half-cells (see top), which are composed of three bending dipoles (MBs), one quadrupole (MQ) and corrector magnets (not shown). The two counter-rotating beams are represented by the blue line (beam 1) and the red line (beam 2), respectively. The beams are crossing each other in four interaction points.}
\label{fig:lhclayout}
\end{figure}

Although dust particle events are observed in all regions of the LHC, the studies presented in this paper focus exclusively on the cryogenic arc sectors (see Fig.~\ref{fig:lhclayout}). The arcs are believed to be the regions where dust-induced losses might have the highest impact in future runs because of the higher risk of magnet quenches. The arc sectors offer ideal conditions for a systematic study of dust events. They are less exposed to other types of beam losses than the insertion regions or dispersion suppressors, which facilitates the identification of such transient loss events. Another advantage is the cumulative arc length of more than 19~km and the cell-by-cell periodicity of the BLM layout, which help in enhancing the statistical significance of observations.

The paper is organized as follows. Section~\ref{sec:eventreconstr} describes the methods for reconstructing the number of nuclear beam-dust particle collisions from BLM measurements in the arc sectors. Section~\ref{sec:beammacroparticleint} explores the characteristics of dust-induced losses at 6.5~TeV by comparing reconstructed distributions (loss rates, loss durations and integral losses) with predictions from dust particle dynamics simulations. By constraining the dust properties, we attempt to find the best agreement between simulations and measurements. Based on these results, Sec.~\ref{sec:beamparamdep} analyses the dependence of observables on beam parameters, while Sec.~\ref{sec:losseshighint} provides predictions of dust-induced losses for beam intensities in the HL-LHC era. Section~\ref{sec:conclusions} summarizes the studies and provides some concluding remarks.

\section{Reconstruction of dust-induced beam losses in the LHC arcs}
\label{sec:eventreconstr}

Dust-induced loss events can only be studied indirectly by analysing beam losses through the shower-induced energy deposition in BLMs. In this section, we discuss the methods for reconstructing the number of inelastic encounters from spatial BLM signal patterns near the collision vertex. The event reconstruction is based on \textsc{FLUKA} \cite{Bohlen2014,Battistoni2015,FlukaWeb} Monte Carlo simulations. General-purpose transport codes like \textsc{FLUKA} can provide an estimate of macroscopic observables like monitor signals by describing the propagation of showers in complex geometries based on microscopic interaction models. The predictive ability of the \textsc{FLUKA} code for BLM response studies in the LHC radiation environment has been demonstrated in Ref.~\cite{Lechner2019}. In the following, we use this simulation technique for reproducing dust-induced signal patterns recorded during 6.5~TeV operation in run~II.

\subsection{Detection of dust events in LHC operation}

The eight LHC arcs are composed of superconducting twin-bore magnets, which host the two counter-rotating proton beams in physically separated apertures \cite{Bruning2004}. Each of the arcs consists of 23 lattice cells. A cell is made up of two 53.45~m long half-cells (see Fig.~\ref{fig:lhclayout}), which are composed of three bending dipoles, one quadrupole and corrector magnets. The mechanical clearance for the beams is defined by racetrack-shaped beam screens inside the vacuum chambers. The beam screens protect the cold magnets from different heat sources such as electron clouds and synchrotron photons. The beam screens are perforated at the top and bottom surfaces and are maintained at a higher temperature (5-20~K) than the vacuum chambers and magnets (1.9~K) \cite{groebner1995}. They are made of stainless steel with a layer of copper on the inner side. The beam screens of neighboring magnet cryostats are connected by means of radiofrequency bridges, which provide a passage for the beam-image current \cite{Veness1999}. The bridges consist of gold-plated copper-beryllium fingers, which can slide along a copper tube. Dust samples collected from the vacuum components of an arc dipole in run II showed that dust contamination is present on all components, including beam screens, vacuum chambers and the plug-in modules containing the radiofrequency bridges \cite{Grob2019}. 

When a dust grain enters the LHC proton beam, it gets ionized by the traversing beam particles and is rapidly ejected due to the repelling force exerted by the electric field of the beam. While a dust particle travels in the beam, a small fraction of the impacting protons will be subject to an inelastic nuclear collision and will be lost from the beam. The energetic collision products, mainly $\pi^\pm$, protons, neutrons, kaons, as well as photons from decaying $\pi^0$, impact on the machine aperture and induce hadronic and electromagnetic showers in the surrounding beam screens, vacuum chambers and magnets. Most of the secondary particles are lost within the same or the neighboring lattice half-cell, i.e. nearby the primary collision vertex. An exception are diffractive protons, which can travel longer distances in the collider rings. Contrary to the inelastic collision products, beam protons undergoing an elastic nuclear collision either stay in the transverse beam acceptance or they are intercepted by collimators in the cleaning or experimental insertions if the deflection angle is large enough \cite{Hempel2012}. 

\begin{figure}[!t]
\centering
\includegraphics[width=0.48\textwidth]{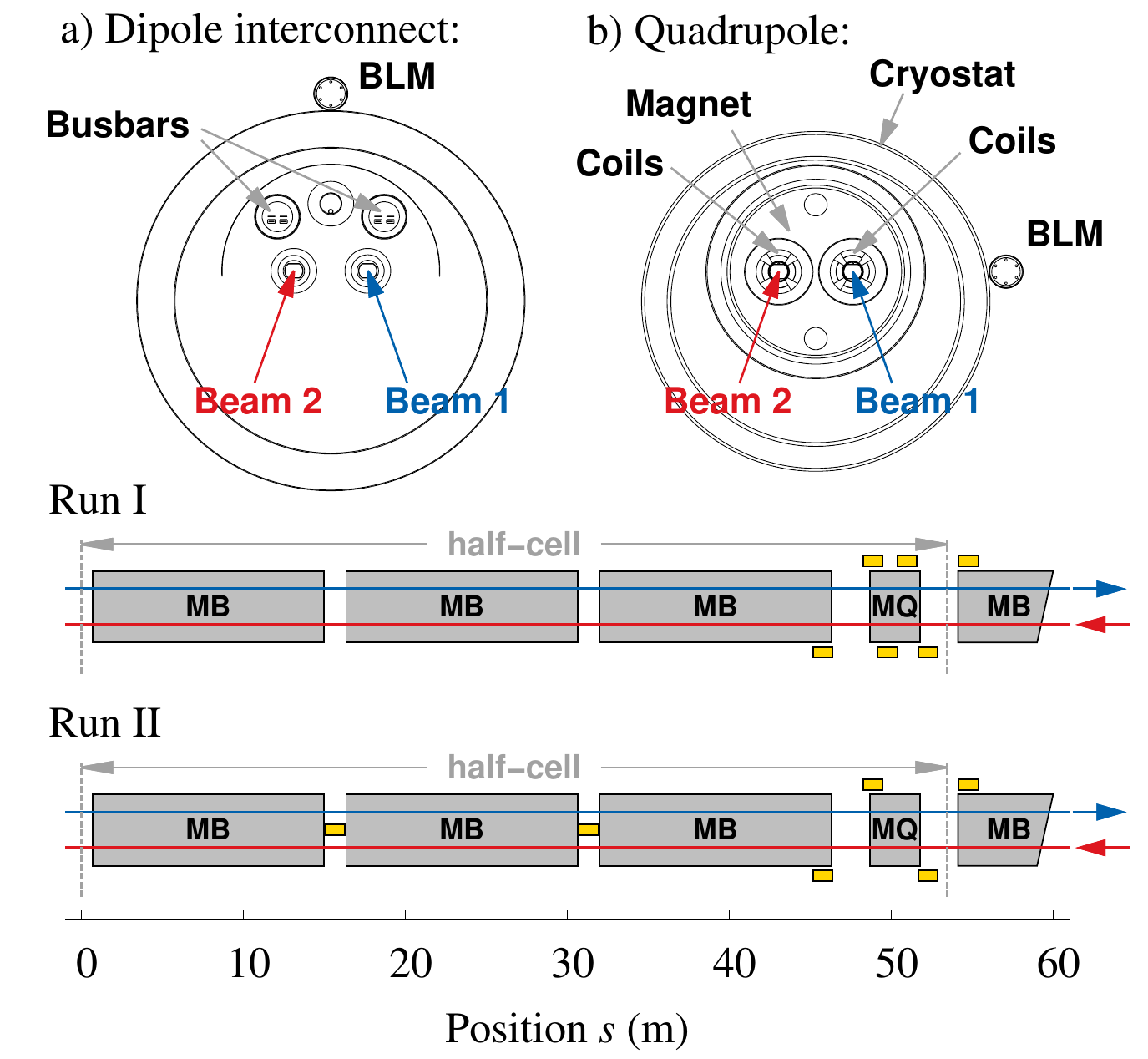}
\caption[minimum]{Illustration of BLM positions in a standard LHC arc half-cell. The two layout plots at the bottom show a top view of the different BLM positions in run~I and run~II, respectively. Gray boxes represent magnets (bending dipoles: MB, quadrupoles: MQ), yellow boxes represent monitors. Monitors around the quadrupole are located on the horizontal plane (figure top right), while monitors between dipoles (run~II only) are located on top of the dipole interconnects (figure top left). The blue and red arrows indicate the beam direction of beam 1 and beam 2, respectively.}
\label{fig:blmlayout}
\end{figure}

The LHC is equipped with almost 4000 ionization chambers which constantly monitor beam losses around the two rings \cite{Holzer2005,Dehning2007}. The chambers are filled with nitrogen gas and have a sensitive volume of $\sim$1500~cm$^3$. About three quarters of these monitors are located in the arc sectors. The arc BLMs are mounted on the outside of magnet cryostats (see Fig.~\ref{fig:blmlayout}) and detect electromagnetic and hadronic particle showers induced by beam losses in magnets and other equipment. The BLM system records the deposited dose in twelve sliding time windows of different length (from $\Delta t=40~\mu$s to 83.9~s). These running sums enable a customized monitoring of beam losses with different time characteristics, from very fast losses up to steady-state losses.

Dust events represent some of the fastest beam losses in the LHC. They are recorded in real time during operation by a dedicated software application developed in run~I \cite{Baer2013}. A beam loss event is classified as a dust particle event if the dose measured in the 640~$\mu$s sliding time window of at least two BLMs exceeds a certain trigger threshold. The two BLMs must be located within 40~m in order to exclude false triggers on uncorrelated signals. In addition, the detection algorithm verifies that dose values recorded in shorter time windows (40-320~$\mu$s) do not exhibit an unphysical correlation. This should minimize false triggers because of spurious noise spikes. The noise suppression parameters were optimized in run~I \cite{Baer2013} and the final settings were retained in run~II. The trigger threshold for the 640~$\mu$s time window was adapted a few times throughout the years, and hence the data must be post-processed to remove any bias.  

Each of the arc half-cells is equipped with six BLMs. In run~I, all six BLMs were installed in the vicinity of quadrupoles, as illustrated in Fig.~\ref{fig:blmlayout}. Quadrupoles were expected to be the bottleneck for beam losses in the arcs due to local restrictions of the effective aperture (maxima in the $\beta$ and dispersion functions), aperture discontinuities (beam position monitors) and possible imperfections (magnet misalignment). Because of the absence of BLMs along the bending dipoles, which account for more than 85\% of the arc length, this configuration provided only a limited resolution for detecting and localizing dust-induced loss events. As dust particles posed the major source of transient beam losses in the arcs in run~I, hundreds of arc BLMs were relocated from quadrupoles to dipoles in the shutdown between run~I and run~II (2013-2015) to improve the detection of such loss events \cite{Kalliokoski2015}. The run~II BLM layout is illustrated at the bottom of Fig.~\ref{fig:blmlayout}. The BLMs were installed on top of dipole-dipole interconnects, as shown in the upper figure. 

\subsection{Methods for reconstructing the number of collisions}

\begin{figure}[!b]
\centering
\includegraphics[width=0.48\textwidth]{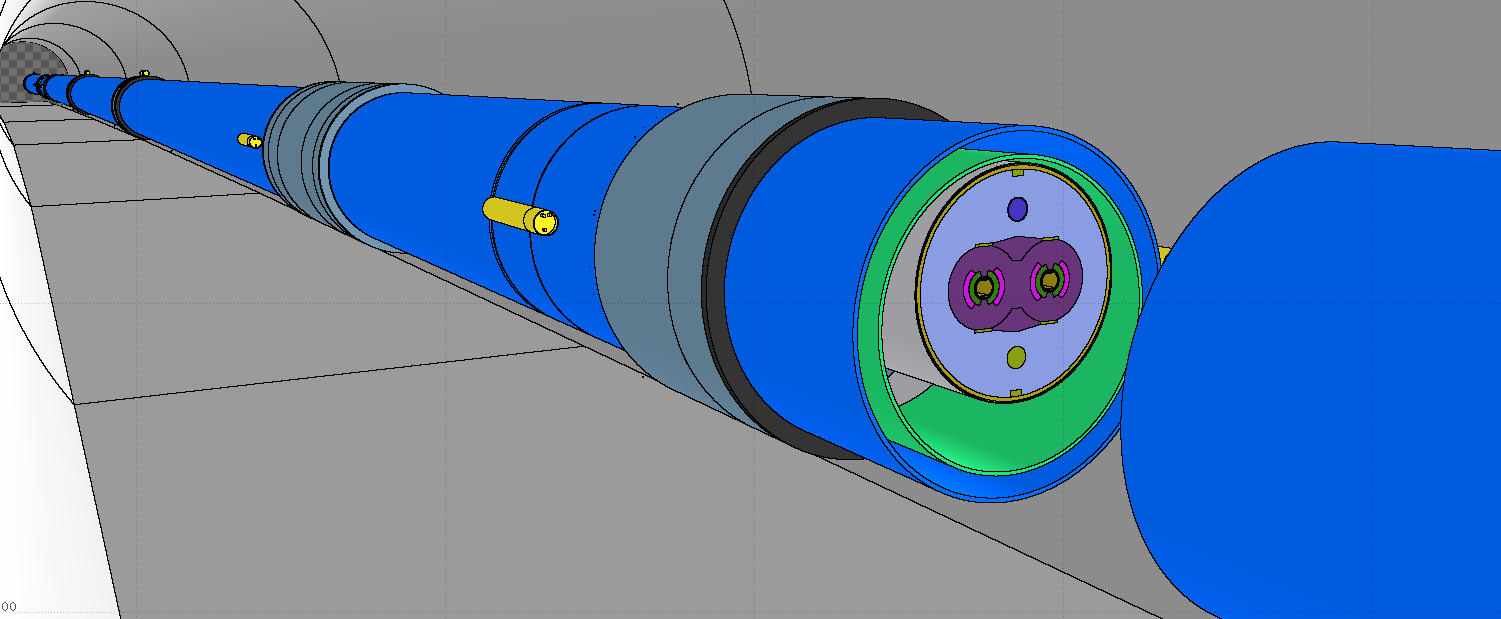}
\caption[]{Geometry model of a LHC arc cell used in the particle shower simulations. The cryostat and a dipole are cut open to show the interior. The beam loss monitors (yellow cylinders) are mounted on the outside of the cryostat.}
\label{fig:flukageomodel}
\end{figure}

The number of inelastic nuclear collisions of beam protons in pointlike obstacles like dust particles can be estimated empirically from BLM signals using particle shower simulations \cite{Lechner2019}. This requires finding the longitudinal position of dust particles since the BLM response per collision depends on the distance to the loss location due to the shower attenuation by nearby magnets. The spatial pattern of BLM signals along an arc half-cell, i.e. the measured dose as a function of the longitudinal BLM position, exhibits a relatively high sensitivity to the longitudinal loss location. The collision vertex can therefore be determined empirically by finding the best match between measured and simulated patterns. This method already proved to be successful for localizing the aforementioned obstacle in the dispersion suppressor \cite{Mirarchi2019} as well as the location of accidental air inflow in one of the arc cells \cite{Lechner2018}. The same approach was also used in run~I for estimating the location of dust events at 3.5~TeV and 4~TeV in a specific arc cell equipped with additional BLMs \cite{Lechner2019}. 

Since dust events appear at any longitudional position in the arcs, BLM patterns were calculated for different loss locations inside a representative arc half-cell consisting of three bending dipoles and a quadrupole. The \textsc{FLUKA} geometry model is illustrated in Fig.~\ref{fig:flukageomodel}. Neighboring magnets in adjacent half-cells were included as well. The BLM signals were calculated as described in Ref.~\cite{Lechner2019}, by recording the energy deposition in the sensitive gas volume between the electrodes of the BLM model. The dust particles were assumed to be static and pointlike \cite{Lechner2018,Lechner2019}.

The loss locations were spaced by 0.5-2 meters in order to achieve a good resolution in the vertex identification. Because of the significant computational requirements, losses in the two upstream dipoles were only studied for the anti-clockwise rotating beam (beam 2) and the obtained signal patterns were then mapped to the other beam (beam 1). This approach is justified since the dipole BLMs, located on top of the dipole interconnects, are equally exposed to losses from both beams. The situation is different for beam-dust particle collisions in or nearby quadrupoles since the relative BLM positions are not fully identical for the two counter-rotating beams. In addition, the layout around quadrupoles differs for the two beams because of the asymmetric sequence of corrector magnets in the quadrupole assemblies \cite{Bruning2004}. Losses in the third dipole and the quadrupole were hence studied separately for the two beams as the different geometry can affect the shower leakage to BLMs.

In all studies, it was assumed that the clockwise rotating beam circulates in the inner aperture of the twin-aperture magnets, while the other beam circulates in the outer aperture. This assumption holds only for half of the arc sectors since the beams change from the inner to the outer aperture and vice versa due to the beam crossing in the four detectors (see Fig.~\ref{fig:lhclayout}).  It was further assumed that the field of the quadrupole in the considered half-cell is defocusing in the horizontal plane. These assumptions may slightly increase the uncertainty of the pattern reconstruction, but it is not expected to impact the general conclusions.

\begin{figure}[!t]
  \centering
  \includegraphics[width=0.48\textwidth]{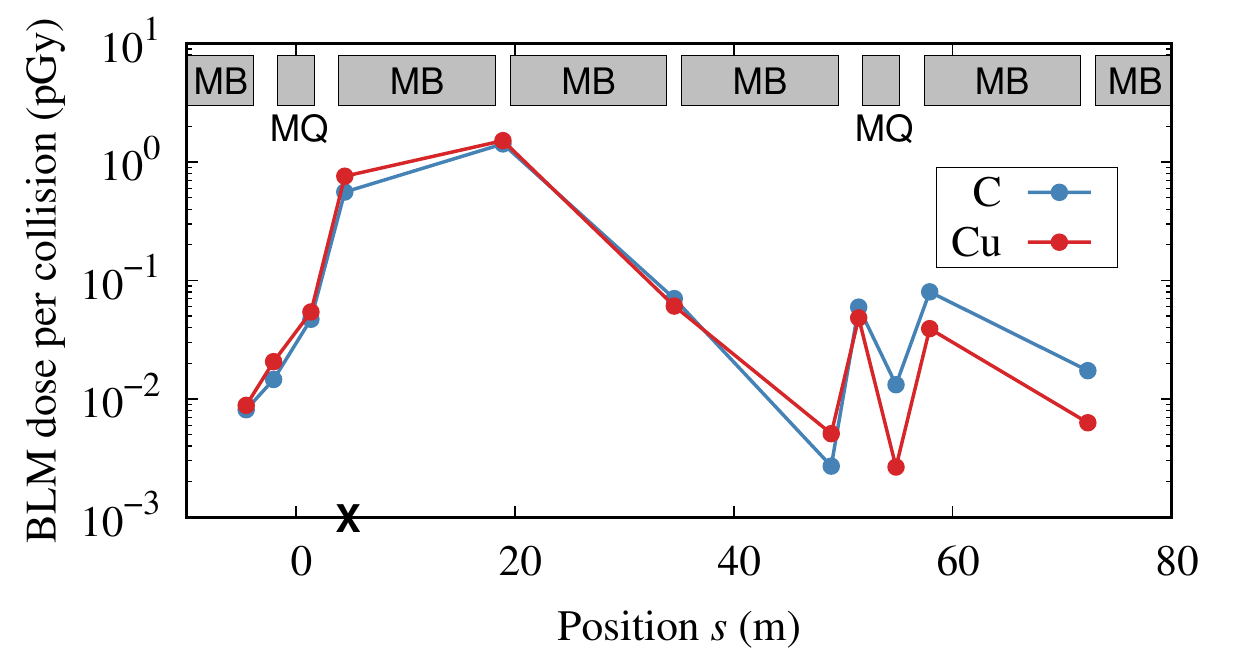}
\caption[]{Comparison of simulated BLM dose patterns for beam collisions with dust particles of different composition. The proton beam energy was 6.5~TeV. The dose is given per proton-nucleus collision. The assumed loss location is indicated by a cross on the $s$-axis. The beam direction is from the left to the right. The statistical error of the simulations is less than 3\% for the highest signal, but can be as large as few 100\% for the smallest signals. The data points are connected by lines to guide the eye.}
\label{fig:blmpatterns_sim}
\end{figure}

The composition of dust particles can possibly vary from event to event. The analysis of dust samples collected in run~II showed that macroparticles of different chemical position are present in the vacuum system \cite{Grob2019}. Figure~\ref{fig:blmpatterns_sim} presents simulated BLM signal patterns for dust particles composed of carbon and copper, respectively. The BLM signals are given per proton-nucleus collision. The different dipoles (MBs) and quadrupoles (MQs) are illustrated by gray boxes. The results show that BLM signals up to 40~m from the loss location agree within 30\% percent for the two compositions (the highest signal agrees within a few percent). Larger discrepancies can be observed for more distant BLMs, which can at least partially be explained by the higher statistical uncertainty of the simulation results because of the much smaller dose values. The good agreement of the BLM patterns around the loss location suggests that the number of inelastic collisions can be reconstructed from the measurements with reasonable accuracy even without having an exact knowledge of the dust particle composition. In the shower simulations presented in the following, dust particles were assumed to be made of carbon.

\subsection{Matching of simulated and measured BLM dose patterns}

In order to estimate the number of inelastic hadronic proton-dust particle collisions $N_i$ for a measured loss event, the following expression was minimized as a function of $N_i$ and the discrete loss location $s_{j}$:  
\begin{equation}
  Q(N_i,s_j) = \sum_k \frac{[D_{tot,k} - N_i d_k(s_{j})]^2}{N_i d_k(s_{j})},
  \label{eq:match}
\end{equation}  
where $D_{tot,k}$ is the time integral of the dose rate $\dot{D_k}(t)$ measured in BLM $k$ during the loss event, 
\begin{equation}
D_{tot,k} = \int \dot{D_k}(t') dt'
\end{equation}
and $d_k(s_{j})$ is the corresponding simulated dose per proton-nucleus collision assuming that the collisions occur at $s_{j}$. The time-integrated dose was obtained from the $\Delta t=2.56$~ms long sliding window of BLMs. This time window is sufficiently long to contain the full loss event. The BLM signals were corrected for the noise floor. For each case, six BLMs were used to find the best match between simulated and measured patterns (typically 1-2 BLMs upstream of the loss location and 4-5 BLMs downstream). The statistical error of the simulated signals was usually between 1-10\%, except for some BLMs upstream of the loss location and for BLMs $>$30-40~m downstream of the loss location. 

\begin{figure*}[!t]
\centering
\includegraphics[width=0.48\textwidth]{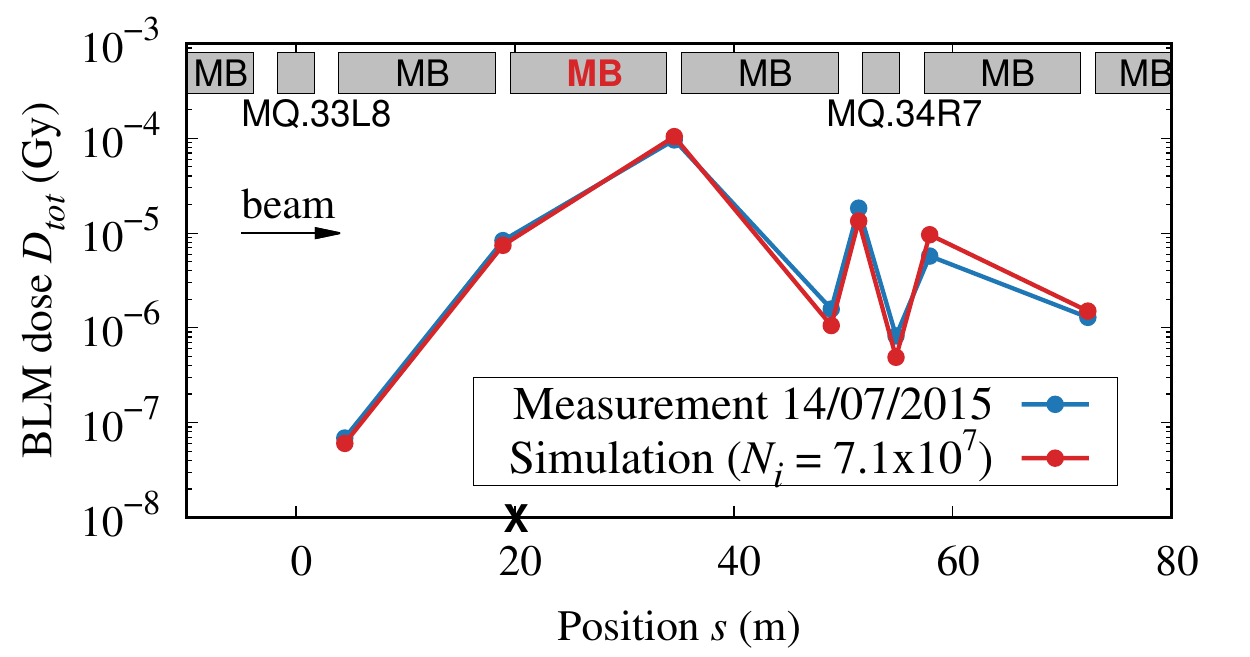}
\includegraphics[width=0.48\textwidth]{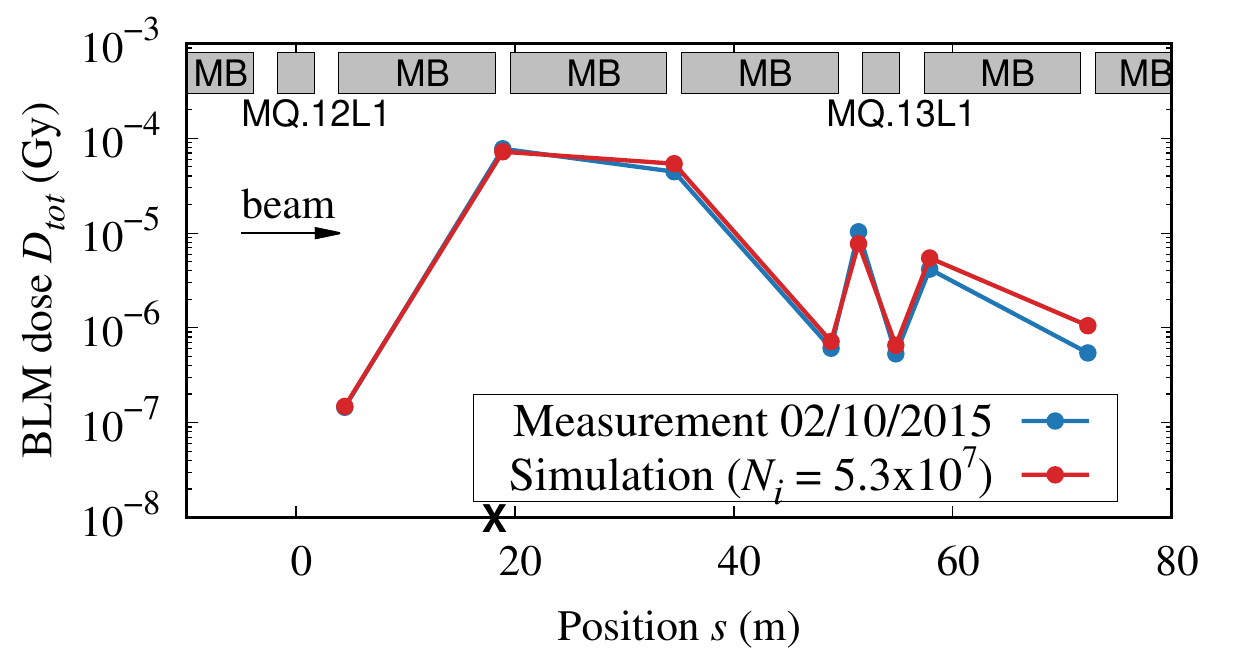}\\
\includegraphics[width=0.48\textwidth]{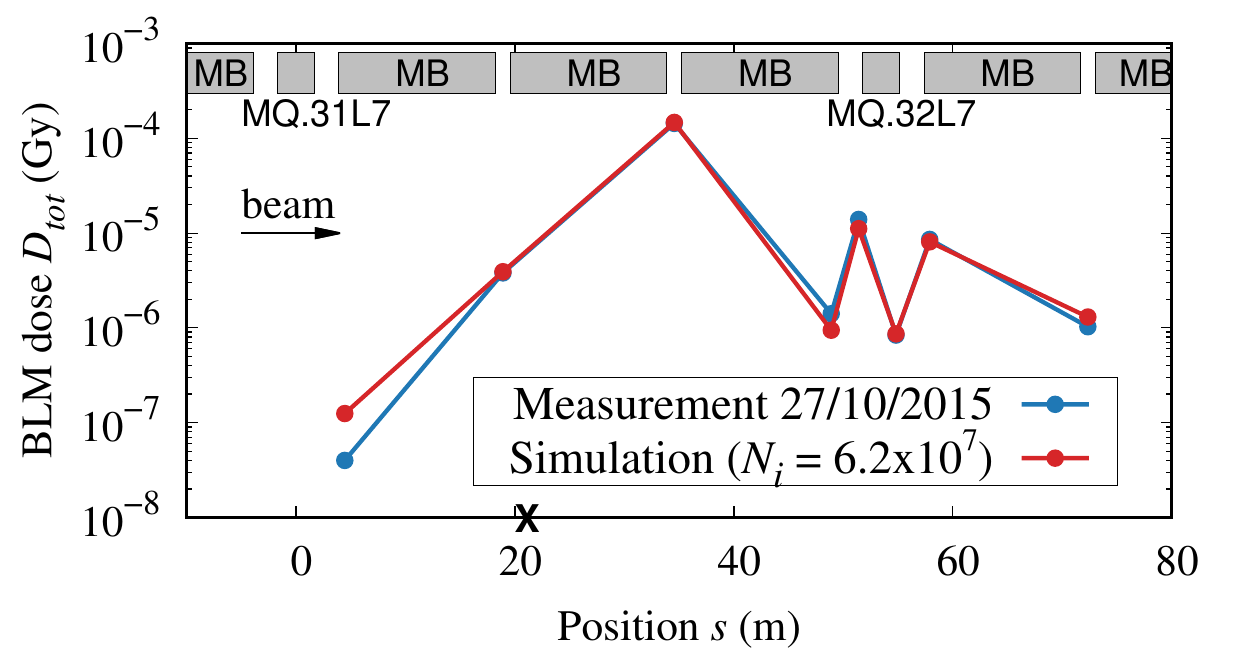}
\includegraphics[width=0.48\textwidth]{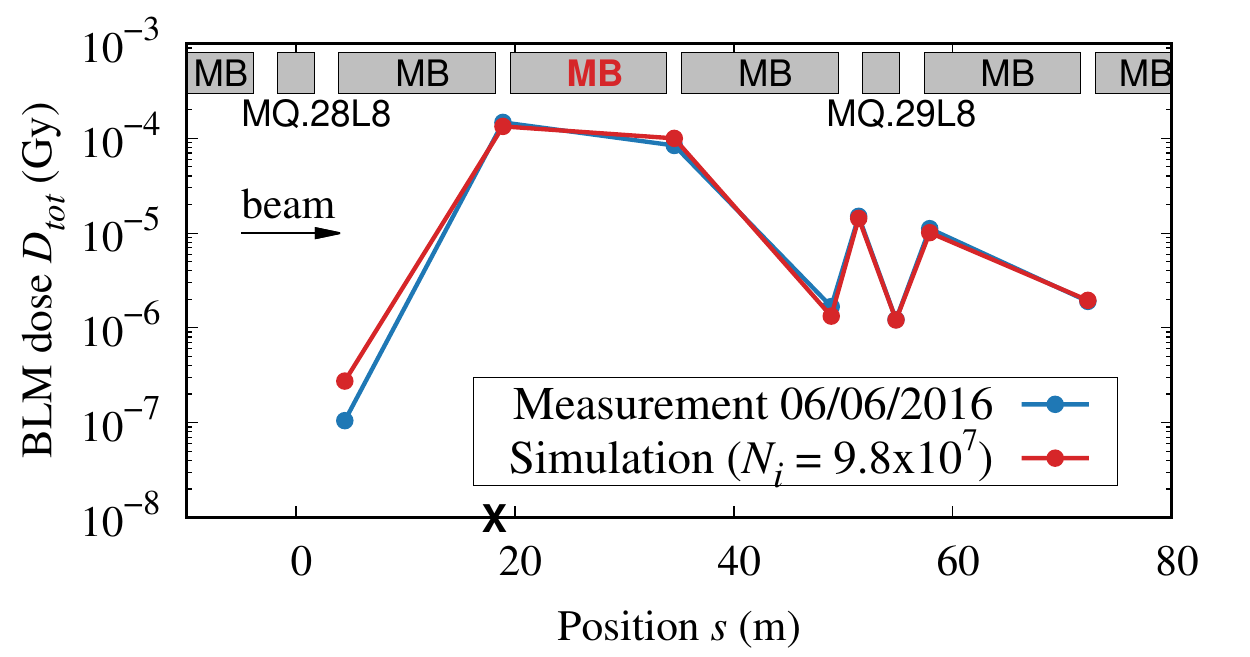}\\
\includegraphics[width=0.48\textwidth]{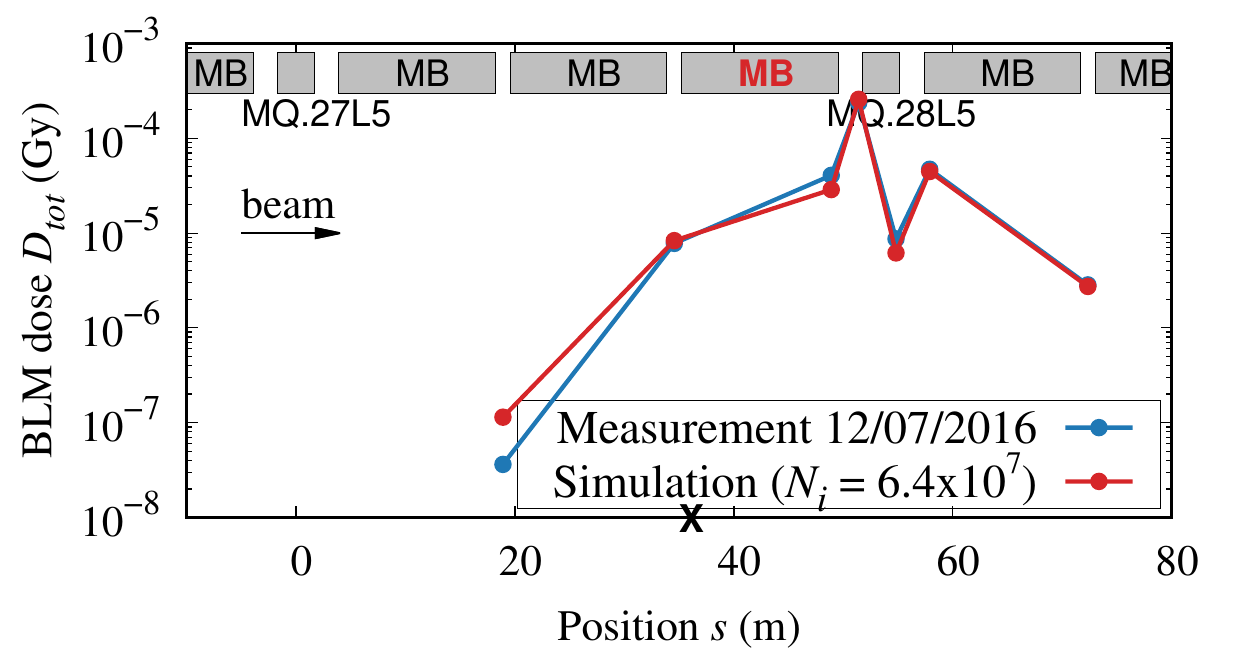}
\includegraphics[width=0.48\textwidth]{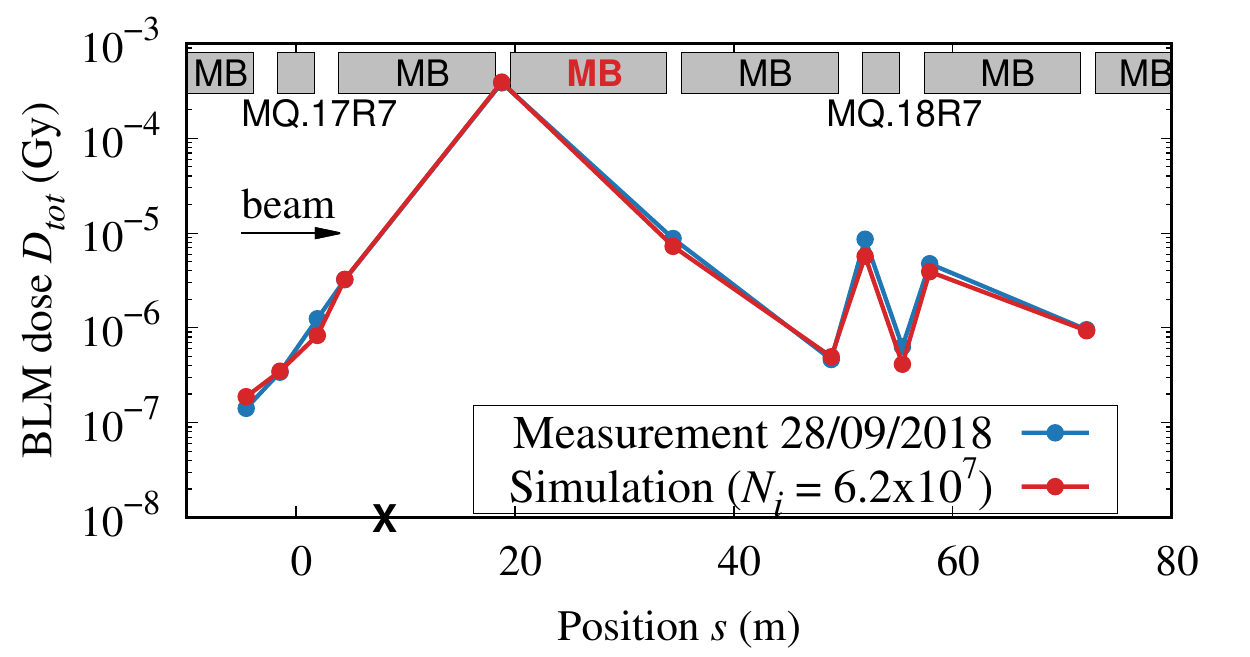}
\caption[]{Comparison of measured and simulated BLM dose patterns along LHC arc cells for dust-induced loss events in run~II physics operation at 6.5~TeV. In four out of the six cases a dipole quench occurred (labels of dipoles which quenched due to showers are in bold red). The simulated patterns were scaled according to Eq.~(\ref{eq:match}) to obtain the best match with the measurements. The scaling factors, which represent estimates of the inelastic proton-dust particle collisions, are indicated in the labels. The data points are connected by lines to guide the eye. The estimated $s$-location of the dust particles is identified by 'x' labels.}
\label{fig:blmpatterns}
\end{figure*}

Figure~\ref{fig:blmpatterns} shows a selection of simulated BLM patterns which were matched to measured loss patterns using Eq.~(\ref{eq:match}). All measurements were recorded at 6.5~TeV in run~II. The x-axis represents the $s$-coordinate of the curvilinear coordinate system of the concerned beam; the origin is arbitrarily set to coincide with the center of the quadrupole upstream of the loss location. The measurements shown in the figure exhibited some of the highest BLM signals of all dust events in run~II. In four out of the six cases, the losses induced a dipole quench, while no quench was observed in the other two events. The simulated patterns generally show a good agreement with the measurements. It is estimated that in most cases analysed in the following section, the number of inelastic collisions per event can be determined with an uncertainty less than a factor of two despite the unknown dust particle composition and the model approximations. 

\section{Characteristics of beam-dust particle interactions}
\label{sec:beammacroparticleint}

\begin{figure}[!t]
\centering
\includegraphics[width=0.48\textwidth]{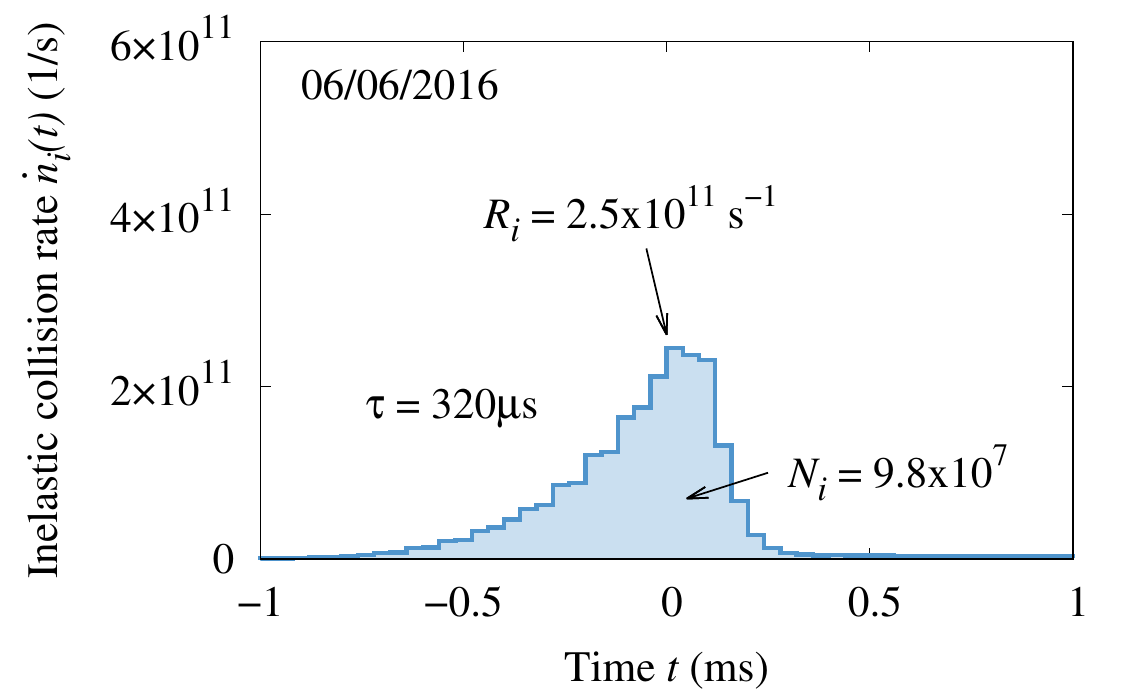}\\
\includegraphics[width=0.48\textwidth]{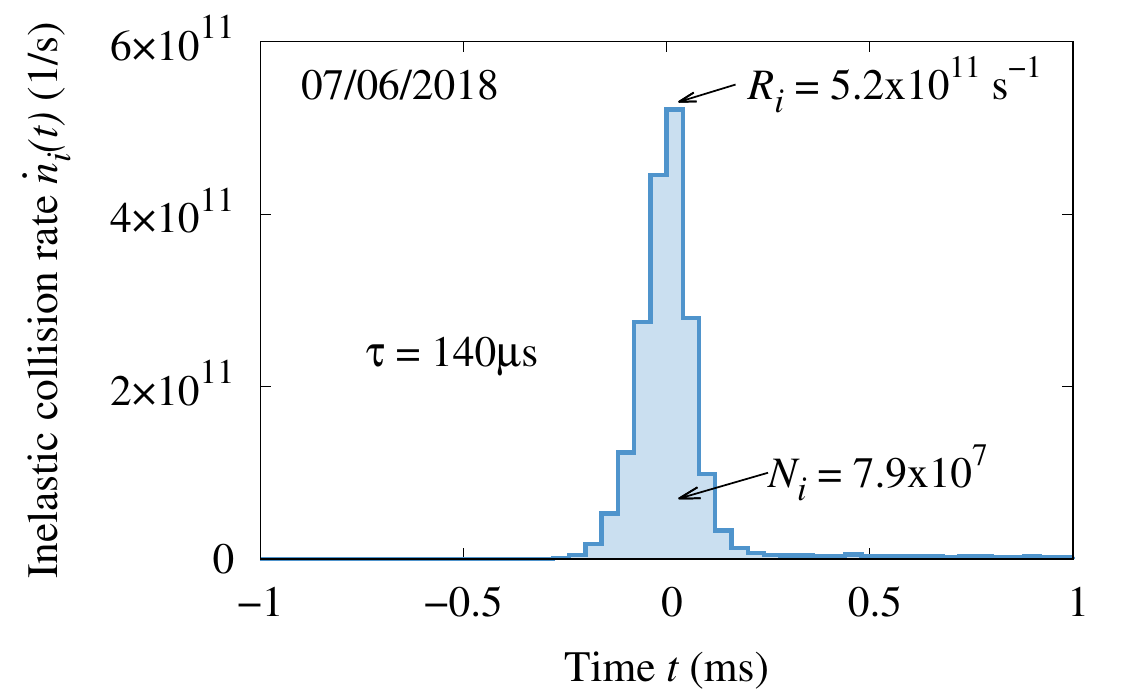}
\caption[]{Rate of inelastic nuclear collisions between 6.5~TeV protons and a dust particle entering the LHC beam. The two events were recorded in the LHC arcs in run~II. In both cases, the losses lead to a magnet quench. The loss rate was reconstructed from BLM measurements using particle shower simulations. The time resolution of the measurements is 40~$\mu$s.}
\label{fig:timeprofiles}
\end{figure}

The characteristics of dust-induced BLM signals have been studied since run~I \cite{Baer2011,Baer2012a,Baer2013,Rowan2015,Rowan2016,Lindstrom2018,Lindstrom2020}. Different observations were made, which provided some insight into the nature of these events. It was found that the distribution of BLM signals in run~I was proportional to 1/$D^2$, where $D$ is the time-integrated BLM dose in the BLM with the highest signal near the collision vertex \cite{Baer2011,Baer2012a,Baer2013}. This observation provided an approximate indication about the size distribution of dust particles in the cold arc sectors, with the caveat that BLM signals depend on the loss position and the dust particle trajectory in the beam. The latter can vary from event to event even for similar-sized dust particles. Hence BLM signals do not uniquely reflect the dust particle size for a given event. The signal distribution nonetheless demonstrates the abundance of smaller dust particulates in the arc vacuum system, which is also confirmed by the dust samples collected in run~II \cite{Grob2019}. It was also observed that the time profiles of events can typically be described by a skewed Gaussian distribution \cite{Nebot2012,Baer2013,Lindstrom2020} (see Fig.~\ref{fig:timeprofiles}). Some of the profiles had a shorter rise time, whereas others exhibited a faster fall time. The latter can be explained by the rapid repulsion of dust particulates once they get ionized in the beam, but the opposite observation (faster rise times) still lacks a theoretical understanding \cite{Baer2013,Lindstrom2020}. 

The trajectory of a dust particle and hence the induced nuclear collisions depend on various parameters like the beam intensity, the transverse beam size, the dust particle radius, the dust particle composition and density, as well as the initial dust particle position. The trajectory is also strongly influenced by the initial charge carried by the dust particle when it enters the beam. A numerical simulation model has been developed for studying the motion of dust particles in the LHC beams as a function of these parameters \cite{Zimmermann2010,Fuster2011}. The model assumes that the dust particle is initally located on the beam screen. Different physics improvements have been incorporated over time, as described in Refs.~\cite{Auchmann2014,Rowan2015,Rowan2016} and more recently in Ref.~\cite{Lindstrom2020}. The simulations suggest that the dust particles are expelled before reaching the beam core \cite{Zimmermann2010,Fuster2011,Rowan2015}. This was confirmed in recent experimental and numerical studies in run~II, where a maximum penetration depth of $\sim$3$\sigma$ from the beam center was found for particular events recorded at 5.5~TeV and 6.5~TeV, respectively \cite{Lindstrom2018,Lindstrom2020}. The model also explained other features of dust events, like the asymmetry of time profiles (in case of faster fall times) \cite{Rowan2015,Rowan2016,Lindstrom2020} and the decrease of the loss duration as a function of beam intensity \cite{Nebot2012}. The model calculations further made it possible to better understand dust particle properties. It could be shown that measured BLM signals in run~I can be reproduced by assuming dust particle radii between 5 and 100~$\mu$m \cite{Rowan2015,Rowan2016}, which was consistent with the dust samples collected in run~II \cite{Grob2019}. In a more recent article, based on BLM data from run~II, it was demonstrated that the rising slope of certain loss profiles can only be explained if dust particles are initially attracted by the beam \cite{Lindstrom2020}. As proposed in this article, this can be explained if dust particles possess a negative charge.

In this section, we extend previous studies by performing an absolute comparison between simulations and beam loss observables. The latter were reconstructed from BLM measurements at 6.5~TeV in run~II. By constraining the material properties of dust particulates, we probe the ability of the simulation model to reproduce the experimental distributions and their dependency on beam parameters. This is an important prerequisite for assessing the predictive ability of model calculations. We also discuss possible implications for future operation with higher-intensity. 

\subsection{Observables under study}

Dust-induced loss events can be characterized by the inelastic nuclear collision rate $\dot{n}_i(t)$ between beam protons and the dust grain. Figure~\ref{fig:timeprofiles} illustrates two typical collision rate profiles for events which lead to a quench at 6.5~TeV. Time profiles, as shown in the figure, are only recorded for a subset of events. In absence of such profiles, we can nonetheless describe dust particle events by different observables, which can be reconstructed from the maximum dose recorded in the different sliding time windows of BLMs. The maximum dose per window is logged by the monitoring application and is therefore available for all events. The observables under study include the integral number of inelastic proton-dust particle collisions per event (as already introduced in the previous section),
\begin{equation}
  N_i=\int \dot{n}_i(t) \mathrm{d}t,
\end{equation}
the maximum collision rate
\begin{equation}
  R_i=\max[{\dot{n}_i(t)}],
\end{equation}
and the loss duration $\tau$. We define the latter as the full width at half maximum (FWHM) of $\dot{n}_i(t)$ profiles,
\begin{equation}
\tau = t_2 - t_1,
\end{equation}
where $t_1 < t_{max}$, $t_2 > t_{max}$ and 
\begin{equation}
\dot{n}_i(t_1) = \dot{n}_i(t_2) = \frac{R_i}{2}.
\end{equation}
$t_{max}$ refers to the time where the collision rate is maximum. These quantities can provide more insight into the nature of dust particle events. As can be seen in Fig.~\ref{fig:timeprofiles}, the two profiles exhibit different features, one being twice as long as the other, but featuring a lower peak loss rate. The integral number of inelastic collisions was similar in both cases. Evidently, the considered observables $N_i$, $R_i$ and $\tau$ are not independent from each other, but we still consider it instructive to study all three together. While $N_i$ can be obtained from Eq.~(\ref{eq:match}), the reconstruction of $R_i$ and $\tau$ entails a few additional approximations, as detailed below. 

The peak collision rate was derived from the following expression,
\begin{equation}
  R_i = \frac{N_i}{D_{tot}} \frac{D^{40\mu \mathrm{s}}_{max}}{\Delta t},
  \label{eq:peaklossrateappr}
\end{equation}
where $D^{40\mu \mathrm{s}}_{max}$ is the maximum dose recorded in the shortest sliding time window of BLMs ($\Delta t=40\mu$s),
\begin{equation}
  D^{40\mu \mathrm{s}}_{max}= \max \bigg[\int_{t_n}^{t_{n+1}} \dot{D}(t')dt' : t_n = n \times 40~\mu \mathrm{s} \bigg],
\end{equation}
and $D_{tot}$ is the time-integrated dose in the same BLM where $D^{40\mu \mathrm{s}}_{max}$ was measured. Eq.~(\ref{eq:peaklossrateappr}) provides only an approximation of the real peak collision rate because of the signal delay introduced by the readout cables and the delayed charge collection in the ionization chambers. The collection time is around 300~ns for electrons, but is around 80~$\mu$s for ions \cite{Holzer2005,Dehning2007}. In general one can assume that about 40\% of the signal is registered within 40~$\mu$s, and about 70\% within 80~$\mu$s. Because of this delay, the actual peak dose rate and hence the peak collision rate might be underestimated. This underestimation is expected to be more pronounced for very fast events, lasting only one to two turns (80-160~$\mu$s), which account for about one third of the considered events. Even without the delayed signal collection, the measurement of ultra-fast events would be intrinsically limited by the 40~$\mu$s time resolution of the BLMs.

\begin{figure}[!t]
  \centering
  \includegraphics[width=0.48\textwidth]{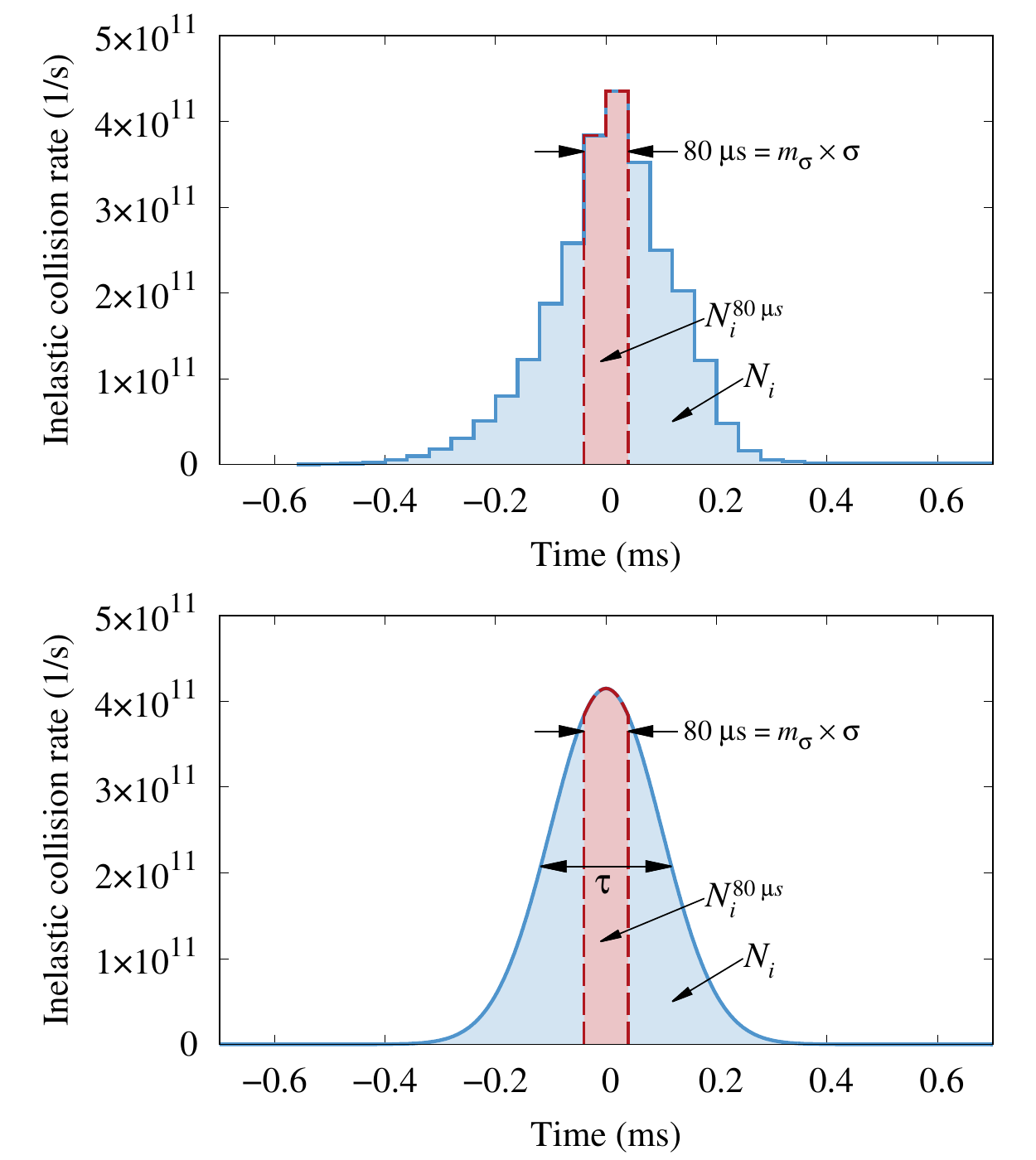}
  \caption[]{
    Measured time profile of a dust event recorded in LHC operation (top) and Gaussian profile used for estimating the loss duration $\tau$ (bottom). The standard deviation of the Gaussian distribution was calculated from the relative number of collisions ${N_i^{80\mu \mathrm{s}}}/{N_i}$ contained in the central 80~$\mu$s time window. The relative number of collisions in this interval was estimated from the measurements (ratio of red area and total area in top graph).} 
\label{fig:tau}
\end{figure}

The loss duration $\tau$ was reconstructed by assuming that the $\dot{n}_i(t)$ profiles are of Gaussian shape, neglecting any possible asymmetry. For consistency, this approximation was applied to all events, even if time profiles were available. The integral of the Gaussian profiles was assumed to be $N_i$. The width $\tau$ of the Gaussian profiles was determined by calculating the number of standard deviations $m_\sigma$ contained in an 80~$\mu$s time window centered around the mean (see bottom graph in Fig.~\ref{fig:tau}); $m_\sigma$ depends on the number of collisions $N_i^{80\mu s}$ in this time interval. We assume that $N_i^{80\mu s}$ corresponds to the maximum number of collisions measured in the 80~$\mu$s sliding time window of BLMs: 
\begin{equation}
  N_i^{80\mu \mathrm{s}} = N_i \frac{D^{80\mu \mathrm{s}}_{max}}{D_{tot}},
  \label{eq:coll80us}
\end{equation}
where $D_{tot}$ is the time-integrated dose and $D^{80\mu \mathrm{s}}_{max}$ is the maximum dose in the 80~$\mu$s sliding time window of the BLM, which is updated every 40~$\mu$s,
\begin{equation}
  D^{80\mu \mathrm{s}}_{max}= \max \bigg[\int_{t_n}^{t_{n+2}} \dot{D}(t')dt' : t_n = n \times 40~\mu\mathrm{s} \bigg].
\end{equation}
Figure~\ref{fig:tau} illustrates the time profile of a typical dust particle event measured in run~II (top graph) together with the derived Gaussian distribution (bottom graph). The red area in the top graph indicates the number of collisions as given by Eq.~(\ref{eq:coll80us}), which was in turn used to calculate the width $\tau$ of the Gaussian. Numerically, the results for $\tau$ would be the same if the total area of the Gaussian in Fig.~\ref{fig:tau} would be $D_{tot}$ and if the area within the central 80~$\mu$s window would be $D^{80\mu \mathrm{s}}_{max}$. For clarity, the formulas and figures were still expressed in terms of collisions rather than in terms of dose.

Although the approach could have also been based on a different window length, 80~$\mu$s was found to be the best compromise between faster and slower events. The calculated $\tau$ values can differ from the actual width of loss rate profiles because of the neglected skewness, which is unknown for events where no time profile was recorded. The approximation is nonetheless suitable for identifying trends. For very fast events, with a duration $\tau<10^{-4}$ seconds, the obtained values become unreliable because of the limited time resolution.

\subsection{Distribution of events as a function of $N_i$, $R_i$ and $\tau$}

About 21000 dust particle events were detected at 6.5~TeV in the LHC arcs in run~II. The smallest of these events gave rise to about $10^4$ inelastic proton-nucleus collisions, while the largest event resulted in more than 10$^8$ collisions. The shower simulations indicate that events, which generated less than $\sim5\times10^5$ collisions within a time interval of $\Delta t=640~\mu$s, were not uniformly detected along arc cells even with the improved BLM layout in run~II. These events were only registered if the loss location was nearby a BLM. In the following, we therefore discard all events with less than $5\times10^5$ collisions ($\sim$15500 out of the 21000) since they represent an incomplete sample of the true event population. The detection limit of $5\times10^5$ collisions within $\Delta t=640~\mu$s depends on the adopted dose threshold in the monitoring software. Although the dose threshold was temporarily raised in 2015 and 2016, it is estimated that $>$98\% of all events with more than $5\times10^5$ collisions in 640~$\mu$s were recorded in run~II. The considered 5500 events therefore represent an almost complete sample of the true number of events.

\begin{figure}[!t]
  \centering
  \includegraphics[width=0.48\textwidth]{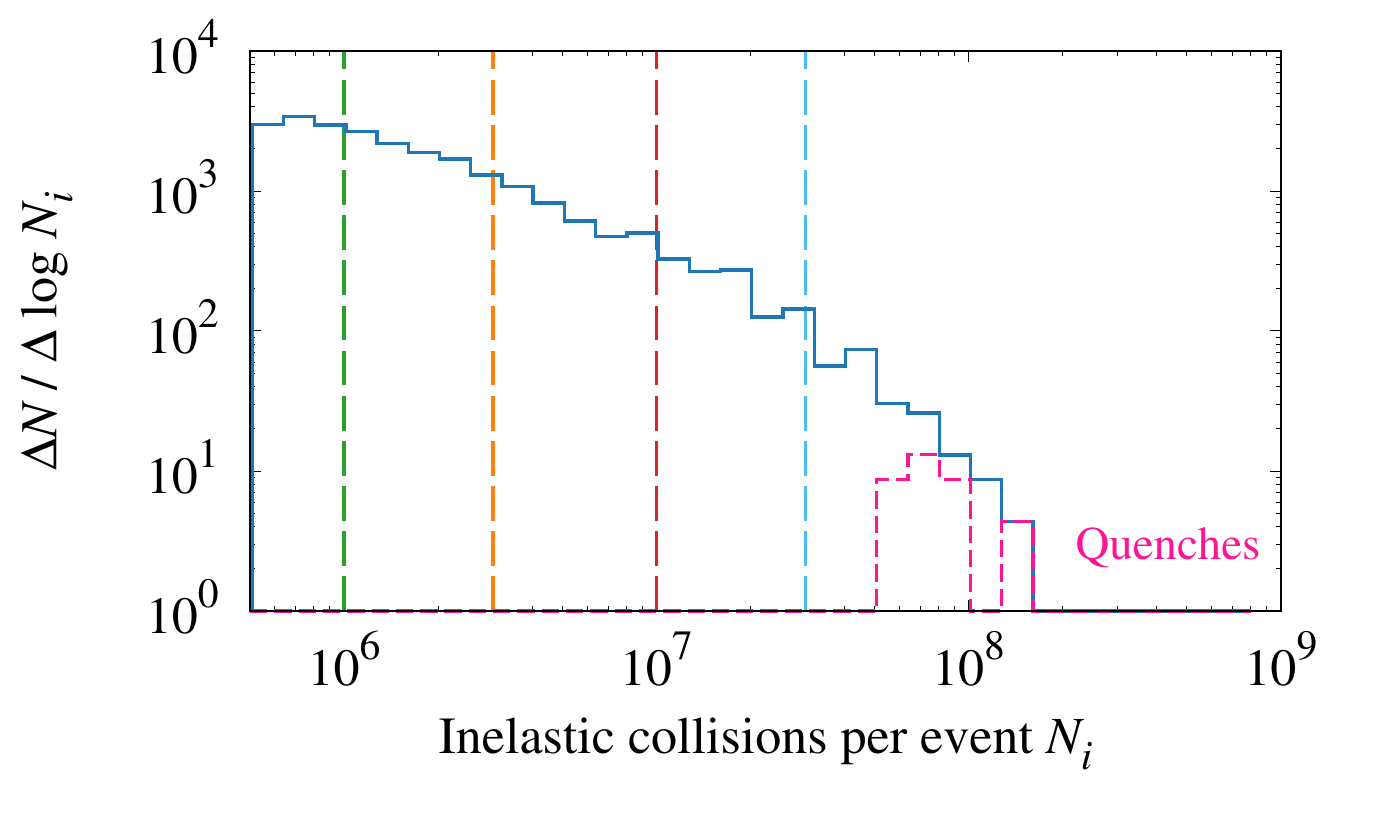}\\
  \includegraphics[width=0.48\textwidth]{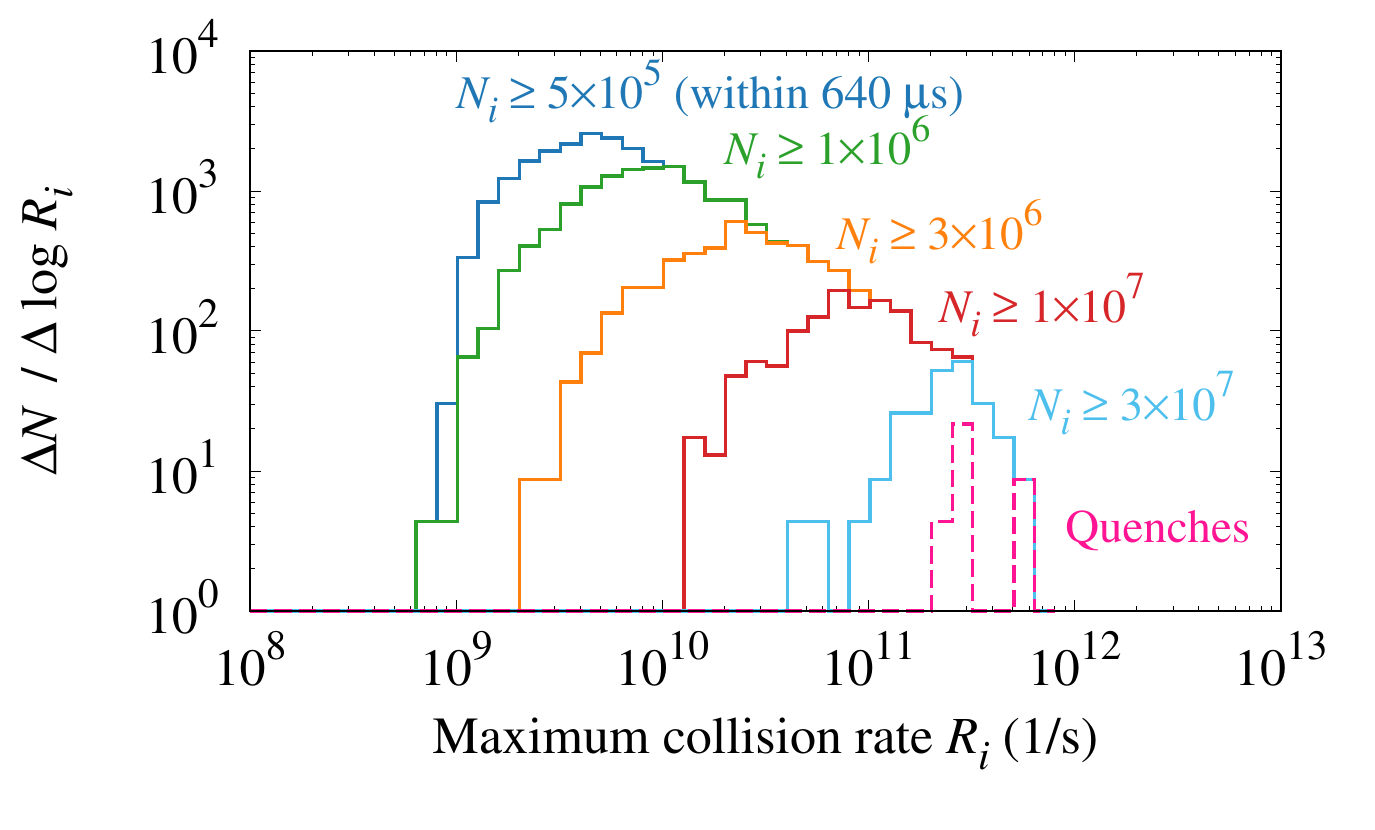}\\
  \includegraphics[width=0.48\textwidth]{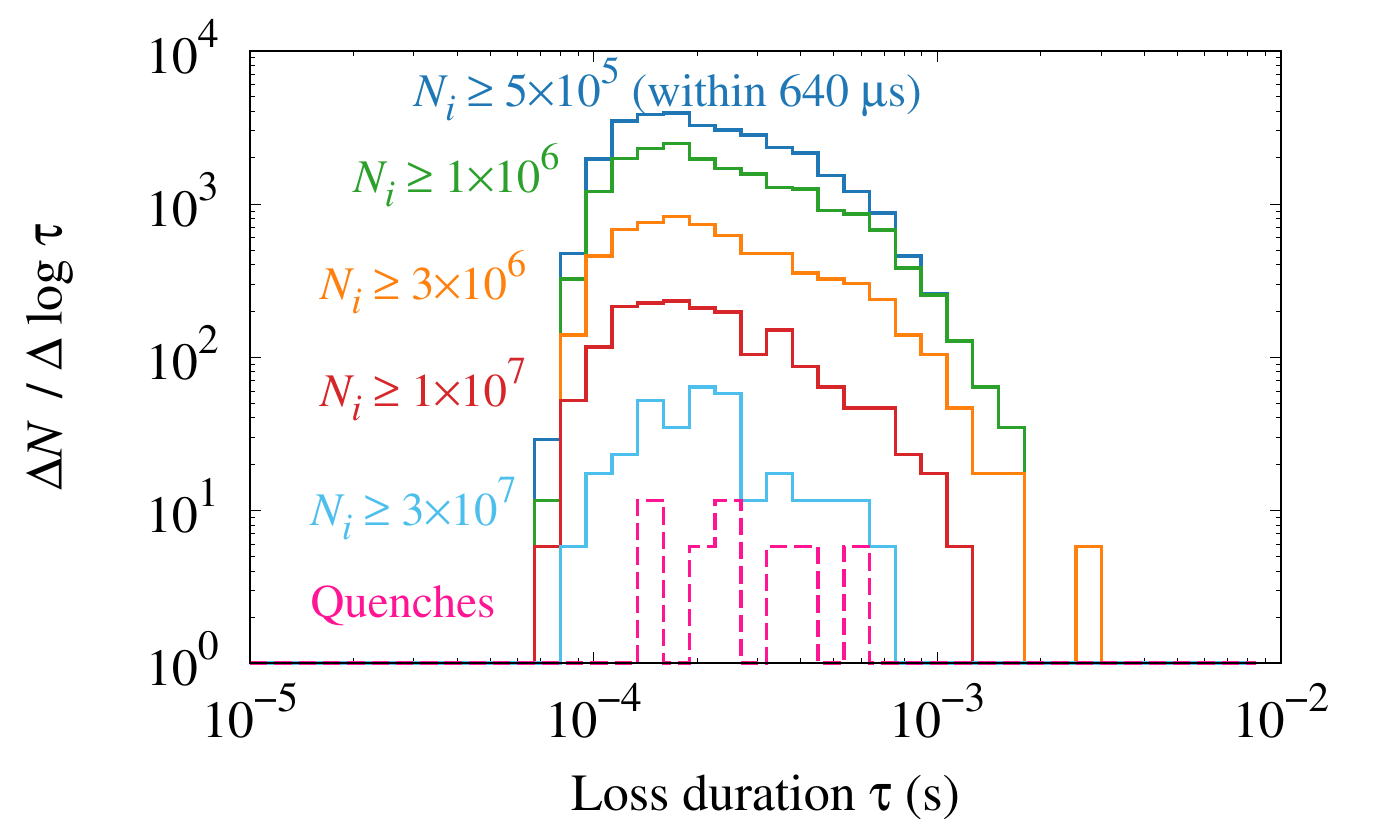}
  \caption[]{
    Reconstructed experimental distributions of dust events as a function of collisions per event (top), maximum collision rate (center), and loss duration (bottom). The figure considers about $N=5500$ loss events, which were recorded in the LHC arcs during 6.5~TeV proton operation in run~II. In addition, one dust-induced quench at 6.39~TeV and one quench in the dispersion suppressor are included (at 6.5~TeV). The two bottom figures show different subsets of events exceeding a minimum  number of collisions indicated by the vertical lines in the top figure. Events leading to a quench are represented by the dashed histograms. } 
\label{fig:eventsvscoll}
\end{figure}

Figure~\ref{fig:eventsvscoll} presents the obtained distributions of inelastic collisions per event $N_i$, peak collision rates $R_i$, and loss durations $\tau$. The different $R_i$ and $\tau$ histograms represent subsets of events, where a minimum number of collisions $N_i$ was exceeded. The results show that events with a higher peak collision rate $R_i$ also yield a higher number of integral collisions $N_i$. Considering the delayed signal registration in BLMs, the actual $R_i$ values can be higher. In case of a fast event, which lasts 40~$\mu$s, $R_i$ is underestimated by a factor of 2.5 since only 40\% of the dose is registered within such a time interval. For faster events, the systematic error can be more significant due to the intrinsic time resolution of BLMs. This concerns however only a fraction of events. In a majority of the considered cases, the underestimation of $R_i$ is estimated to be less than a factor of 2.5 since the events last longer than 40~$\mu$s. 

The large majority of events shown in Fig.~\ref{fig:eventsvscoll} did not cause any disruption of LHC operation, i.e. neither a beam-induced quench nor a BLM abort. Events, which resulted in a quench, are represented by separate histograms (dashed lines). Besides the six dipole quenches in the arcs at 6.5~TeV, the figure also includes one dust-induced quench in the energy ramp (at 6.39~TeV) and one quench in the dispersion suppressor (at 6.5~TeV). In both cases, the quenched magnet was also a bending dipole. In all events where a quench occurred, the estimated number of inelastic proton-dust particle collisions was higher than $\sim6\times10^7$. This corresponds to a fraction of $\sim2\times10^{-7}$ of the maximum beam intensity in run~II, or to a fraction of $5\times10^{-4}$ of the intensity of a single nominal bunch ($1.2\times10^{11}$ protons). The occurrence of a quench depends not only on the number of lost protons, but also on the longitudinal loss location and the resulting energy deposition density in magnet coils. The quench level depends also on the loss duration and on the local temperature margin in the volume heated by the showers \cite{Auchmann2015}. This explains why not all events with $>6\times10^7$ collisions resulted in a quench. The maximum number of collisions observed in a single event was $\sim1.6\times10^8$. 

\begin{figure*}[!t]
  \centering
  \includegraphics[width=0.32\textwidth]{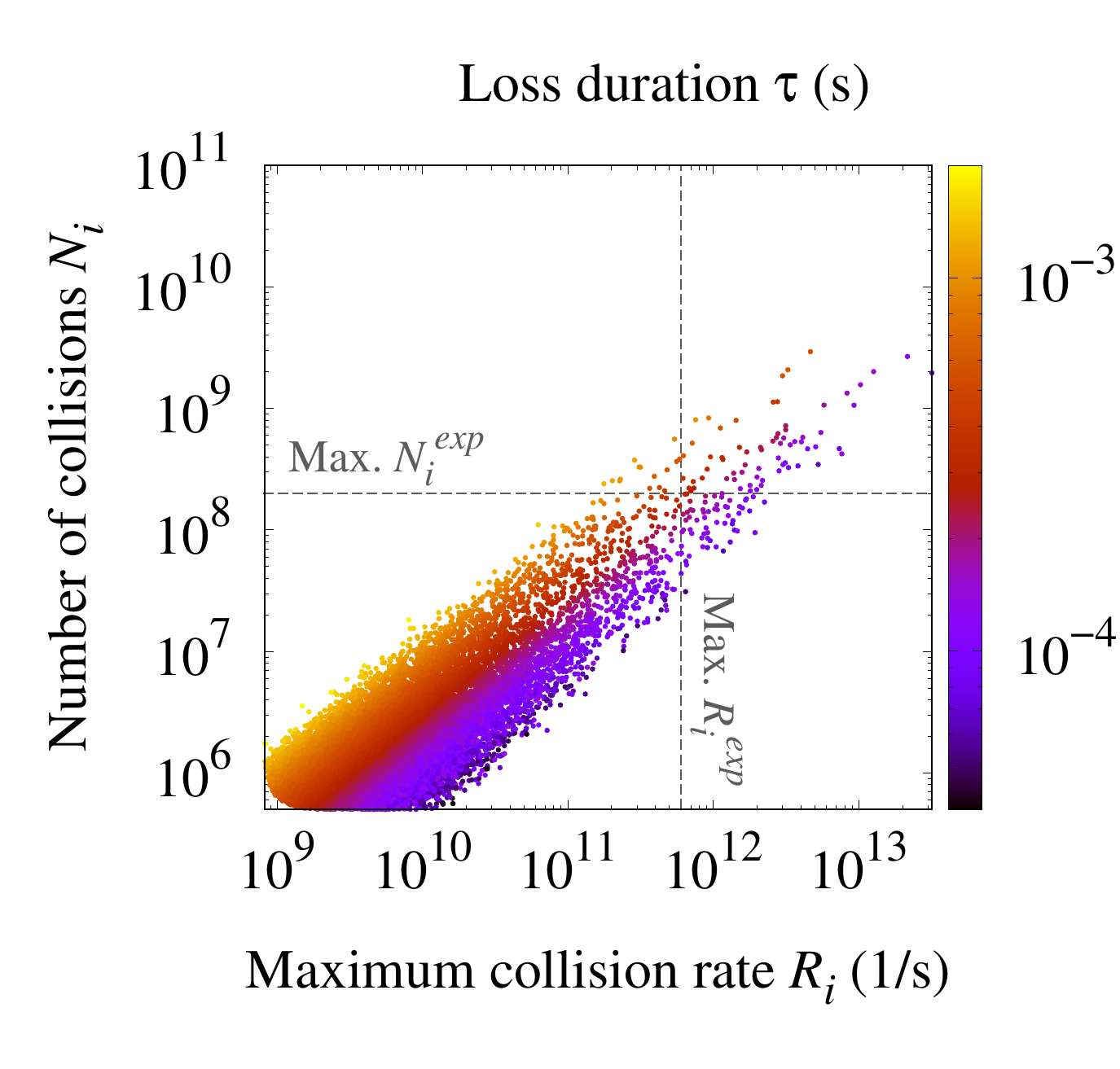}
  \includegraphics[width=0.32\textwidth]{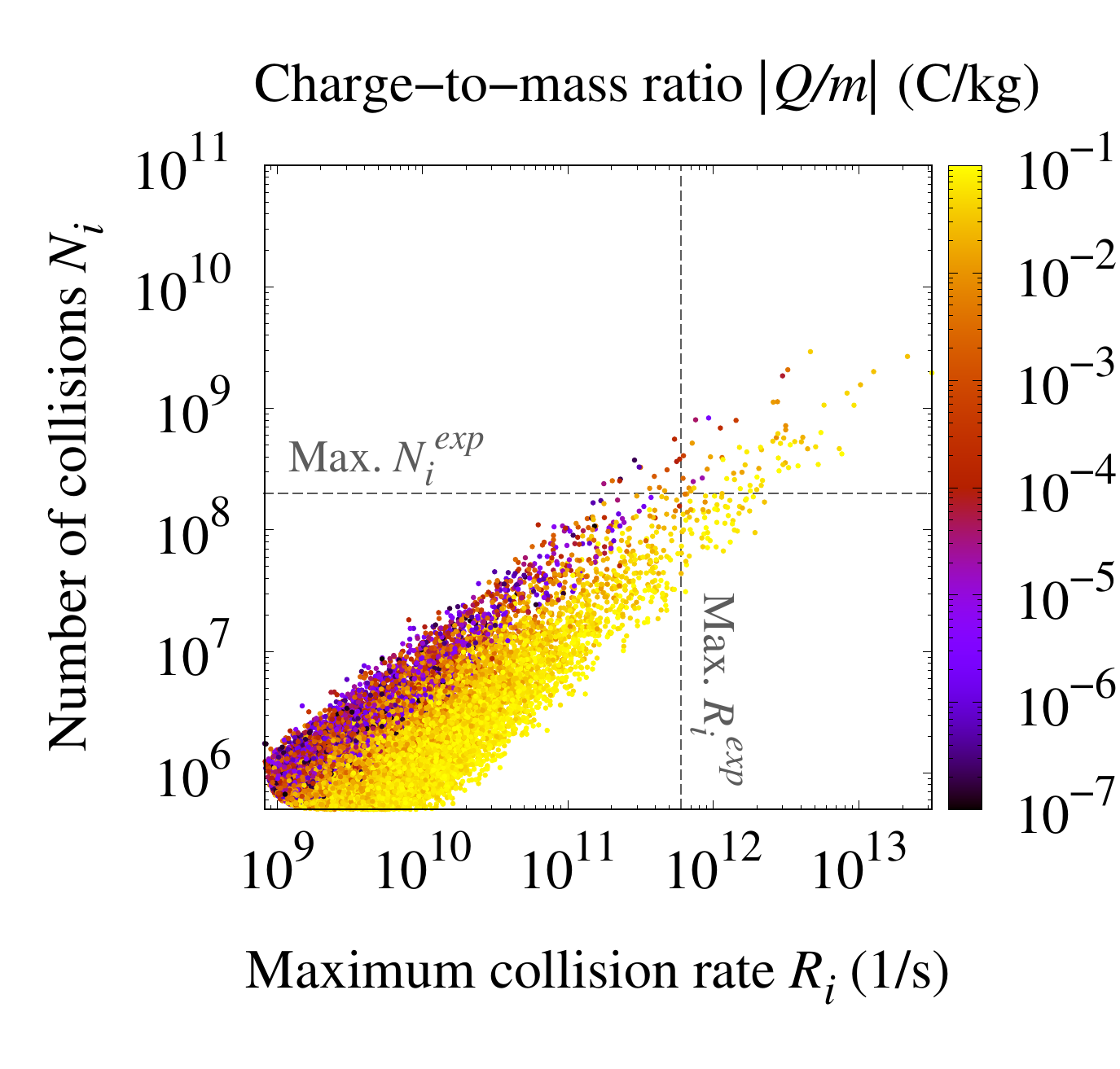}
  \includegraphics[width=0.32\textwidth]{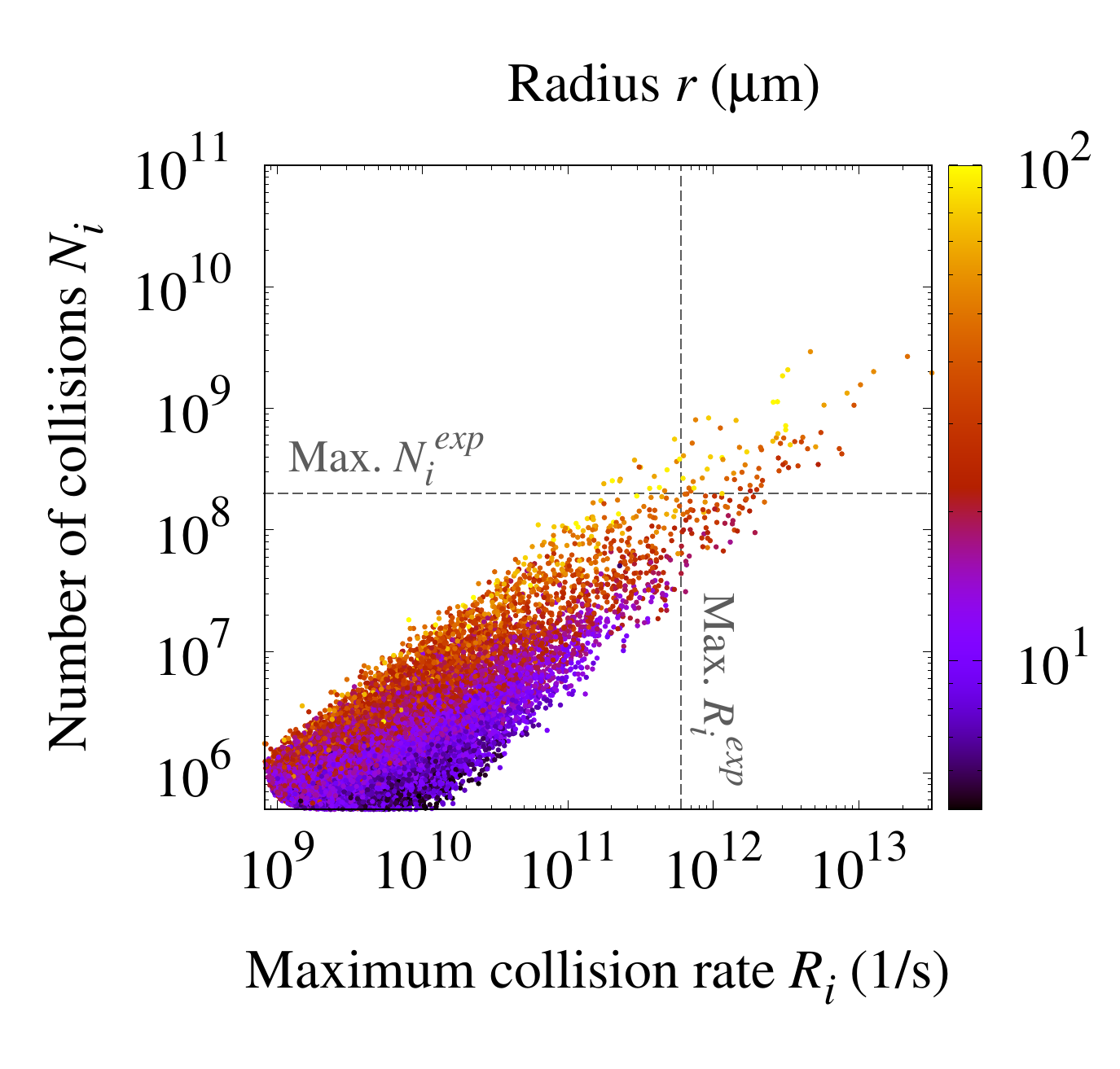}
  \caption[]{Scatter plots of randomly sampled dust events based upon the simulation model described in Ref.~\cite{Lindstrom2020}. The graphs show the number of inelastic proton-dust particle collisions per event versus the maximum collision rate. The color coding indicates the loss duration (left), the charge-to-mass ratio (center) and the radius (right). Only events with $\geq5\times10^5$ collisions within 640~$\mu$s are shown. See text for more details about the simulation parameters. The horizontal and vertical lines indicate the maximum $N_i$ and $R_i$ values observed in the experimental distributions in Fig.~\ref{fig:eventsvscoll}.}
\label{fig:simscattercmratioradius}
\end{figure*}

In case of a magnet quench, the beam abort does not shorten the loss event since the delay until a quench is detected is generally much longer [O(10~ms)] than the duration of dust events. On the other hand, BLM abort triggers can be fast enough to induce the extraction of the beams while the circulating protons still interact with the dust grain. In such cases, the number of beam-dust particle collisions could have been higher if the event would have been unperturbed. The histograms in Fig.~\ref{fig:eventsvscoll} include 17 events, where the BLM abort thresholds were exceeded (14 events without quench and three events where a quench developed despite the BLM abort trigger). We estimate that in four out of these 17 cases, $N_i$ could have possibly exceeded $1.6\times10^8$ collisions if the beams would not have been extracted. In another four to six cases, we can neither exclude that the abort trigger shortened the events, but the number of collisions would have likely stayed below $1.6\times10^8$ collisions. In the remaining seven to nine cases, the time profiles of the events suggest that the losses had already ceased at the moment of beam extraction \cite{Auchmann2015a}.

\subsection{Constraints on the properties of dust particles}

The reconstructed $N_i$, $R_i$ and $\tau$ distributions can be used to constrain some of the properties of dust particles. A similar approach was adopted in the study of event rise times \cite{Lindstrom2020}, where evidence of a negative pre-charge of dust particles was found. The amount of negative pre-charge, together with the composition and mass of a dust grain, are the key quantities, which govern the induced beam losses. Using the latest simulation model as described in Ref.~\cite{Lindstrom2020}, a random sample of loss events was generated for comparison with the distributions presented in the previous section. For simplicity, the dust particles were assumed to be of spherical shape. The volumes $V$ were sampled according to a $1/V^2$ distribution. As discussed in Ref.~\cite{Baer2013}, this describes well the measured distribution of dust particle sizes in the magnet test hall. It was also hypothesized in \cite{Baer2013} that such a distribution can explain the $1/D^2$ BLM dose distribution although there is no unique relationship between $V$ and $D$. The minimum and maximum radii adopted in the volume sampling were 5~$\mu$m and 100~$\mu$m, respectively. An analysis of dust samples showed that an abundance of particulates with $r<5$~$\mu$m is present in the vacuum chamber \cite{Grob2019}, but these are estimated to be irrelevant for the comparison with the measurement sample. Such small dust grains are still expected to cause copious smaller events. In many cases, the beam losses would, however, be too small to be detected by the BLMs.

Other assumptions about the dust particle properties were similar to the ones used in Ref.~\cite{Lindstrom2020}, which are briefly summarized in the following. The dust particle composition was randomly selected from four chemical elements (C, Al, Si, Cu) since the actual composition may vary from event to event. This accounts for the observation that dust particles of different chemical composition are present in the vacuum system, which can possibly be explained by different fabrications methods and by different materials used for vacuum system components \cite{Grob2019}. The dust samples collected in run~II showed also the presence of other chemical elements \cite{Grob2019}, but the four selected materials are considered to be a representative subset of the actual dust constituents. For simplicity, no mixtures of multiple elements were considered and each element was sampled with equal probability. The dust particles were assumed to carry a negative charge when entering the beam. The charge-to-mass ratio $|Q/m|$ was sampled from a log-uniform distribution between 10$^{-7}$~C/kg and 10$^{-1}$~C/kg, albeit different subintervals were studied as discussed below. The dust particles were assumed to be initially attached to the top surface of the beam screen. The horizontal offset of the initial position with respect to the beam was assumed to be uniformly distributed up to two millimeters from the center, which is sufficient to cover the horizontal extent of the beam. 

Figure~\ref{fig:simscattercmratioradius} presents scatter plots of events sampled within the defined parameter space. The two axes show the maximum collision rate $R_i$ and integral number of collisions per event $N_i$, respectively. Only events which generated more than $5\times10^5$ collisions within 640~$\mu$s are shown. The color coding indicates the distribution of loss durations, charge-to-mass ratios and radii. The loss duration was calculated in the same way as for the experimental data. For this purpose, the simulated collision rate profiles were discretized in time, using 80~$\mu$s intervals. The time window with the highest fraction of $N_i$ was used for the calculation of $\tau$. The figures illustrate that for a given number of collisions $N_i$, the peak collision rate and hence the loss duration can vary by more than one order of magnitude. The highest peak collision rate and therefore the shortest loss duration for a given $N_i$ can be attributed to smaller dust particles with a high $|Q/m|$ ratio. The figure also shows that the simulation results include events which exceed the maximum $N_i$ and $R_i$ values observed in the experimental distribution (see horizontal and vertical lines in Fig.~\ref{fig:simscattercmratioradius}). The dust properties hence need to be further constrained.

\begin{figure*}[!ht]
  \centering
  \includegraphics[width=0.98\textwidth]{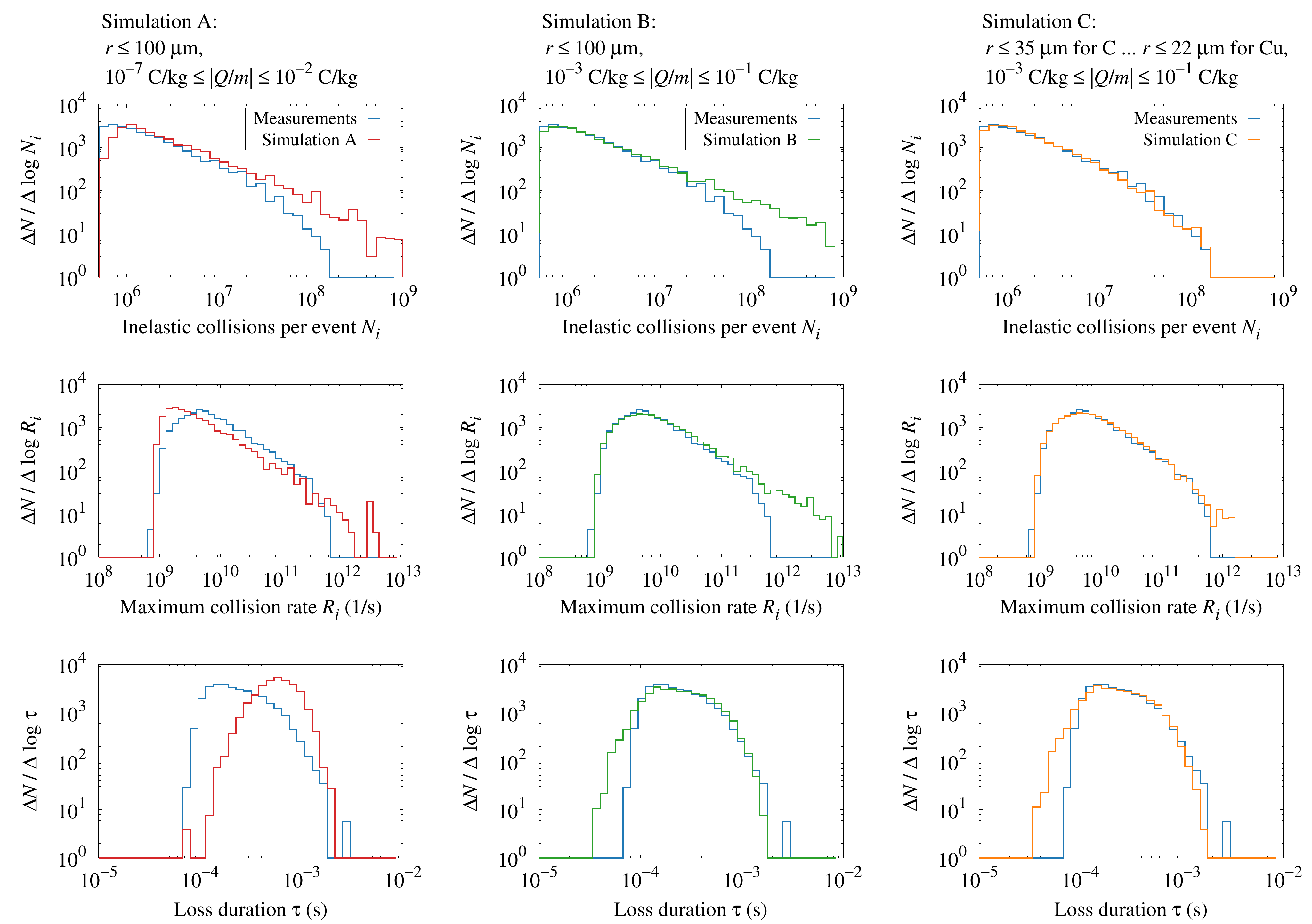}
  \caption[]{Measured and simulated distributions of dust-induced loss events as a function of collisions per event (top), maximum collision rate (center), and loss duration (bottom). The measurements are the same as in Fig.~\ref{fig:simscattercmratioradius}, including about $N=5500$ events. The simulation results correspond to different intervals of $|Q/m|$ and different maximum radii (see top of the figure). }
\label{fig:lossfeatvssim}
\end{figure*}

Figure~\ref{fig:lossfeatvssim} compares measured and simulated distributions of $N_i$, $R_i$ and $\tau$, for events with $\ge5\times10^5$ collisions within 640~$\mu$s. The distribution of beam intensities was modeled to be the same as in the measurement sample to remove any bias due to the intensity dependence of the considered quantities. About two third of the events occurred at a stored beam intensity between $1\times10^{14}$ and $2.5\times10^{14}$ protons. The figure includes different simulation sub-samples from Fig.~\ref{fig:simscattercmratioradius}, with different constraints on $|Q/m|$ and $r$. Each simulation sample was normalized such that the total number of events was the same as in the measurements. The first sample (simulation ``A'') excludes high charge-to-mass ratios ($|Q/m|>10^{-2}$~C/kg), whereas the second sample (simulation ``B'') excludes lower charge-to-mass ratios ($|Q/m|<10^{-3}$~C/kg). In both cases, the upper radius was 100~$\mu$m. In the third sample (simulation ``C''), the range of charge-to-mass ratios was the same as in ``B'', but the mass of dust particles was restricted to less than 0.4~$\mu$g. This corresponds to a maximum radius of 35~$\mu$m for spherical carbon grains and 22~$\mu$m for spherical copper grains. As discussed in the following, sample ``C'' reproduces best the features of the measured distributions.

The comparison between simulated and measured $R_i$ and $\tau$ distributions suggests that $|Q/m|$ must have been higher than 10$^{-2}$~C/kg in at least a fraction of the events, otherwise the most probable peak loss rate (loss duration) is underestimated (overestimated). The most probable $R_i$ and $\tau$ values can be best reproduced if $|Q/m|$ ranges from 10$^{-3}$~C/kg to 10$^{-1}$~C/kg, as in the samples ``B'' and ``C''. This finding is in good agreement with the charge-to-mass ratio inferred from the rise time of loss events~\cite{Lindstrom2020}, where $|Q/m|$ was found to be higher than $5\times10^{-3}$~C/kg. 

The results also show that the assumed distribution of dust volumes ($1/V^2$) reproduces well the measured distribution of $N_i$, but overestimates the actual number of higher-loss events ($N_i>3\times10^7$) if $r_{max}=100~\mu$m (as in sample ``A'' and ``B''). In particular, the simulation predicts events with $N_i>1.6\times10^8$ collisions, which are absent in the measurement sample, apart from possibly four cases where the BLM abort trigger shortened the event. A similar observation can be made for the distribution of maximum collision rates, which extends to higher values in the simulations than in the measurements. The absence of larger events in the measurements was already observed in BLM signal distributions in run~I \cite{Baer2013}. This observation was solely attributed to the shortening of events due BLM abort triggers, which, according to our analysis, does not apply to the run~II data. It is also unlikely that the absence of large events in the run~II measurements is due to the limited sample size. Assuming that dust volumes are distributed as $1/V^2$ and that $r_{max}$=100~$\mu$m, then one would have expected with 95\% confidence to observe $\geq$12 events with $N_i>1.6\times10^8$ in run~II. A more likely explanation is that the maximum radius of dust grains which interact with the beam is smaller than 100~$\mu$m, or that the propensity of reaching a high charge-to-mass ratio $|Q/m|$ diminishes for larger dust grains. As shown in Fig.~\ref{fig:lossfeatvssim}, the measured distribution can be well reproduced by sample ``C'', where the mass of dust particles was limited to 0.4~$\mu$g. Such dust grains can still generate a higher number of losses $N_i$ than observed in the measurements, but the likelihood of such events diminishes significantly. The remaining discrepancies between simulated and measured $\tau$ and $R_i$ distributions can be explained by the limited time resolution of the measurements and by delayed charge collection in BLMs. 

The reason that dust particles heavier than 0.4~$\mu$g would not interact with the beam could be twofold. One possibility could be that more massive dust particles are less susceptible to being detached from the cold beam screen in presence of the beams. An alternative explanation could be that the assumed $1/V^2$ distribution does not correctly describe the population of larger dust grains in the vacuum system. If the population of dust grains with larger radii is sufficiently smaller than a $1/V^2$ distribution, then the absence of higher-loss events could be of statistical nature. The dust samples extracted from a dipole in run~II showed that dust grains with masses larger than 0.4~$\mu$g are present in the vacuum system \cite{Grob2019}. However, the analysis also indicated that the size distribution can vary depending on the location where the dust sample was taken. We can therefore not entirely exclude that the global size distribution deviates from the assumed $1/V^2$ dependence. This might in particular apply to dust grains adhering to the top surface of the beam screen.

The measured distributions can also be reproduced in a similar way as in Fig.~\ref{fig:lossfeatvssim} by constraining the absolute dust charge $|Q|$ instead of the mass. This assumption implies that larger dust grains cannot reach the same charge-to-mass ratio $|Q/m|$ as smaller ones. Considering a maximum $|Q|$ of $2\times 10^{-11}$~C, similar curves would be obtained as if the mass $m$ is limited to 0.4~$\mu$g. Based on these findings, it cannot be established with certainty if the population of larger dust particles is smaller than expected, if larger dust particles are not released into the beam even if they possess a high charge-to-mass ratio, or if the maximum charge dust particles can acquire is limited. The latter would reduce their ability to induce higher losses. 

\section{Dependence of proton losses on beam parameters}
\label{sec:beamparamdep}

The correlation between dust-induced beam losses and beam parameters has been studied both experimentally (based on data from LHC run~I operation) \cite{Nebot2011,Holzer2010} and through simulations \cite{Fuster2011,Rowan2016}. However, no direct comparison of simulations and measurements was carried out so far, mainly due to a lack of a common observable. The experimental studies primarily relied on BLM signals, which could not be compared directly with the number of inelastic collisions. In this section, we use the experimental and simulated distributions from the previous section to study the dependence of observables on the beam intensity and transverse emittance. Since the reconstructed distributions contain the full event population above a given $N_i$ threshold, an absolute comparison can be performed. As dust properties, we assume the same as in simulation ``C'' (see Fig.~\ref{fig:lossfeatvssim}), i.e. $|Q/m|$ ranging from $10^{-3}$~C/kg to $10^{-1}$~C/kg and $m$ being smaller than $0.4~\mu$g.

\subsection{Correlation between $N_i$, $R_i$ and $\tau$ and the beam intensity}

\begin{figure}[!t]
  \centering
  \includegraphics[width=0.48\textwidth]{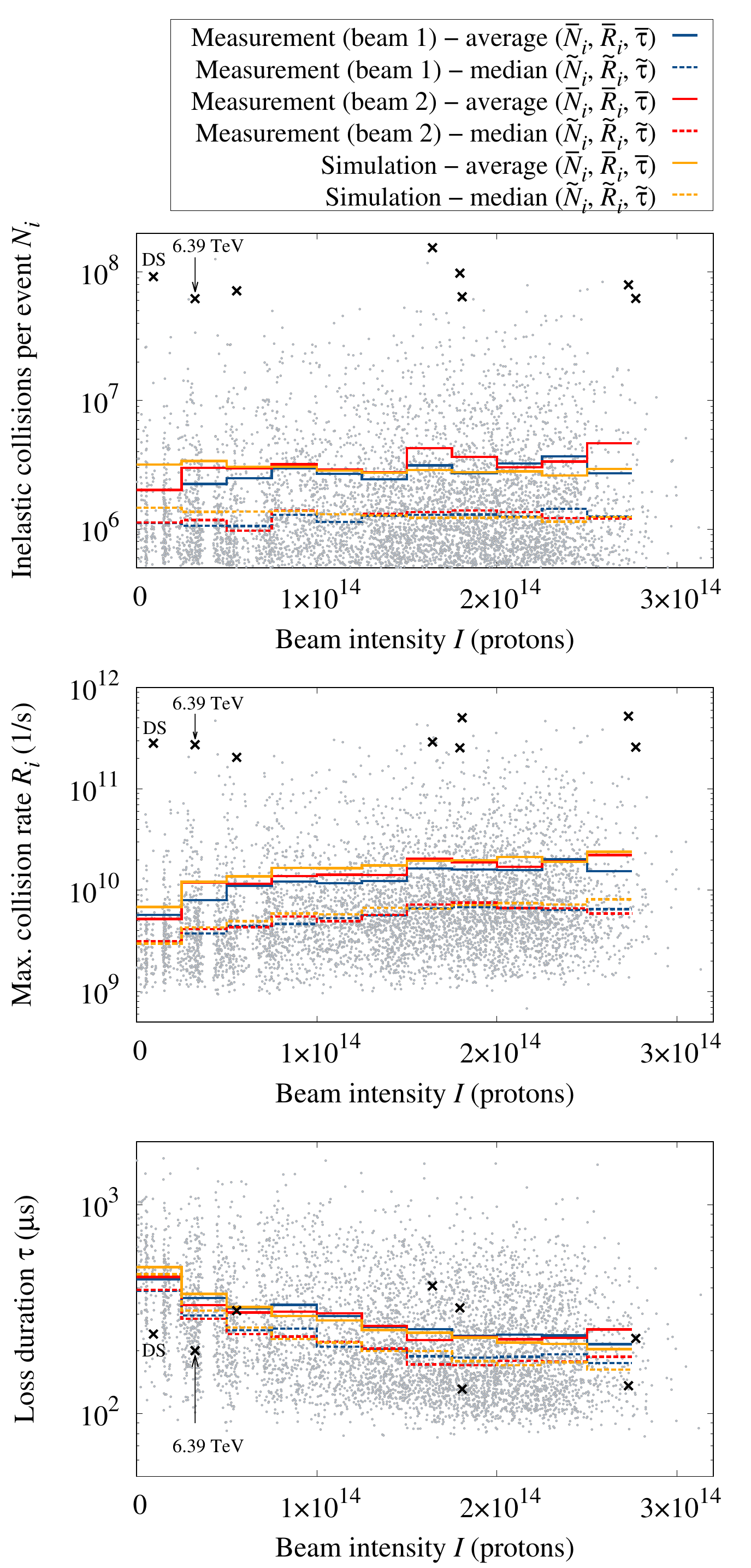}
  \caption[]{Number of inelastic proton-nucleus collisions (top), maximum collision rate (center) and loss duration (bottom) of dust particle events as a function of the beam intensity. Every dot represents a loss event reconstructed from BLM measurements in run~II. All considered events occurred at 6.5~TeV, except one event at 6.39~TeV. Only events with more than 5$\times$10$^{5}$ collisions within 640~$\mu$s were considered. The crosses indicate events, which resulted in a magnet quench. The average and median values are indicated by the solid and dashed lines, respectively (beam 1 in blue and beam 2 in red). The yellow lines represent the simulation results (simulation ``C'' in Fig.~\ref{fig:lossfeatvssim}).}
\label{fig:collvsint}
\end{figure}

Figure~\ref{fig:collvsint} shows scatter plots of $N_i$, $R_i$ and $\tau$ as a function of the beam intensity $I$. Each point represents a reconstructed loss event from run~II. The measurements are the same as in Fig.~\ref{fig:lossfeatvssim}, i.e. only events with more than $5\times10^{5}$ collisions within 640~$\mu$s were considered. The data points exhibit a large spread at all beam intensities. Relatively high losses and loss rates can occur already with low-intensity beams. In particular, three of the eight dust-induced quenches in run~II occurred at $I=9.4\times10^{12}$, $3.2\times10^{13}$ and $5.6\times10^{13}$ protons, which corresponds to 3\%, 11\% and 18\% of the maximal intensity achieved in run~II, respectively. 

The solid and dashed lines represent the average and median values of $N_i$, $R_i$ and $\tau$ for beam~1 (blue) and beam~2 (red), respectively. The intensity dependence is very similar for both beams. While the average and median peak collision rates, $\overline{R}_i$ and $\widetilde{R}_i$, show a gradual increase as a function of $I$, the opposite trend can be observed for the average and median loss duration, $\overline{\tau}$ and $\widetilde{\tau}$. The behaviour is compatible with the rather weak dependence of $\overline{N}_i$ and $\widetilde{N}_i$ on $I$. The decrease of $\overline{\tau}$ as a function of $I$ has already been observed in previous empirical studies, which were based on data from 3.5~TeV operation in run~I \cite{Nebot2011,Holzer2010}; the same trend has also been predicted by the theoretical model \cite{Fuster2011}. In the present data set, the average duration $\overline{\tau}$ is found to be around 500~$\mu$s for low beam intensities and decreases to $\sim$200~$\mu$s for intensities higher than $1.5\times10^{14}$ protons. 

The simulation results, represented by the yellow lines, reproduce well the absolute intensity dependence of the different observables, in particular the increase of the maximum collision rate and the decrease of the loss duration. The simulation results suggest that the average number of collisions is rather independent of $I$, while a very slight increase can be observed in the measurements. The absolute agreement between simulations and measurements is nevertheless satisfactory.

Previous simulation studies \cite{Fuster2011,Auchmann2014,Rowan2015,Rowan2016} suggested that $\overline{R}_i$ would decline with increasing $I$ if the gravitational force would be the sole force acting on the dust particle before entering the beam, i.e. if dust particles were not pre-charged. This decline of $\overline{R}_i$ with $I$ can be explained by the smaller degree of ionization needed to reach the point of repulsion in presence of higher-intensity beams, which then results in a reduced inelastic collision rate at the turning point of the dust particle trajectory. Observing the opposite trend in the present data provides another strong indication that dust particles are initially attracted by the beam. The increasing attraction negatively pre-charged dust particles experience in case of higher-intensity beams outweighs the aforementioned effect, i.e. the dust particles can penetrate deeper into the beam and hence the peak collision rate increases as a function of $I$.

The results also suggest that the charge-to-mass ratio of dust particles interacting with the beam is similar at all beam intensities. The physical mechanism, which causes dust particles to acquire a negative charge, is beyond the scope of this study and is investigated in another paper \cite{Belanger2021}. The results nonetheless suggest that, if the charging mechanism depends on the presence of the proton beams, the amount of charge picked up by the dust particle does not or only moderately depend on the number of circulating protons. 

\subsection{Correlation between $N_i$, $R_i$ and $\tau$ and the transverse beam emittance}

\begin{figure}[!t]
  \centering
  \includegraphics[width=0.48\textwidth]{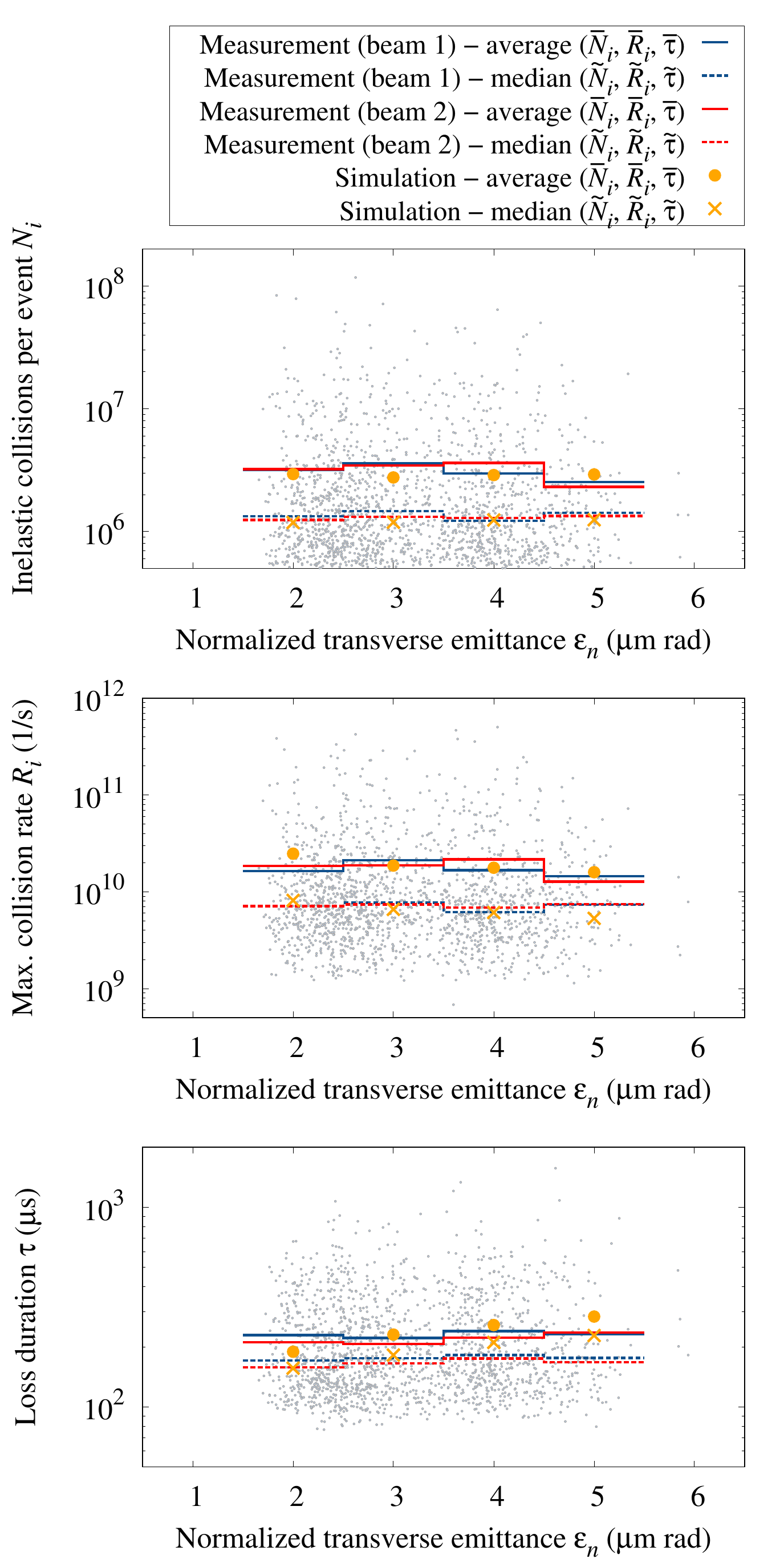}
  \caption[]{Number of inelastic proton-nucleus collisions (top), maximum collision rate (center) and loss duration (bottom) of dust particle events as a function of the normalized transverse beam emittance. The measurements are a subset of the data shown in Fig.~\ref{fig:collvsint}. The average and median values of the measurements are indicated by the solid and dashed lines, respectively (beam~1 in blue and beam~2 in red). Only events from 2016-2018 at beam intensities $I\geq1.5\times 10^{14}$ protons were considered. Simulations (yellow symbols) were performed for discrete emittance values only (average values are given by circles, median values are given by crosses).}
\label{fig:collvsemit}
\end{figure}

Figure~\ref{fig:collvsemit} shows scatter plots of $N_i$, $R_i$ and $\tau$ as a function of the normalized transverse emittance $\varepsilon_n$. Like in the previous section, the average and median values of the measured distributions are represented by solid and dashed lines, respectively. Only events which occurred at beam intensities $I\geq1.5\times 10^{14}$ protons were considered due to the weaker dependence on $I$. This shall reduce any intensity-related bias when studying the dependence on the transverse emittance. The emittance at the time of each the dust event was approximated by
\begin{equation}
  \varepsilon_n = \varepsilon_{n,i} + k \times \Delta t,
  \label{eq:emit}
\end{equation}
where $\varepsilon_{n,i}$ is the averaged convoluted emittance at the start of stable proton-proton collisions for physics data taking, $k$ is the emittance growth rate and $\Delta t$ is the time which elapsed between the start of stable collisions and the dust particle event. The initial emittance $\varepsilon_{n,i}$ was reconstructed considering the luminosity measurements in the ATLAS and CMS experiments and represents the convoluted emittance in the horizontal and vertical plane, averaged over all bunches. Experimental studies showed that the emittance growth in 2018 was about $k=0.07-0.08~ (0.04-0.06)~\mu$m/h in the horizontal (vertical) plane \cite{Papadopoulou2019}. In this study, we consider $k=0.05~\mu$m/h as the average convoluted emittance growth for all years, accepting that this represents only a rough estimate of the actual emittance at the moment of dust particle events. Describing the emittance growth by a single constant value is considered sufficient for identifying trends. Since Eq.~(\ref{eq:emit}) relies on luminosity measurements, only dust events during stable beam collisions were included in this analysis, and only for the years 2016-2018 of run~II.

The yellow symbols in Fig.~\ref{fig:collvsemit} represent the simulation results. The emittances were sampled from discrete values ($\varepsilon_n=2,3,4,5~\mu$m). The corresponding beam size was calculated by using $\beta$- and dispersion functions at randomly selected dust particle positions inside a standard arc cell \cite{Auchmann2014}. The dust particles were assumed to be uniformly distributed along the cell. The minimum and maximum $\beta$-functions were 30~m and 180~m, respectively. 

In general, the simulations and measurements are in good agreement, although some discrepancies are visible. For example, the simulation predicts a slight increase of $\overline{\tau}$ and $\widetilde{\tau}$ with $\varepsilon_n$, while no clear trend can be identified in the measurements. The latter can at least partially be attributed to fluctuations in the experimental data resulting from the limited sample size. The results nevertheless indicate that the average and median $N_i$, $R_i$ and $\tau$ values vary at most by a few ten percent in the considered emittance interval between 2~$\mu$m\,rad and 5~$\mu$m\,rad.

\section{Proton losses at higher beam intensities}
\label{sec:losseshighint}

\begin{figure}[!t]
\centering
\includegraphics[width=0.48\textwidth]{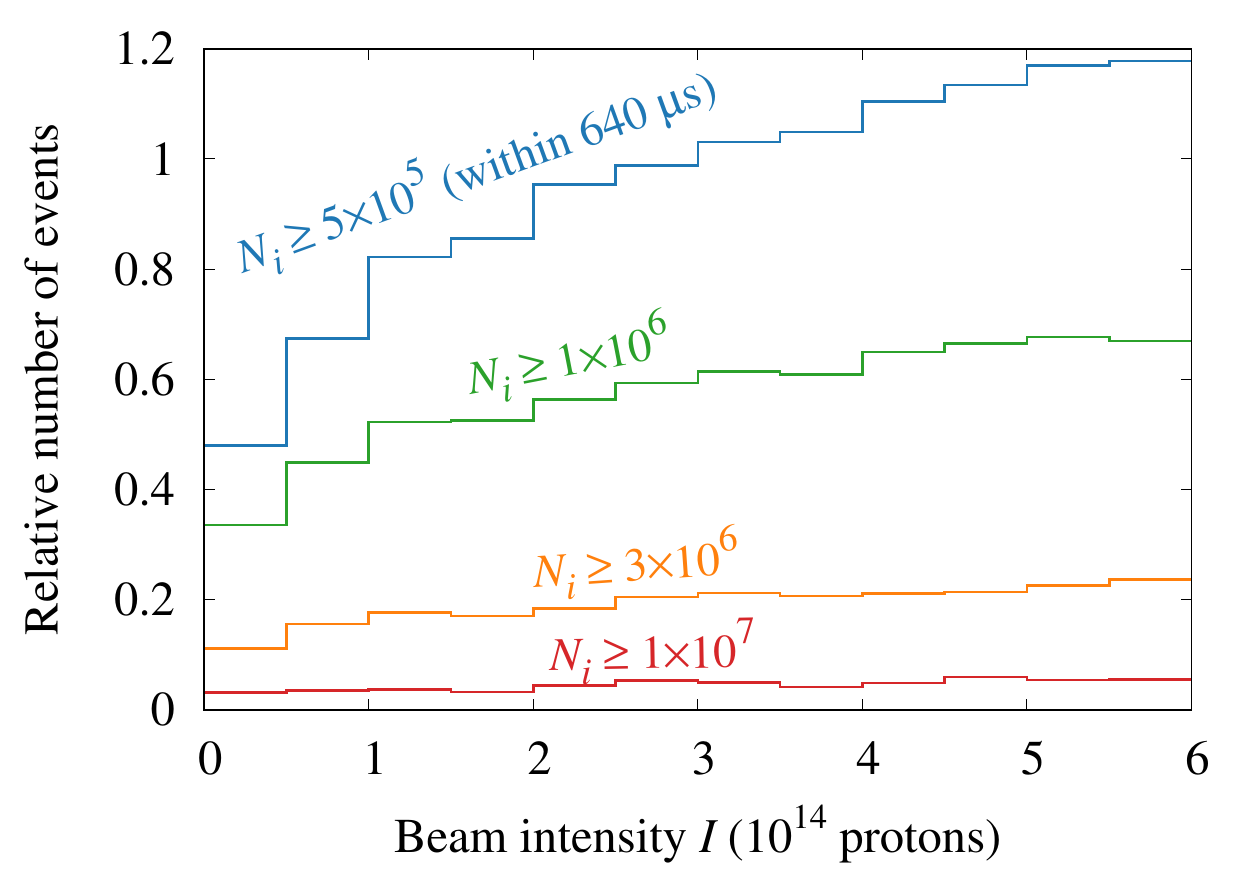}
\caption[minimum]{Simulation predictions of the relative number of dust events exceeding a given number of collisions $N_i$. The absolute number of events was assumed to be independent of the intensity $I$.}
\label{fig:relnumbereventsvsint}
\end{figure}

The proton beam intensity in the HL-LHC era will be twice as high as in run~II, whereas beam emittances as small as the nominal HL-LHC emittance (2.5~$\mu$m\,rad) have already been achieved in run~II (see Table~\ref{tab:beamparam}). A possible worsening of dust-induced losses can therefore be mainly expected due to the higher number of circulating protons. Based on the dust properties found in Sec.~\ref{sec:beammacroparticleint}, we can derive predictions about the distribution of $N_i$, $R_i$ and $\tau$ at higher intensities. As in the previous sections, we consider it instructive to study the distribution of events exceeding a minimum number of collisions $N_i$. The dust dynamics simulations predict that the number of events above a given $N_i$ threshold grows with increasing $I$. This is illustrated in Fig.~\ref{fig:relnumbereventsvsint}, which shows the relative increase of events for different $N_i$ thresholds. The results were arbitrarily normalized to the number of events, which result in more than $5\times 10^5$ collisions ($\Delta t =640~\mu s$) at a beam intensity of $I=3\times 10^{14}$ protons. As indicated by the blue line in Fig.~\ref{sec:beammacroparticleint}, the number of events with $N_i$ above $5\times 10^5$ collisions increases by about 17\% at HL-LHC beam intensities ($\sim 6\times 10^{14}$ protons) compared to the highest beam intensity in run~II ($\sim$$3\times 10^{14}$ protons). 

\begin{figure}[!t]
  \centering
  \includegraphics[width=0.48\textwidth]{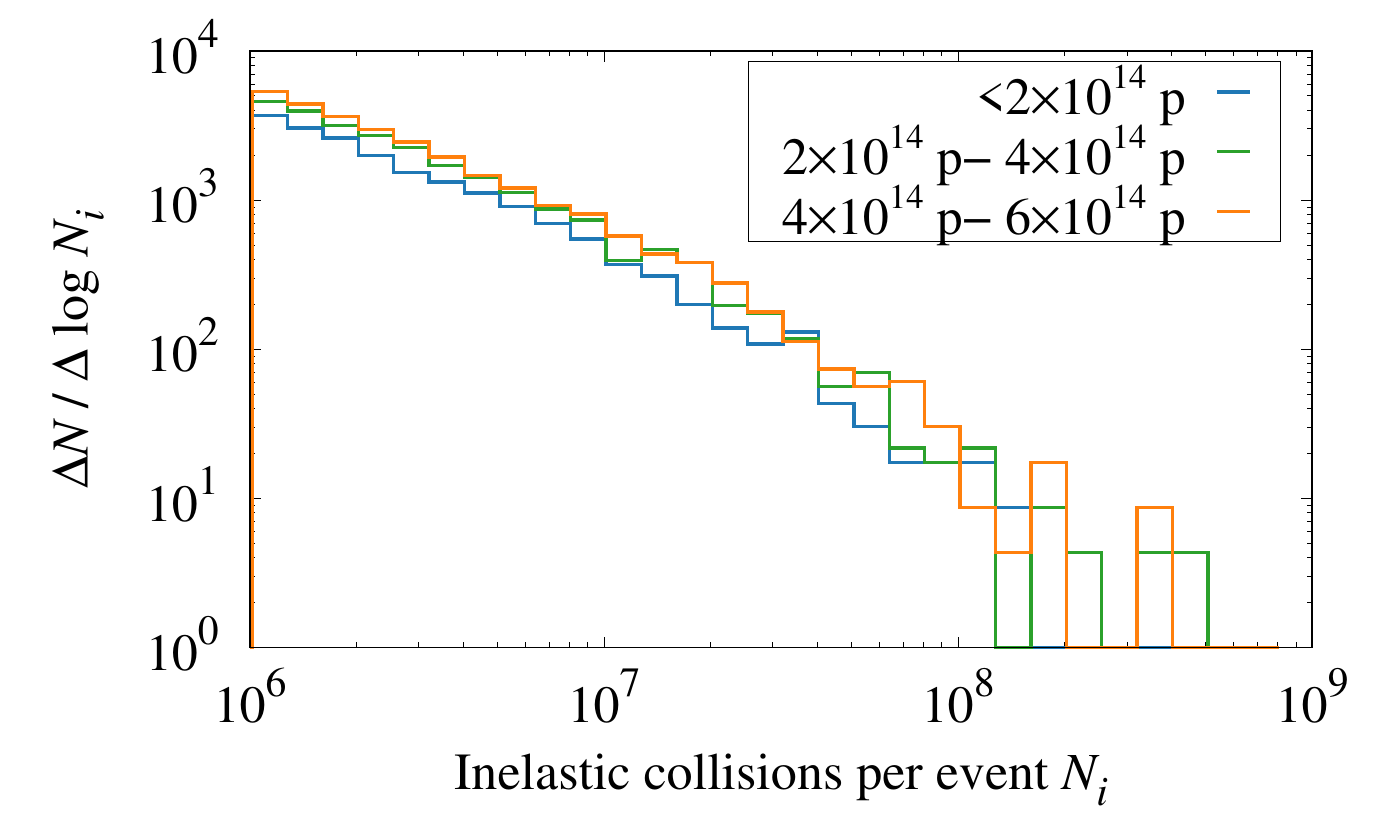}\\
  \includegraphics[width=0.48\textwidth]{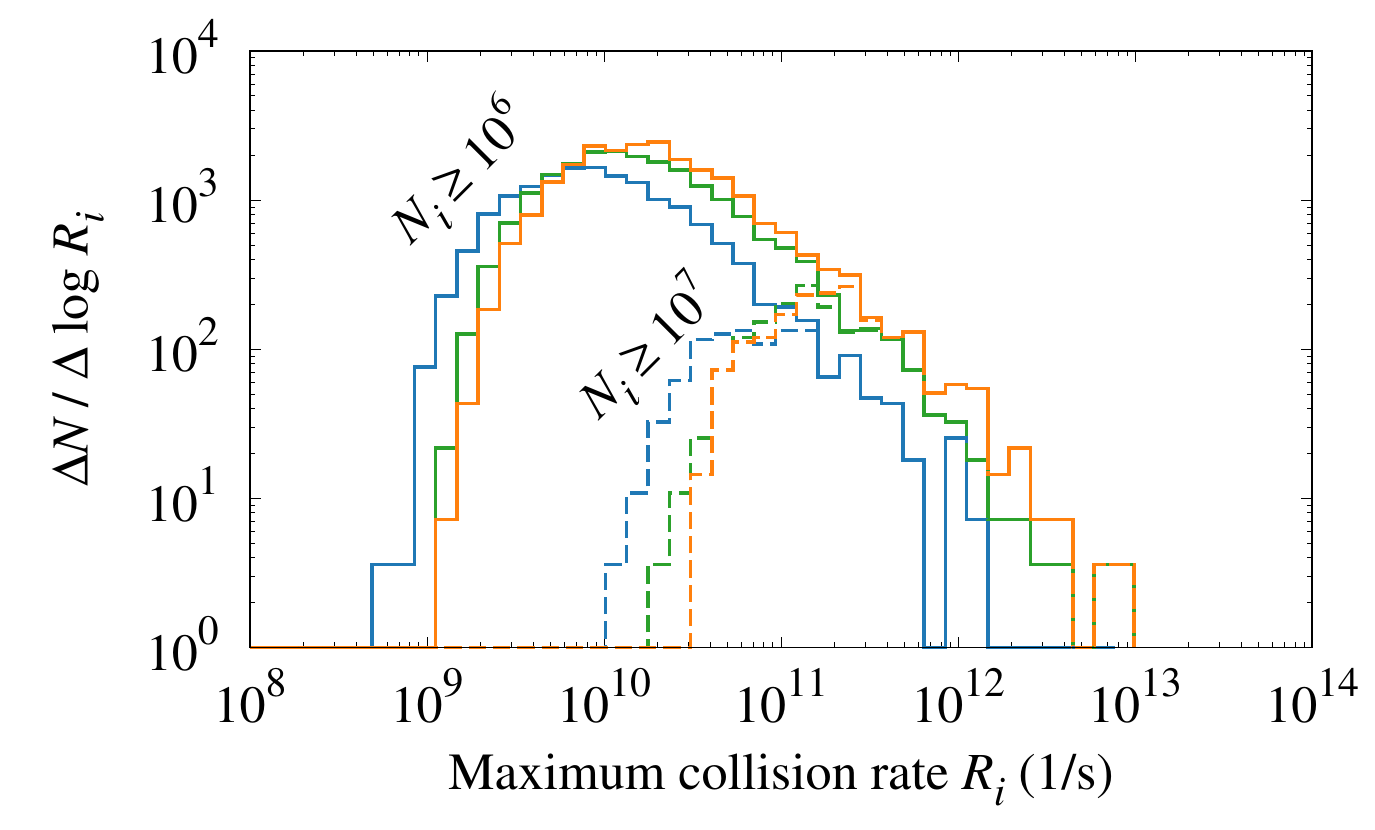}\\
  \includegraphics[width=0.48\textwidth]{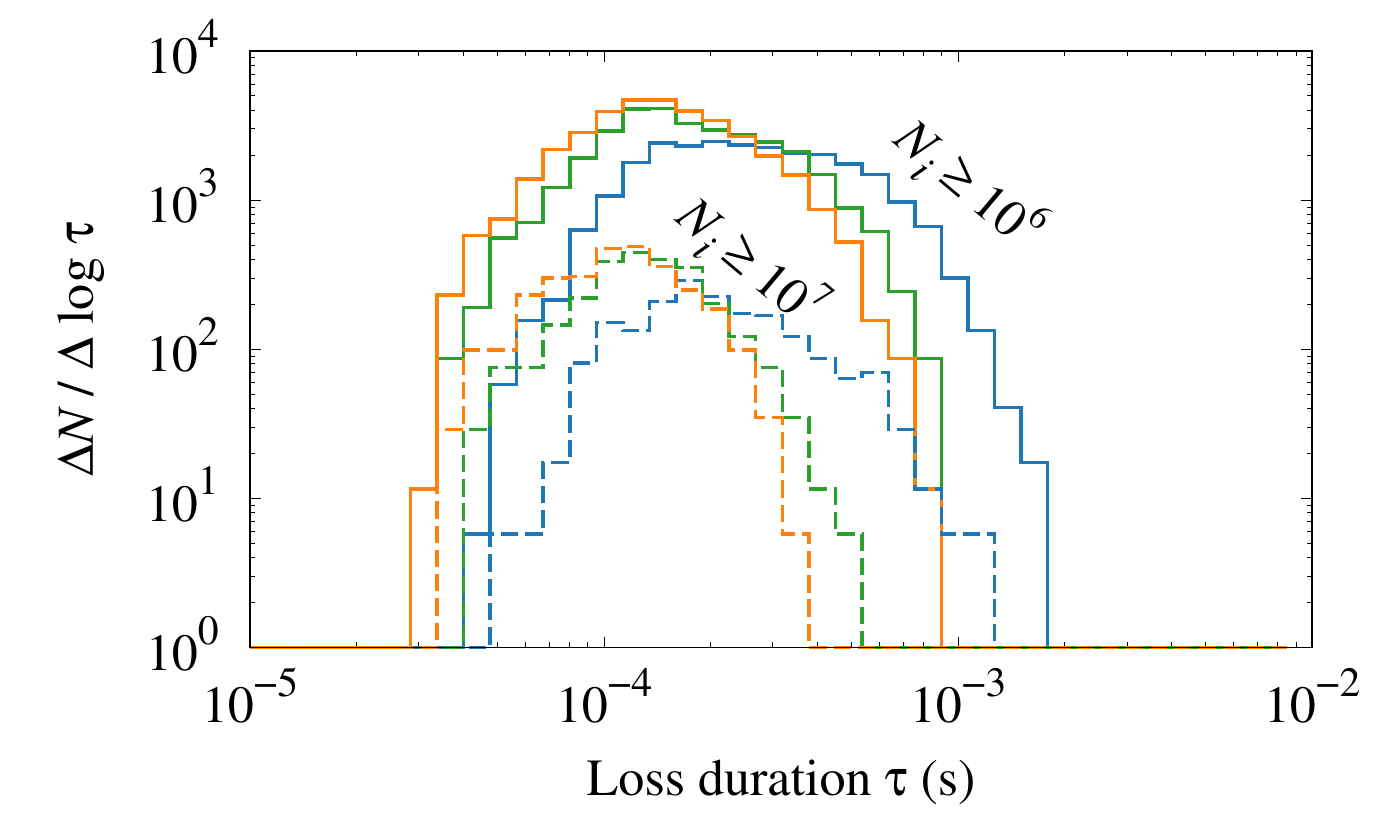}
  \caption[]{Simulated distributions of $N_i$, $R_i$ and $\tau$ for different beam intensity intervals; $N$ indicates the number of events. The different $R_i$ and $\tau$ distributions correspond to different minimum $N_i$ thresholds as indicated in the figures ($N_i\geq 10^6$: solid histograms, $N_i\geq 10^7$: dashed histograms).}
\label{fig:disthigherint}
\end{figure}

Figure~\ref{fig:disthigherint} presents the distribution of $N_i$, $R_i$ and $\tau$ for three different beam intensity intervals, up to the maximum intensity expected in HL-LHC. The stored beam intensity declines throughout physics fills due to the proton burn-off in the experiments and due to beam losses in the collimation system. Dust events may therefore exhibit different characteristics depending on the time when they occur in a fill. The intensity intervals in the figure do not represent specific  operational scenarios, but provide a general comparison of event characteristics for different intensity regimes. For each interval, an equal number of events was generated. Two different $N_i$ thresholds were adopted for the $R_i$ and $\tau$ distributions, $N_i=10^6$ inelastic collisions (solid histograms) and $N_i=10^7$ inelastic collisions (dashed histograms). With increasing intensity, the distribution of peak collision rates shifts towards higher values while the $\tau$ distribution shifts towards smaller values. This behaviour could already be observed in the average and median values of the distributions shown in Sec.~\ref{sec:beamparamdep}. Comparing the distributions for the two upper intensity intervals, the characteristics of dust particle events are not expected to get significantly worse in the HL-LHC era. The minimum energy deposition density for inducing a quench decreases for shorter heating times, but this decrease is small for $\tau<10^{-4}$~s \cite{Auchmann2015,Bottura2019}. The shift of the $\tau$ distribution is therefore not expected to considerably increase the risk of quenches.

As shown in Sec.~\ref{sec:beammacroparticleint}, the minimum number of inelastic proton-dust collisions for inducing a quench at 6.5~TeV was $6\times10^7$. This loss threshold will decrease in 7~TeV operation because of the reduced quench margin and the higher energy density deposited in coils. Electro-thermal model calculations suggest that the quench level of arc dipoles will decrease by about 20\% to 30\% for fast beam losses between 10$^{-4}$~s and 10$^{-3}$~s \cite{Bottura2019}. On the other hand, \textsc{FLUKA} simulations show that the average energy density per proton lost increases by about 14\% because of the higher particle energy and the narrower angular distribution of secondary collision products. We therefore estimate that about $4\times$10$^7$ inelastic collisions can lead to a quench at 7~TeV. The simulation results in Fig.~\ref{fig:disthigherint} show that, independently of the considered $I$ interval, the number of events with $N_i\geq 4\times$10$^7$ collisions is about a factor of two higher than the number of events with $N_i\geq 6\times$10$^7$ collisions. This suggests that an increase of the beam energy from 6.5~TeV to 7~TeV will presumably have a larger impact on the likelihood of dust-induced quenches than the increase of the beam intensity. 

Apart from the higher fraction of events which can induce a quench, the impact of dust events in future runs will strongly depend on the frequency of events. The physical mechanism, which governs the rate at which dust particles are released into the beam, still lacks a theoretical understanding. Much of what is known about the occurrence of these events derives from experimental observations. In particular, it was observed that the rate of dust events gradually decreases during operational years, while the situation could deteriorate after winter shutdowns \cite{Baer2013}. The largest increase of the event rate occurred after the two year-long shutdown between run~I and run~II. The increase of the event rate after future shutdowns will therefore be decisive for the number of dust-induced quenches.

\section{Conclusions}
\label{sec:conclusions}

The interaction of the LHC proton beams with dust particles was the dominant source of beam-induced magnet quenches in the second physics run of the LHC (6.5~TeV). Besides these detrimental events, thousands of harmless beam-dust particle encounters have been observed in the cryogenic arcs every year. In this paper, we studied the characteristics of dust events by reconstructing the number of inelastic nuclear collisions between the beam and dust grains. The paper demonstrated that the experimental distribution of peak collision rates and loss durations can be consistently reproduced by dust dynamics simulations if dust particles are negatively pre-charged and if the charge-to-mass ratio $|Q/m|$ ranges from $10^{-3}$~C/kg to $10^{-1}$~C/kg. This finding is in good agreement with recent studies of the rise time of loss profiles \cite{Lindstrom2020}. 

The assumed range of charge-to-mass ratios also describes well the beam intensity-dependence of observables. In particular, the observed increase of the average peak collision rate with the number of circulating protons can only be explained if dust particles carry sufficient negative charge when entering the beam. The opposite trend would be expected if dust particles were neutral or only weakly charged ($|Q/m|<10^{-5}$~C/kg). Within the resolution achieved in this study, we could not find any evidence that the charge acquired by dust particles depends significantly on the number of circulating protons, i.e. the experimental data could be well reproduced by assuming the same $|Q/m|$ distribution at all beam intensities, up to the maximum intensity achieved in run~II ($\sim 3\times 10^{14}$ protons).

The paper also illustrated that dust events with more than $1.6\times10^8$ inelastic collisions were absent in the run~II measurements, although events with higher losses were theoretically not excluded by the simulation model. The absence of these events can have several reasons. It could indicate that the mass of dust particles interacting with the beam is limited ($\lesssim$0.4~$\mu$g), either because more massive dust grains are less likely to detach from the cold aperture or because the population of larger dust particles is overestimated by the assumed $1/V^2$ volume distribution. In the latter case, the absence of higher-loss events could be of statistical nature. An alternative hypothesis is that dust particles can only acquire a limited negative charge ($|Q|\lesssim2\times 10^{-11}$~C), which would constrain the attainable charge-to-mass ratio for larger dust grains. This in turn diminishes their ability to penetrate deeper into the beam and hence their ability to induce higher losses. A better understanding of the charging mechanism and the detachment of the dust grains from the cold beam screen is needed to conclusively explain the absence of higher-loss events. Independently of the underlaying mechanism, the absence of such loss events was one of the main reason why dust particles did not have a higher impact on the operational performance in run~II. 

Dust-induced loss events remain a concern for future LHC runs. This applies in particular to the HL-LHC era, when the operational energy will be raised to 7~TeV and the stored beam intensity will increase by a factor of two compared to previous years. We showed that already $6\times10^7$ inelastic proton-dust particle collisions could lead to a dipole quench at 6.5~TeV. The loss threshold for inducing a quench will decrease to 4$\times$10$^7$ inelastic collisions at nominal energy (7~TeV) because of the reduced quench margin and the higher energy density induced by the particle showers in superconducting coils. Hence, smaller dust particles will have the ability to provoke a quench. The data from run~II suggests that the number of detrimental events could increase by about a factor of two to three compared to 6.5~TeV, even if the total rate of events remains unchanged.

The simulations also showed that the number of dust events exceeding a certain particle loss threshold increases with beam intensity. This increase is, however, expected to be less than 20\% at the HL-LHC design intensity ($\sim 6\times 10^{14}$ protons) compared to the maximum beam intensity in run~II. Hence, the increase in beam energy is expected to be the most important parameter change compared to previous runs. The dust dynamics simulations also showed that the time profiles of dust events will become shorter at higher intensities, while the peak collision rate will increase. 

Dust-induced quenches may also pose a challenge for the operational efficiency of future high-energy hadron colliders like the FCC-hh, which are being designed to operate with unprecedented stored beam energies. The results presented in this paper, which provide a first quantitative analysis of dust-induced beam losses in a high-intensity hadron storage ring, can serve as a basis for estimating the impact of dust events in such future colliders. The studies demonstrated that the nature of beam-dust particle interactions can be well reproduced by modeling the ionization of dust particles and the resulting repelling force exerted on the dust grain. A similar simulation approach can therefore be used to predict dust-induced beam losses at higher beam energies and intensities. Such studies can provide critical information about the tolerable dust contamination in the vacuum system of energy-frontier hadron machines.

\section*{Acknowledgements}

We would like to thank the LHC operations team, in particular M.~Albert, and the LHC Beam Instrumentation group, in particular C. Zamantzas, for their support concerning the monitoring software. 

\bibliography{refs}

\begin{thebibliography}{51}%
\makeatletter
\providecommand \@ifxundefined [1]{%
 \@ifx{#1\undefined}
}%
\providecommand \@ifnum [1]{%
 \ifnum #1\expandafter \@firstoftwo
 \else \expandafter \@secondoftwo
 \fi
}%
\providecommand \@ifx [1]{%
 \ifx #1\expandafter \@firstoftwo
 \else \expandafter \@secondoftwo
 \fi
}%
\providecommand \natexlab [1]{#1}%
\providecommand \enquote  [1]{``#1''}%
\providecommand \bibnamefont  [1]{#1}%
\providecommand \bibfnamefont [1]{#1}%
\providecommand \citenamefont [1]{#1}%
\providecommand \href@noop [0]{\@secondoftwo}%
\providecommand \href [0]{\begingroup \@sanitize@url \@href}%
\providecommand \@href[1]{\@@startlink{#1}\@@href}%
\providecommand \@@href[1]{\endgroup#1\@@endlink}%
\providecommand \@sanitize@url [0]{\catcode `\\12\catcode `\$12\catcode
  `\&12\catcode `\#12\catcode `\^12\catcode `\_12\catcode `\%12\relax}%
\providecommand \@@startlink[1]{}%
\providecommand \@@endlink[0]{}%
\providecommand \url  [0]{\begingroup\@sanitize@url \@url }%
\providecommand \@url [1]{\endgroup\@href {#1}{\urlprefix }}%
\providecommand \urlprefix  [0]{URL }%
\providecommand \Eprint [0]{\href }%
\providecommand \doibase [0]{https://doi.org/}%
\providecommand \selectlanguage [0]{\@gobble}%
\providecommand \bibinfo  [0]{\@secondoftwo}%
\providecommand \bibfield  [0]{\@secondoftwo}%
\providecommand \translation [1]{[#1]}%
\providecommand \BibitemOpen [0]{}%
\providecommand \bibitemStop [0]{}%
\providecommand \bibitemNoStop [0]{.\EOS\space}%
\providecommand \EOS [0]{\spacefactor3000\relax}%
\providecommand \BibitemShut  [1]{\csname bibitem#1\endcsname}%
\let\auto@bib@innerbib\@empty
\bibitem [{\citenamefont {Br{\"u}ning}\ \emph {et~al.}(2004)\citenamefont
  {Br{\"u}ning}, \citenamefont {Collier}, \citenamefont {Lebrun}, \citenamefont
  {Myers}, \citenamefont {Ostojic}, \citenamefont {Poole},\ and\ \citenamefont
  {Proudlock}}]{Bruning2004}%
  \BibitemOpen
  \bibfield  {author} {\bibinfo {author} {\bibfnamefont {O.~S.}\ \bibnamefont
  {Br{\"u}ning}}, \bibinfo {author} {\bibfnamefont {P.}~\bibnamefont
  {Collier}}, \bibinfo {author} {\bibfnamefont {P.}~\bibnamefont {Lebrun}},
  \bibinfo {author} {\bibfnamefont {S.}~\bibnamefont {Myers}}, \bibinfo
  {author} {\bibfnamefont {R.}~\bibnamefont {Ostojic}}, \bibinfo {author}
  {\bibfnamefont {J.}~\bibnamefont {Poole}},\ and\ \bibinfo {author}
  {\bibfnamefont {P.}~\bibnamefont {Proudlock}},\ }\href
  {https://doi.org/http://dx.doi.org/10.5170/CERN-2004-003-V-1} {\emph
  {\bibinfo {title} {{LHC Design Report}}}},\ CERN Yellow Reports: Monographs\
  (\bibinfo  {publisher} {CERN},\ \bibinfo {address} {Geneva},\ \bibinfo {year}
  {2004})\BibitemShut {NoStop}%
\bibitem [{\citenamefont {Baer}\ \emph {et~al.}(2011)\citenamefont {Baer},
  \citenamefont {Barnes}, \citenamefont {Goddard}, \citenamefont {Holzer},
  \citenamefont {Jimenez}, \citenamefont {Lechner}, \citenamefont {Mertens},
  \citenamefont {Nebot Del~Busto}, \citenamefont {Nordt}, \citenamefont
  {Uythoven}, \citenamefont {Velghe}, \citenamefont {Wenninger},\ and\
  \citenamefont {Zimmermann}}]{Baer2011}%
  \BibitemOpen
  \bibfield  {author} {\bibinfo {author} {\bibfnamefont {T.}~\bibnamefont
  {Baer}}, \bibinfo {author} {\bibfnamefont {M.}~\bibnamefont {Barnes}},
  \bibinfo {author} {\bibfnamefont {B.}~\bibnamefont {Goddard}}, \bibinfo
  {author} {\bibfnamefont {E.~B.}\ \bibnamefont {Holzer}}, \bibinfo {author}
  {\bibfnamefont {J.~M.}\ \bibnamefont {Jimenez}}, \bibinfo {author}
  {\bibfnamefont {A.}~\bibnamefont {Lechner}}, \bibinfo {author} {\bibfnamefont
  {V.}~\bibnamefont {Mertens}}, \bibinfo {author} {\bibfnamefont
  {E.}~\bibnamefont {Nebot Del~Busto}}, \bibinfo {author} {\bibfnamefont
  {A.}~\bibnamefont {Nordt}}, \bibinfo {author} {\bibfnamefont
  {J.}~\bibnamefont {Uythoven}}, \bibinfo {author} {\bibfnamefont
  {B.}~\bibnamefont {Velghe}}, \bibinfo {author} {\bibfnamefont
  {J.}~\bibnamefont {Wenninger}},\ and\ \bibinfo {author} {\bibfnamefont
  {F.}~\bibnamefont {Zimmermann}},\ }\bibfield  {title} {\bibinfo {title}
  {{UFOs in the LHC}},\ }in\ \href@noop {} {\emph {\bibinfo {booktitle}
  {Proceedings of the 2nd International Particle Accelerator Conference,
  TUPC137, San Sebasti\'{a}n, Spain}}}\ (\bibinfo  {publisher} {EPS-AG,
  Spain},\ \bibinfo {year} {2011})\ pp.\ \bibinfo {pages}
  {1347--1349}\BibitemShut {NoStop}%
\bibitem [{\citenamefont {Nebot}\ \emph {et~al.}(2011)\citenamefont {Nebot},
  \citenamefont {Velghe}, \citenamefont {Holzer}, \citenamefont {Dehning},
  \citenamefont {Nordt}, \citenamefont {Sapinski}, \citenamefont {Emery},
  \citenamefont {Zamantzas}, \citenamefont {Effinger}, \citenamefont {Marsili},
  \citenamefont {Wenninger}, \citenamefont {Baer}, \citenamefont {Schmidt},
  \citenamefont {Yang}, \citenamefont {Zimmerman},\ and\ \citenamefont
  {Fuster}}]{Nebot2011}%
  \BibitemOpen
  \bibfield  {author} {\bibinfo {author} {\bibfnamefont {E.}~\bibnamefont
  {Nebot}}, \bibinfo {author} {\bibfnamefont {B.}~\bibnamefont {Velghe}},
  \bibinfo {author} {\bibfnamefont {E.}~\bibnamefont {Holzer}}, \bibinfo
  {author} {\bibfnamefont {B.}~\bibnamefont {Dehning}}, \bibinfo {author}
  {\bibfnamefont {A.}~\bibnamefont {Nordt}}, \bibinfo {author} {\bibfnamefont
  {M.}~\bibnamefont {Sapinski}}, \bibinfo {author} {\bibfnamefont
  {J.}~\bibnamefont {Emery}}, \bibinfo {author} {\bibfnamefont
  {C.}~\bibnamefont {Zamantzas}}, \bibinfo {author} {\bibfnamefont
  {E.}~\bibnamefont {Effinger}}, \bibinfo {author} {\bibfnamefont
  {A.}~\bibnamefont {Marsili}}, \bibinfo {author} {\bibfnamefont
  {J.}~\bibnamefont {Wenninger}}, \bibinfo {author} {\bibfnamefont
  {T.}~\bibnamefont {Baer}}, \bibinfo {author} {\bibfnamefont {R.}~\bibnamefont
  {Schmidt}}, \bibinfo {author} {\bibfnamefont {Z.}~\bibnamefont {Yang}},
  \bibinfo {author} {\bibfnamefont {F.}~\bibnamefont {Zimmerman}},\ and\
  \bibinfo {author} {\bibfnamefont {N.}~\bibnamefont {Fuster}},\ }\bibfield
  {title} {\bibinfo {title} {{Analysis of fast losses in the LHC with the BLM
  system}},\ }in\ \href@noop {} {\emph {\bibinfo {booktitle} {Proceedings of
  the 2nd International Particle Accelerator Conference, TUPC136, San
  Sebasti\'{a}n, Spain}}}\ (\bibinfo  {publisher} {EPS-AG, Spain},\ \bibinfo
  {year} {2011})\ pp.\ \bibinfo {pages} {1344--1346}\BibitemShut {NoStop}%
\bibitem [{\citenamefont {Busto}\ \emph {et~al.}(2013)\citenamefont {Busto},
  \citenamefont {Baer}, \citenamefont {Day}, \citenamefont {Dehning},
  \citenamefont {Holzer}, \citenamefont {Lechner}, \citenamefont {Schmidt},
  \citenamefont {Wenninger}, \citenamefont {Zamantzas}, \citenamefont
  {Zerlauth}, \citenamefont {Zimmermann},\ and\ \citenamefont
  {Hempel}}]{Nebot2012}%
  \BibitemOpen
  \bibfield  {author} {\bibinfo {author} {\bibfnamefont {E.~N.~D.}\
  \bibnamefont {Busto}}, \bibinfo {author} {\bibfnamefont {T.}~\bibnamefont
  {Baer}}, \bibinfo {author} {\bibfnamefont {F.}~\bibnamefont {Day}}, \bibinfo
  {author} {\bibfnamefont {B.}~\bibnamefont {Dehning}}, \bibinfo {author}
  {\bibfnamefont {E.}~\bibnamefont {Holzer}}, \bibinfo {author} {\bibfnamefont
  {A.}~\bibnamefont {Lechner}}, \bibinfo {author} {\bibfnamefont
  {R.}~\bibnamefont {Schmidt}}, \bibinfo {author} {\bibfnamefont
  {J.}~\bibnamefont {Wenninger}}, \bibinfo {author} {\bibfnamefont
  {C.}~\bibnamefont {Zamantzas}}, \bibinfo {author} {\bibfnamefont
  {M.}~\bibnamefont {Zerlauth}}, \bibinfo {author} {\bibfnamefont
  {F.}~\bibnamefont {Zimmermann}},\ and\ \bibinfo {author} {\bibnamefont
  {Hempel}},\ }\bibfield  {title} {\bibinfo {title} {Detection of {Unidentified
  Falling Objects at LHC}},\ }in\ \href@noop {} {\emph {\bibinfo {booktitle}
  {Proceedings of the 52nd ICFA Advanced Beam Dynamics Workshop on
  High-Intensity and High-Brightness Hadron Beams (HB2012), TUO1C04, Beijing,
  China}}}\ (\bibinfo  {publisher} {JACoW, Geneva, Switzerland},\ \bibinfo
  {year} {2013})\ pp.\ \bibinfo {pages} {305--309}\BibitemShut {NoStop}%
\bibitem [{\citenamefont {Baer}\ \emph
  {et~al.}(2012{\natexlab{a}})\citenamefont {Baer}, \citenamefont {Barnes},
  \citenamefont {Cerutti}, \citenamefont {Ferrari}, \citenamefont {Fuster},
  \citenamefont {Garrel}, \citenamefont {Goddard}, \citenamefont {Holzer},
  \citenamefont {Jackson}, \citenamefont {Lechner}, \citenamefont {Mertens},
  \citenamefont {Misiowiec}, \citenamefont {Nebot}, \citenamefont {Nordt},
  \citenamefont {Uythoven}, \citenamefont {Vlachoudis}, \citenamefont
  {Wenninger}, \citenamefont {Zamantzas},\ and\ \citenamefont
  {Zimmermann}}]{Baer2012}%
  \BibitemOpen
  \bibfield  {author} {\bibinfo {author} {\bibfnamefont {T.}~\bibnamefont
  {Baer}}, \bibinfo {author} {\bibfnamefont {M.~J.}\ \bibnamefont {Barnes}},
  \bibinfo {author} {\bibfnamefont {F.}~\bibnamefont {Cerutti}}, \bibinfo
  {author} {\bibfnamefont {A.}~\bibnamefont {Ferrari}}, \bibinfo {author}
  {\bibfnamefont {N.}~\bibnamefont {Fuster}}, \bibinfo {author} {\bibfnamefont
  {N.}~\bibnamefont {Garrel}}, \bibinfo {author} {\bibfnamefont
  {B.}~\bibnamefont {Goddard}}, \bibinfo {author} {\bibfnamefont {E.~B.}\
  \bibnamefont {Holzer}}, \bibinfo {author} {\bibfnamefont {S.}~\bibnamefont
  {Jackson}}, \bibinfo {author} {\bibfnamefont {A.}~\bibnamefont {Lechner}},
  \bibinfo {author} {\bibfnamefont {V.}~\bibnamefont {Mertens}}, \bibinfo
  {author} {\bibfnamefont {M.}~\bibnamefont {Misiowiec}}, \bibinfo {author}
  {\bibfnamefont {E.}~\bibnamefont {Nebot}}, \bibinfo {author} {\bibfnamefont
  {A.}~\bibnamefont {Nordt}}, \bibinfo {author} {\bibfnamefont
  {J.}~\bibnamefont {Uythoven}}, \bibinfo {author} {\bibfnamefont
  {V.}~\bibnamefont {Vlachoudis}}, \bibinfo {author} {\bibfnamefont
  {J.}~\bibnamefont {Wenninger}}, \bibinfo {author} {\bibfnamefont
  {C.}~\bibnamefont {Zamantzas}},\ and\ \bibinfo {author} {\bibfnamefont
  {F.}~\bibnamefont {Zimmermann}},\ }\bibfield  {title} {\bibinfo {title}
  {{UFOs in the LHC: Observations, studies, and extrapolations}},\ }in\
  \href@noop {} {\emph {\bibinfo {booktitle} {Proceedings of the 3rd
  International Particle Accelerator Conference, THPPP086, New Orleans, LA,
  USA}}}\ (\bibinfo  {publisher} {IEEE, Piscataway, NJ},\ \bibinfo {year}
  {2012})\ pp.\ \bibinfo {pages} {3936--3938}\BibitemShut {NoStop}%
\bibitem [{\citenamefont {Baer}(2013)}]{Baer2013}%
  \BibitemOpen
  \bibfield  {author} {\bibinfo {author} {\bibfnamefont {T.}~\bibnamefont
  {Baer}},\ }\emph {\bibinfo {title} {Very Fast Losses of the Circulating LHC
  Beam, their Mitigation and Machine Protection}},\ \href@noop {} {Ph.D.
  thesis},\ \bibinfo  {school} {CERN and Hamburg University} (\bibinfo {year}
  {2013})\BibitemShut {NoStop}%
\bibitem [{\citenamefont {Goddard}\ \emph {et~al.}(2012)\citenamefont
  {Goddard}, \citenamefont {Adraktas}, \citenamefont {Baer}, \citenamefont
  {Barnes}, \citenamefont {Cerutti}, \citenamefont {Ferrari}, \citenamefont
  {Garrel}, \citenamefont {Gerardin}, \citenamefont {Guinchard}, \citenamefont
  {Lechner}, \citenamefont {Masi}, \citenamefont {Mertens}, \citenamefont
  {Mor\'{o}n~Ballester}, \citenamefont {Redaelli}, \citenamefont {Uythoven},
  \citenamefont {Vlachoudis},\ and\ \citenamefont {Zimmermann}}]{Goddard2012}%
  \BibitemOpen
  \bibfield  {author} {\bibinfo {author} {\bibfnamefont {B.}~\bibnamefont
  {Goddard}}, \bibinfo {author} {\bibfnamefont {P.}~\bibnamefont {Adraktas}},
  \bibinfo {author} {\bibfnamefont {T.}~\bibnamefont {Baer}}, \bibinfo {author}
  {\bibfnamefont {M.~J.}\ \bibnamefont {Barnes}}, \bibinfo {author}
  {\bibfnamefont {F.}~\bibnamefont {Cerutti}}, \bibinfo {author} {\bibfnamefont
  {A.}~\bibnamefont {Ferrari}}, \bibinfo {author} {\bibfnamefont
  {N.}~\bibnamefont {Garrel}}, \bibinfo {author} {\bibfnamefont
  {A.}~\bibnamefont {Gerardin}}, \bibinfo {author} {\bibfnamefont
  {M.}~\bibnamefont {Guinchard}}, \bibinfo {author} {\bibfnamefont
  {A.}~\bibnamefont {Lechner}}, \bibinfo {author} {\bibfnamefont
  {A.}~\bibnamefont {Masi}}, \bibinfo {author} {\bibfnamefont {V.}~\bibnamefont
  {Mertens}}, \bibinfo {author} {\bibfnamefont {R.}~\bibnamefont
  {Mor\'{o}n~Ballester}}, \bibinfo {author} {\bibfnamefont {S.}~\bibnamefont
  {Redaelli}}, \bibinfo {author} {\bibfnamefont {J.}~\bibnamefont {Uythoven}},
  \bibinfo {author} {\bibfnamefont {V.}~\bibnamefont {Vlachoudis}},\ and\
  \bibinfo {author} {\bibfnamefont {F.}~\bibnamefont {Zimmermann}},\ }\bibfield
   {title} {\bibinfo {title} {{Transient beam losses in the LHC injection
  kickers from micron scale dust particles}},\ }in\ \href@noop {} {\emph
  {\bibinfo {booktitle} {Proceedings of the 3rd International Particle
  Accelerator Conference, TUPPR092, New Orleans, LA, USA}}}\ (\bibinfo
  {publisher} {IEEE, Piscataway, NJ},\ \bibinfo {year} {2012})\ pp.\ \bibinfo
  {pages} {2044--2046}\BibitemShut {NoStop}%
\bibitem [{\citenamefont {Auchmann}\ \emph
  {et~al.}(2015{\natexlab{a}})\citenamefont {Auchmann}, \citenamefont {Ghini},
  \citenamefont {Grob}, \citenamefont {Iadarola}, \citenamefont {Lechner},\
  and\ \citenamefont {Papotti}}]{Auchmann2015a}%
  \BibitemOpen
  \bibfield  {author} {\bibinfo {author} {\bibfnamefont {B.}~\bibnamefont
  {Auchmann}}, \bibinfo {author} {\bibfnamefont {J.}~\bibnamefont {Ghini}},
  \bibinfo {author} {\bibfnamefont {L.}~\bibnamefont {Grob}}, \bibinfo {author}
  {\bibfnamefont {G.}~\bibnamefont {Iadarola}}, \bibinfo {author}
  {\bibfnamefont {A.}~\bibnamefont {Lechner}},\ and\ \bibinfo {author}
  {\bibfnamefont {G.}~\bibnamefont {Papotti}},\ }\bibfield  {title} {\bibinfo
  {title} {{How to survive an UFO attack?}},\ }in\ \href@noop {} {\emph
  {\bibinfo {booktitle} {Proceedings of the 6th Evian Workshop on LHC Beam
  Operation, Evian Les Bains, France}}}\ (\bibinfo  {publisher} {CERN, Geneva,
  Switzerland},\ \bibinfo {year} {2015})\ pp.\ \bibinfo {pages}
  {81--86}\BibitemShut {NoStop}%
\bibitem [{\citenamefont {Papotti}\ \emph {et~al.}(2016)\citenamefont
  {Papotti}, \citenamefont {Albert}, \citenamefont {Auchmann}, \citenamefont
  {Holzer}, \citenamefont {Kalliokoski},\ and\ \citenamefont
  {Lechner}}]{Papotti2016}%
  \BibitemOpen
  \bibfield  {author} {\bibinfo {author} {\bibfnamefont {G.}~\bibnamefont
  {Papotti}}, \bibinfo {author} {\bibfnamefont {M.}~\bibnamefont {Albert}},
  \bibinfo {author} {\bibfnamefont {B.}~\bibnamefont {Auchmann}}, \bibinfo
  {author} {\bibfnamefont {E.~B.}\ \bibnamefont {Holzer}}, \bibinfo {author}
  {\bibfnamefont {M.}~\bibnamefont {Kalliokoski}},\ and\ \bibinfo {author}
  {\bibfnamefont {A.}~\bibnamefont {Lechner}},\ }\bibfield  {title} {\bibinfo
  {title} {{Macroparticle-induced losses during 6.5 TeV LHC operation}},\ }in\
  \href {https://doi.org/http://dx.doi.org/10.18429/JACoW-IPAC2016-TUPMW023}
  {\emph {\bibinfo {booktitle} {Proceedings of the 7th International Particle
  Accelerator Conference, TUPMW023, Busan, South Korea}}}\ (\bibinfo
  {publisher} {JACoW, Geneva, Switzerland},\ \bibinfo {year} {2016})\ pp.\
  \bibinfo {pages} {1481--1484}\BibitemShut {NoStop}%
\bibitem [{\citenamefont {Lechner}\ \emph {et~al.}(2016)\citenamefont
  {Lechner}, \citenamefont {Albert}, \citenamefont {Auchmann}, \citenamefont
  {Castro}, \citenamefont {Grob}, \citenamefont {Holzer}, \citenamefont
  {Jowett}, \citenamefont {Kalliokoski}, \citenamefont {Naour}, \citenamefont
  {Lunt}, \citenamefont {Mereghetti}, \citenamefont {Papotti}, \citenamefont
  {Schmidt}, \citenamefont {Veness}, \citenamefont {Verweij}, \citenamefont
  {Willering}, \citenamefont {Wollmann}, \citenamefont {Xu},\ and\
  \citenamefont {Zerlauth}}]{Lechner2016}%
  \BibitemOpen
  \bibfield  {author} {\bibinfo {author} {\bibfnamefont {A.}~\bibnamefont
  {Lechner}}, \bibinfo {author} {\bibfnamefont {M.}~\bibnamefont {Albert}},
  \bibinfo {author} {\bibfnamefont {B.}~\bibnamefont {Auchmann}}, \bibinfo
  {author} {\bibfnamefont {C.~B.}\ \bibnamefont {Castro}}, \bibinfo {author}
  {\bibfnamefont {L.}~\bibnamefont {Grob}}, \bibinfo {author} {\bibfnamefont
  {E.~B.}\ \bibnamefont {Holzer}}, \bibinfo {author} {\bibfnamefont
  {J.}~\bibnamefont {Jowett}}, \bibinfo {author} {\bibfnamefont
  {M.}~\bibnamefont {Kalliokoski}}, \bibinfo {author} {\bibfnamefont {S.~L.}\
  \bibnamefont {Naour}}, \bibinfo {author} {\bibfnamefont {A.}~\bibnamefont
  {Lunt}}, \bibinfo {author} {\bibfnamefont {A.}~\bibnamefont {Mereghetti}},
  \bibinfo {author} {\bibfnamefont {G.}~\bibnamefont {Papotti}}, \bibinfo
  {author} {\bibfnamefont {R.}~\bibnamefont {Schmidt}}, \bibinfo {author}
  {\bibfnamefont {R.}~\bibnamefont {Veness}}, \bibinfo {author} {\bibfnamefont
  {A.}~\bibnamefont {Verweij}}, \bibinfo {author} {\bibfnamefont
  {G.}~\bibnamefont {Willering}}, \bibinfo {author} {\bibfnamefont
  {D.}~\bibnamefont {Wollmann}}, \bibinfo {author} {\bibfnamefont
  {C.}~\bibnamefont {Xu}},\ and\ \bibinfo {author} {\bibfnamefont
  {M.}~\bibnamefont {Zerlauth}},\ }\bibfield  {title} {\bibinfo {title} {{BLM
  Thresholds and UFOs}},\ }in\ \href@noop {} {\emph {\bibinfo {booktitle}
  {Proceedings of the 7th Evian Workshop on LHC Beam Operation, Evian Les
  Bains, France}}}\ (\bibinfo  {publisher} {CERN, Geneva, Switzerland},\
  \bibinfo {year} {2016})\ pp.\ \bibinfo {pages} {209--214}\BibitemShut
  {NoStop}%
\bibitem [{\citenamefont {Saeki}\ \emph {et~al.}(1991)\citenamefont {Saeki},
  \citenamefont {Momose},\ and\ \citenamefont {Ishimaru}}]{Hiroshi1991}%
  \BibitemOpen
  \bibfield  {author} {\bibinfo {author} {\bibfnamefont {H.}~\bibnamefont
  {Saeki}}, \bibinfo {author} {\bibfnamefont {T.}~\bibnamefont {Momose}},\ and\
  \bibinfo {author} {\bibfnamefont {H.}~\bibnamefont {Ishimaru}},\ }\bibfield
  {title} {\bibinfo {title} {{Observations of dust trapping phenomena in the
  TRISTAN accumulation ring and a study of dust removal in a beam chamber}},\
  }\href {https://doi.org/https://doi.org/10.1063/1.1142024} {\bibfield
  {journal} {\bibinfo  {journal} {Rev. Sci. Instrum.}\ }\textbf {\bibinfo
  {volume} {62}},\ \bibinfo {pages} {874} (\bibinfo {year} {1991})}\BibitemShut
  {NoStop}%
\bibitem [{\citenamefont {Sagan}(1993)}]{Sagan1993}%
  \BibitemOpen
  \bibfield  {author} {\bibinfo {author} {\bibfnamefont {D.}~\bibnamefont
  {Sagan}},\ }\bibfield  {title} {\bibinfo {title} {{Mass and charge
  measurement of trapped dust in the CESR storage ring}},\ }\href
  {https://doi.org/https://doi.org/10.1016/0168-9002(93)90566-Z} {\bibfield
  {journal} {\bibinfo  {journal} {Nucl. Instr. Meth. Phys. Res. A}\ }\textbf
  {\bibinfo {volume} {330}},\ \bibinfo {pages} {371} (\bibinfo {year}
  {1993})}\BibitemShut {NoStop}%
\bibitem [{\citenamefont {Zimmermann}\ \emph {et~al.}(1995)\citenamefont
  {Zimmermann}, \citenamefont {Seeman}, \citenamefont {Zolotorev},\ and\
  \citenamefont {Stoeffl}}]{Zimmermann1995}%
  \BibitemOpen
  \bibfield  {author} {\bibinfo {author} {\bibfnamefont {F.}~\bibnamefont
  {Zimmermann}}, \bibinfo {author} {\bibfnamefont {J.~T.}\ \bibnamefont
  {Seeman}}, \bibinfo {author} {\bibfnamefont {M.}~\bibnamefont {Zolotorev}},\
  and\ \bibinfo {author} {\bibfnamefont {W.}~\bibnamefont {Stoeffl}},\
  }\bibfield  {title} {\bibinfo {title} {{Trapped macroparticles in electron
  storage rings}},\ }in\ \href
  {https://doi.org/https://doi.org/10.1109/PAC.1995.504705} {\emph {\bibinfo
  {booktitle} {Proceedings of the 16th Particle Accelerator Conference, WPG10,
  Dallas, Texas, USA}}}\ (\bibinfo  {publisher} {IEEE, Piscataway, NJ},\
  \bibinfo {year} {1995})\ pp.\ \bibinfo {pages} {517--519}\BibitemShut
  {NoStop}%
\bibitem [{\citenamefont {Kelly}(1997)}]{Kelly1997}%
  \BibitemOpen
  \bibfield  {author} {\bibinfo {author} {\bibfnamefont {D.~R.}\ \bibnamefont
  {Kelly}},\ }\bibfield  {title} {\bibinfo {title} {{Dust in accelerator vacuum
  systems}},\ }in\ \href
  {https://doi.org/https://doi.org/10.1109/PAC.1997.753270} {\emph {\bibinfo
  {booktitle} {Proceedings of the 17th Particle Accelerator Conference,
  Vancouver, B.C., Canada}}}\ (\bibinfo  {publisher} {IEEE, Piscataway, NJ},\
  \bibinfo {year} {1997})\ pp.\ \bibinfo {pages} {3547--3550}\BibitemShut
  {NoStop}%
\bibitem [{\citenamefont {Kling}(2006)}]{Kling2006}%
  \BibitemOpen
  \bibfield  {author} {\bibinfo {author} {\bibfnamefont {A.}~\bibnamefont
  {Kling}},\ }\bibfield  {title} {\bibinfo {title} {{Dust macroparticles in
  HERA and DORIS}},\ }in\ \href@noop {} {\emph {\bibinfo {booktitle}
  {Proceedings of the 10th European Particle Accelerator Conference, TUPLS002,
  Edinburgh, Scotland}}}\ (\bibinfo  {publisher} {EPS-AG, Edinburgh,
  Scotland},\ \bibinfo {year} {2006})\ pp.\ \bibinfo {pages}
  {1486--1488}\BibitemShut {NoStop}%
\bibitem [{\citenamefont {Tanimoto}\ \emph {et~al.}(2009)\citenamefont
  {Tanimoto}, \citenamefont {Honda},\ and\ \citenamefont
  {Sakanaka}}]{Tanimoto2009}%
  \BibitemOpen
  \bibfield  {author} {\bibinfo {author} {\bibfnamefont {Y.}~\bibnamefont
  {Tanimoto}}, \bibinfo {author} {\bibfnamefont {T.}~\bibnamefont {Honda}},\
  and\ \bibinfo {author} {\bibfnamefont {S.}~\bibnamefont {Sakanaka}},\
  }\bibfield  {title} {\bibinfo {title} {{Experimental demonstration and visual
  observation of dust trapping in an electron storage ring}},\ }\href
  {https://doi.org/https://doi.org/10.1103/PhysRevSTAB.12.110702} {\bibfield
  {journal} {\bibinfo  {journal} {Phys. Rev. ST Accel. Beams}\ }\textbf
  {\bibinfo {volume} {12}},\ \bibinfo {pages} {110702} (\bibinfo {year}
  {2009})}\BibitemShut {NoStop}%
\bibitem [{\citenamefont {Suetsugu}\ \emph {et~al.}(2016)\citenamefont
  {Suetsugu}, \citenamefont {Shibata}, \citenamefont {Ishibashi}, \citenamefont
  {Kanazawa}, \citenamefont {Shirai}, \citenamefont {Terui},\ and\
  \citenamefont {Hisamatsu}}]{Suetsugu2016}%
  \BibitemOpen
  \bibfield  {author} {\bibinfo {author} {\bibfnamefont {Y.}~\bibnamefont
  {Suetsugu}}, \bibinfo {author} {\bibfnamefont {K.}~\bibnamefont {Shibata}},
  \bibinfo {author} {\bibfnamefont {T.}~\bibnamefont {Ishibashi}}, \bibinfo
  {author} {\bibfnamefont {K.}~\bibnamefont {Kanazawa}}, \bibinfo {author}
  {\bibfnamefont {M.}~\bibnamefont {Shirai}}, \bibinfo {author} {\bibfnamefont
  {S.}~\bibnamefont {Terui}},\ and\ \bibinfo {author} {\bibfnamefont
  {H.}~\bibnamefont {Hisamatsu}},\ }\bibfield  {title} {\bibinfo {title}
  {{First commissioning of the SuperKEKB vacuum system}},\ }\href
  {https://doi.org/https://doi.org/10.1103/PhysRevAccelBeams.19.121001}
  {\bibfield  {journal} {\bibinfo  {journal} {Phys. Rev. Accel. Beams}\
  }\textbf {\bibinfo {volume} {19}},\ \bibinfo {pages} {121001} (\bibinfo
  {year} {2016})}\BibitemShut {NoStop}%
\bibitem [{\citenamefont {Auchmann}\ \emph {et~al.}(2014)\citenamefont
  {Auchmann}, \citenamefont {Baer}, \citenamefont {Lechner}, \citenamefont
  {Riegler}, \citenamefont {Rowan}, \citenamefont {Schindler}, \citenamefont
  {Schmidt},\ and\ \citenamefont {Zimmermann}}]{Auchmann2014}%
  \BibitemOpen
  \bibfield  {author} {\bibinfo {author} {\bibfnamefont {B.}~\bibnamefont
  {Auchmann}}, \bibinfo {author} {\bibfnamefont {T.~M.}\ \bibnamefont {Baer}},
  \bibinfo {author} {\bibfnamefont {A.}~\bibnamefont {Lechner}}, \bibinfo
  {author} {\bibfnamefont {W.}~\bibnamefont {Riegler}}, \bibinfo {author}
  {\bibfnamefont {S.}~\bibnamefont {Rowan}}, \bibinfo {author} {\bibfnamefont
  {H.}~\bibnamefont {Schindler}}, \bibinfo {author} {\bibfnamefont
  {R.}~\bibnamefont {Schmidt}},\ and\ \bibinfo {author} {\bibfnamefont
  {F.}~\bibnamefont {Zimmermann}},\ }\href@noop {} {\emph {\bibinfo {title}
  {Proton-beam macro-particle interaction: beam dumps and quenches}}},\
  \bibinfo {type} {Tech. Rep.}\ \bibinfo {number} {CERN-ACC-NOTE-2020-0041}\
  (\bibinfo  {institution} {CERN},\ \bibinfo {year} {2014})\BibitemShut
  {NoStop}%
\bibitem [{\citenamefont {Rowan}\ \emph {et~al.}(2015)\citenamefont {Rowan},
  \citenamefont {Apollonio}, \citenamefont {Auchmann}, \citenamefont {Lechner},
  \citenamefont {Picha}, \citenamefont {Riegler}, \citenamefont {Schindler},
  \citenamefont {Schmidt},\ and\ \citenamefont {Zimmermann}}]{Rowan2015}%
  \BibitemOpen
  \bibfield  {author} {\bibinfo {author} {\bibfnamefont {S.}~\bibnamefont
  {Rowan}}, \bibinfo {author} {\bibfnamefont {A.}~\bibnamefont {Apollonio}},
  \bibinfo {author} {\bibfnamefont {B.}~\bibnamefont {Auchmann}}, \bibinfo
  {author} {\bibfnamefont {A.}~\bibnamefont {Lechner}}, \bibinfo {author}
  {\bibfnamefont {O.}~\bibnamefont {Picha}}, \bibinfo {author} {\bibfnamefont
  {W.}~\bibnamefont {Riegler}}, \bibinfo {author} {\bibfnamefont
  {H.}~\bibnamefont {Schindler}}, \bibinfo {author} {\bibfnamefont
  {R.}~\bibnamefont {Schmidt}},\ and\ \bibinfo {author} {\bibfnamefont
  {F.}~\bibnamefont {Zimmermann}},\ }\bibfield  {title} {\bibinfo {title}
  {{Interactions between macroparticles and high-energy proton beams}},\ }in\
  \href {https://doi.org/https://doi.org/10.18429/JACoW-IPAC2015-TUPTY045}
  {\emph {\bibinfo {booktitle} {Proceedings of the 6th International Particle
  Accelerator Conference, TUPTY045, Richmond, VA, USA}}}\ (\bibinfo
  {publisher} {JACoW, Geneva, Switzerland},\ \bibinfo {year} {2015})\ pp.\
  \bibinfo {pages} {2112--2115}\BibitemShut {NoStop}%
\bibitem [{\citenamefont {Rowan}(2016)}]{Rowan2016}%
  \BibitemOpen
  \bibfield  {author} {\bibinfo {author} {\bibfnamefont {S.}~\bibnamefont
  {Rowan}},\ }\emph {\bibinfo {title} {{LHC} main dipole magnet circuits:
  sustaining near-nominal beam energies}},\ \href@noop {} {Ph.D. thesis},\
  \bibinfo  {school} {CERN and Glasgow University} (\bibinfo {year}
  {2016})\BibitemShut {NoStop}%
\bibitem [{\citenamefont {Lindstrom}\ \emph {et~al.}(2018)\citenamefont
  {Lindstrom}, \citenamefont {Apollonio}, \citenamefont {B\'{e}langer},
  \citenamefont {Dziadosz}, \citenamefont {Gorzawski}, \citenamefont {Grob},
  \citenamefont {Holzer}, \citenamefont {Lechner}, \citenamefont {Schmidt},
  \citenamefont {Valette}, \citenamefont {Valuch},\ and\ \citenamefont
  {Wollmann}}]{Lindstrom2018}%
  \BibitemOpen
  \bibfield  {author} {\bibinfo {author} {\bibfnamefont {B.}~\bibnamefont
  {Lindstrom}}, \bibinfo {author} {\bibfnamefont {A.}~\bibnamefont
  {Apollonio}}, \bibinfo {author} {\bibfnamefont {P.}~\bibnamefont
  {B\'{e}langer}}, \bibinfo {author} {\bibfnamefont {M.}~\bibnamefont
  {Dziadosz}}, \bibinfo {author} {\bibfnamefont {A.}~\bibnamefont {Gorzawski}},
  \bibinfo {author} {\bibfnamefont {L.}~\bibnamefont {Grob}}, \bibinfo {author}
  {\bibfnamefont {E.}~\bibnamefont {Holzer}}, \bibinfo {author} {\bibfnamefont
  {A.}~\bibnamefont {Lechner}}, \bibinfo {author} {\bibfnamefont
  {R.}~\bibnamefont {Schmidt}}, \bibinfo {author} {\bibfnamefont
  {M.}~\bibnamefont {Valette}}, \bibinfo {author} {\bibfnamefont
  {D.}~\bibnamefont {Valuch}},\ and\ \bibinfo {author} {\bibfnamefont
  {D.}~\bibnamefont {Wollmann}},\ }\bibfield  {title} {\bibinfo {title}
  {{R}esults of {UFO} {D}ynamics {S}tudies with {B}eam in the {LHC}},\ }in\
  \href {https://doi.org/https://doi.org/10.18429/JACoW-IPAC2018-THYGBD2}
  {\emph {\bibinfo {booktitle} {Proceedings of the 9th International Particle
  Accelerator Conference, THYGBD2, Vancouver, Canada}}}\ (\bibinfo  {publisher}
  {JACoW, Geneva, Switzerland},\ \bibinfo {year} {2018})\ pp.\ \bibinfo {pages}
  {2914--2917}\BibitemShut {NoStop}%
\bibitem [{\citenamefont {Lindstrom}\ \emph {et~al.}(2020)\citenamefont
  {Lindstrom}, \citenamefont {B\'elanger}, \citenamefont {Gorzawski},
  \citenamefont {Kral}, \citenamefont {Lechner}, \citenamefont {Salvachua},
  \citenamefont {Schmidt}, \citenamefont {Siemko}, \citenamefont {Vaananen},
  \citenamefont {Valuch}, \citenamefont {Wiesner}, \citenamefont {Wollmann},\
  and\ \citenamefont {Zamantzas}}]{Lindstrom2020}%
  \BibitemOpen
  \bibfield  {author} {\bibinfo {author} {\bibfnamefont {B.}~\bibnamefont
  {Lindstrom}}, \bibinfo {author} {\bibfnamefont {P.}~\bibnamefont
  {B\'elanger}}, \bibinfo {author} {\bibfnamefont {A.}~\bibnamefont
  {Gorzawski}}, \bibinfo {author} {\bibfnamefont {J.}~\bibnamefont {Kral}},
  \bibinfo {author} {\bibfnamefont {A.}~\bibnamefont {Lechner}}, \bibinfo
  {author} {\bibfnamefont {B.}~\bibnamefont {Salvachua}}, \bibinfo {author}
  {\bibfnamefont {R.}~\bibnamefont {Schmidt}}, \bibinfo {author} {\bibfnamefont
  {A.}~\bibnamefont {Siemko}}, \bibinfo {author} {\bibfnamefont
  {M.}~\bibnamefont {Vaananen}}, \bibinfo {author} {\bibfnamefont
  {D.}~\bibnamefont {Valuch}}, \bibinfo {author} {\bibfnamefont
  {C.}~\bibnamefont {Wiesner}}, \bibinfo {author} {\bibfnamefont
  {D.}~\bibnamefont {Wollmann}},\ and\ \bibinfo {author} {\bibfnamefont
  {C.}~\bibnamefont {Zamantzas}},\ }\bibfield  {title} {\bibinfo {title}
  {Dynamics of the interaction of dust particles with the lhc beam},\ }\href
  {https://doi.org/https://doi.org/10.1103/PhysRevAccelBeams.23.124501}
  {\bibfield  {journal} {\bibinfo  {journal} {Phys. Rev. Accel. Beams}\
  }\textbf {\bibinfo {volume} {23}},\ \bibinfo {pages} {124501} (\bibinfo
  {year} {2020})}\BibitemShut {NoStop}%
\bibitem [{\citenamefont {Mirarchi}\ \emph {et~al.}(2016)\citenamefont
  {Mirarchi}, \citenamefont {Bruce}, \citenamefont {Giovannozzi}, \citenamefont
  {Hermes}, \citenamefont {Redaelli}, \citenamefont {Salvachua}, \citenamefont
  {Valentino},\ and\ \citenamefont {Wenninger}}]{Mirarchi2015}%
  \BibitemOpen
  \bibfield  {author} {\bibinfo {author} {\bibfnamefont {D.}~\bibnamefont
  {Mirarchi}}, \bibinfo {author} {\bibfnamefont {R.}~\bibnamefont {Bruce}},
  \bibinfo {author} {\bibfnamefont {M.}~\bibnamefont {Giovannozzi}}, \bibinfo
  {author} {\bibfnamefont {P.}~\bibnamefont {Hermes}}, \bibinfo {author}
  {\bibfnamefont {S.}~\bibnamefont {Redaelli}}, \bibinfo {author}
  {\bibfnamefont {B.}~\bibnamefont {Salvachua}}, \bibinfo {author}
  {\bibfnamefont {G.}~\bibnamefont {Valentino}},\ and\ \bibinfo {author}
  {\bibfnamefont {J.}~\bibnamefont {Wenninger}},\ }\bibfield  {title} {\bibinfo
  {title} {{LHC aperture and ULO restrictions: are they a possible limitation
  in 2016?}},\ }in\ \href@noop {} {\emph {\bibinfo {booktitle} {Proceedings of
  the 6th Evian Workshop on LHC Beam Operation, Evian Les Bains, France}}}\
  (\bibinfo  {publisher} {CERN, Geneva, Switzerland},\ \bibinfo {year} {2016})\
  pp.\ \bibinfo {pages} {87--94}\BibitemShut {NoStop}%
\bibitem [{\citenamefont {Mirarchi}\ \emph {et~al.}(2019)\citenamefont
  {Mirarchi}, \citenamefont {Arduini}, \citenamefont {Giovannozzi},
  \citenamefont {Lechner}, \citenamefont {Redaelli},\ and\ \citenamefont
  {J}}]{Mirarchi2019}%
  \BibitemOpen
  \bibfield  {author} {\bibinfo {author} {\bibfnamefont {D.}~\bibnamefont
  {Mirarchi}}, \bibinfo {author} {\bibfnamefont {G.}~\bibnamefont {Arduini}},
  \bibinfo {author} {\bibfnamefont {M.}~\bibnamefont {Giovannozzi}}, \bibinfo
  {author} {\bibfnamefont {A.}~\bibnamefont {Lechner}}, \bibinfo {author}
  {\bibfnamefont {S.}~\bibnamefont {Redaelli}},\ and\ \bibinfo {author}
  {\bibfnamefont {W.}~\bibnamefont {J}},\ }\bibfield  {title} {\bibinfo {title}
  {{Special losses during LHC Run~2}},\ }in\ \href@noop {} {\emph {\bibinfo
  {booktitle} {Proceedings of the 9th Evian Workshop on LHC Beam Operation,
  Evian Les Bains, France}}}\ (\bibinfo  {publisher} {CERN, Geneva,
  Switzerland},\ \bibinfo {year} {2019})\ pp.\ \bibinfo {pages}
  {213--220}\BibitemShut {NoStop}%
\bibitem [{\citenamefont {Mether}\ \emph {et~al.}(2017)\citenamefont {Mether},
  \citenamefont {Amorim}, \citenamefont {Arduini}, \citenamefont {Buffat},
  \citenamefont {Iadarola}, \citenamefont {Lechner}, \citenamefont {Métral},
  \citenamefont {Mirarchi}, \citenamefont {Rumolo},\ and\ \citenamefont
  {Salvant}}]{Mether2017}%
  \BibitemOpen
  \bibfield  {author} {\bibinfo {author} {\bibfnamefont {L.}~\bibnamefont
  {Mether}}, \bibinfo {author} {\bibfnamefont {D.}~\bibnamefont {Amorim}},
  \bibinfo {author} {\bibfnamefont {G.}~\bibnamefont {Arduini}}, \bibinfo
  {author} {\bibfnamefont {X.}~\bibnamefont {Buffat}}, \bibinfo {author}
  {\bibfnamefont {G.}~\bibnamefont {Iadarola}}, \bibinfo {author}
  {\bibfnamefont {A.}~\bibnamefont {Lechner}}, \bibinfo {author} {\bibfnamefont
  {E.}~\bibnamefont {Métral}}, \bibinfo {author} {\bibfnamefont
  {D.}~\bibnamefont {Mirarchi}}, \bibinfo {author} {\bibfnamefont
  {G.}~\bibnamefont {Rumolo}},\ and\ \bibinfo {author} {\bibfnamefont
  {B.}~\bibnamefont {Salvant}},\ }\bibfield  {title} {\bibinfo {title} {{16L2:
  Operation, observations and physics aspects}},\ }in\ \href@noop {} {\emph
  {\bibinfo {booktitle} {Proceedings of the 8th Evian Workshop on LHC Beam
  Operation, Evian Les Bains, France}}}\ (\bibinfo  {publisher} {CERN, Geneva,
  Switzerland},\ \bibinfo {year} {2017})\ pp.\ \bibinfo {pages}
  {99--105}\BibitemShut {NoStop}%
\bibitem [{\citenamefont {Jim\'{e}nez}\ \emph {et~al.}(2018)\citenamefont
  {Jim\'{e}nez} \emph {et~al.}}]{Jimenez2018}%
  \BibitemOpen
  \bibfield  {author} {\bibinfo {author} {\bibfnamefont {J.}~\bibnamefont
  {Jim\'{e}nez}} \emph {et~al.},\ }\bibfield  {title} {\bibinfo {title}
  {{Observations, analysis and mitigation of recurrent LHC beam dumps caused by
  fast losses in arc half-cell 16L2}},\ }in\ \href
  {https://doi.org/http://dx.doi.org/10.18429/JACoW-IPAC2018-MOPMF053} {\emph
  {\bibinfo {booktitle} {Proceedings of the 9th International Particle
  Accelerator Conference, MOPMF053, Vancouver, Canada}}}\ (\bibinfo
  {publisher} {JACoW, Geneva, Switzerland},\ \bibinfo {year} {2018})\ pp.\
  \bibinfo {pages} {228--231}\BibitemShut {NoStop}%
\bibitem [{\citenamefont {Lechner}\ \emph {et~al.}(2018)\citenamefont {Lechner}
  \emph {et~al.}}]{Lechner2018}%
  \BibitemOpen
  \bibfield  {author} {\bibinfo {author} {\bibfnamefont {A.}~\bibnamefont
  {Lechner}} \emph {et~al.},\ }\bibfield  {title} {\bibinfo {title} {{Beam loss
  measurements for recurring fast loss events during 2017 operation possibly
  caused by macroparticles}},\ }in\ \href
  {https://doi.org/http://dx.doi.org/10.18429/JACoW-IPAC2018-TUPAF040} {\emph
  {\bibinfo {booktitle} {Proceedings of the 9th International Particle
  Accelerator Conference, TUPAF040, Vancouver, Canada}}}\ (\bibinfo
  {publisher} {JACoW, Geneva, Switzerland},\ \bibinfo {year} {2018})\ pp.\
  \bibinfo {pages} {780--784}\BibitemShut {NoStop}%
\bibitem [{\citenamefont {Béjar~Alonso}\ \emph {et~al.}(2020)\citenamefont
  {Béjar~Alonso}, \citenamefont {Br{\"u}ning}, \citenamefont {Fessia},
  \citenamefont {Lamont}, \citenamefont {Rossi}, \citenamefont {Tavian},\ and\
  \citenamefont {Zerlauth}}]{HL2020}%
  \BibitemOpen
  \bibinfo {editor} {\bibfnamefont {I.}~\bibnamefont {Béjar~Alonso}}, \bibinfo
  {editor} {\bibfnamefont {O.}~\bibnamefont {Br{\"u}ning}}, \bibinfo {editor}
  {\bibfnamefont {P.}~\bibnamefont {Fessia}}, \bibinfo {editor} {\bibfnamefont
  {M.}~\bibnamefont {Lamont}}, \bibinfo {editor} {\bibfnamefont
  {L.}~\bibnamefont {Rossi}}, \bibinfo {editor} {\bibfnamefont
  {L.}~\bibnamefont {Tavian}},\ and\ \bibinfo {editor} {\bibfnamefont
  {M.}~\bibnamefont {Zerlauth}},\ eds.,\ \href
  {https://doi.org/https://doi.org/10.23731/CYRM-2020-0010} {\emph {\bibinfo
  {title} {{High-Luminosity Large Hadron Collider (HL-LHC): Technical design
  report}}}},\ CERN Yellow Reports: Monographs\ (\bibinfo  {publisher} {CERN,
  Geneva, Switzerland},\ \bibinfo {year} {2020})\BibitemShut {NoStop}%
\bibitem [{\citenamefont {Lamont}(2013)}]{Lamont2013}%
  \BibitemOpen
  \bibfield  {author} {\bibinfo {author} {\bibfnamefont {M.}~\bibnamefont
  {Lamont}},\ }\bibfield  {title} {\bibinfo {title} {{The first years of LHC
  operation for luminosity production}},\ }in\ \href@noop {} {\emph {\bibinfo
  {booktitle} {Proceedings of the 4th International Particle Accelerator
  Conference, MOYAB101, Shanghai, China}}}\ (\bibinfo  {publisher} {JACoW,
  Geneva, Switzerland},\ \bibinfo {year} {2013})\ pp.\ \bibinfo {pages}
  {6--10}\BibitemShut {NoStop}%
\bibitem [{\citenamefont {Steerenberg}\ \emph {et~al.}(2019)\citenamefont
  {Steerenberg} \emph {et~al.}}]{Steerenberg2019}%
  \BibitemOpen
  \bibfield  {author} {\bibinfo {author} {\bibfnamefont {R.}~\bibnamefont
  {Steerenberg}} \emph {et~al.},\ }\bibfield  {title} {\bibinfo {title}
  {{O}peration and {P}erformance of the {C}ern {L}arge {H}adron {C}ollider
  {D}uring {P}roton {R}un 2},\ }in\ \href
  {https://doi.org/https://doi.org/10.18429/JACoW-IPAC2019-MOPMP031} {\emph
  {\bibinfo {booktitle} {Proc. 10th International Particle Accelerator
  Conference (IPAC'19), Melbourne, Australia, 19-24 May 2019}}}\ (\bibinfo
  {publisher} {JACoW, Geneva, Switzerland},\ \bibinfo {address} {Geneva,
  Switzerland},\ \bibinfo {year} {2019})\ pp.\ \bibinfo {pages}
  {504--507}\BibitemShut {NoStop}%
\bibitem [{\citenamefont {Karastathis}\ \emph {et~al.}(2019)\citenamefont
  {Karastathis} \emph {et~al.}}]{Karastathis2019}%
  \BibitemOpen
  \bibfield  {author} {\bibinfo {author} {\bibfnamefont {N.}~\bibnamefont
  {Karastathis}} \emph {et~al.},\ }\bibfield  {title} {\bibinfo {title} {{LHC
  Run 3 Configuration Working Group Report}},\ }in\ \href
  {http://cds.cern.ch/record/2750302} {\emph {\bibinfo {booktitle} {Proceedings
  of the 9th Evian Workshop on LHC Beam Operation, Evian Les Bains, France}}}\
  (\bibinfo {address} {Geneva, Switzerland},\ \bibinfo {year} {2019})\ pp.\
  \bibinfo {pages} {273--284}\BibitemShut {NoStop}%
\bibitem [{\citenamefont {Grob}\ \emph {et~al.}(2019)\citenamefont {Grob},
  \citenamefont {Apollonio}, \citenamefont {Busom}, \citenamefont {Charvet},
  \citenamefont {Fontenla}, \citenamefont {Valdivieso}, \citenamefont {Kos},
  \citenamefont {Schmidt},\ and\ \citenamefont {Neves}}]{Grob2019}%
  \BibitemOpen
  \bibfield  {author} {\bibinfo {author} {\bibfnamefont {L.}~\bibnamefont
  {Grob}}, \bibinfo {author} {\bibfnamefont {A.}~\bibnamefont {Apollonio}},
  \bibinfo {author} {\bibfnamefont {J.~D.}\ \bibnamefont {Busom}}, \bibinfo
  {author} {\bibfnamefont {C.}~\bibnamefont {Charvet}}, \bibinfo {author}
  {\bibfnamefont {A.~P.}\ \bibnamefont {Fontenla}}, \bibinfo {author}
  {\bibfnamefont {E.~G.-T.}\ \bibnamefont {Valdivieso}}, \bibinfo {author}
  {\bibfnamefont {H.}~\bibnamefont {Kos}}, \bibinfo {author} {\bibfnamefont
  {R.}~\bibnamefont {Schmidt}},\ and\ \bibinfo {author} {\bibfnamefont
  {C.}~\bibnamefont {Neves}},\ }\bibfield  {title} {\bibinfo {title} {Dust
  analysis from {LHC} vacuum system to identify the source of
  macro-particle-beam-interactions},\ }in\ \href
  {https://doi.org/https://doi.org/10.18429/JACoW-IPAC2019-MOPTS094} {\emph
  {\bibinfo {booktitle} {Proceedings of the 10th International Particle
  Accelerator Conference, MOPTS094, Melbourne, Australia}}}\ (\bibinfo
  {publisher} {JACoW, Geneva, Switzerland},\ \bibinfo {year} {2019})\ pp.\
  \bibinfo {pages} {1082--1085}\BibitemShut {NoStop}%
\bibitem [{\citenamefont {Zimmermann}\ \emph {et~al.}(2010)\citenamefont
  {Zimmermann}, \citenamefont {Giovannozzi},\ and\ \citenamefont
  {Xagkoni}}]{Zimmermann2010}%
  \BibitemOpen
  \bibfield  {author} {\bibinfo {author} {\bibfnamefont {F.}~\bibnamefont
  {Zimmermann}}, \bibinfo {author} {\bibfnamefont {M.}~\bibnamefont
  {Giovannozzi}},\ and\ \bibinfo {author} {\bibfnamefont {A.}~\bibnamefont
  {Xagkoni}},\ }\bibfield  {title} {\bibinfo {title} {{Interaction of
  macro-particles with LHC proton beam}},\ }in\ \href@noop {} {\emph {\bibinfo
  {booktitle} {Proceedings of the 1st International Particle Accelerator
  Conference, MOPEC016, Kyoto, Japan}}}\ (\bibinfo  {publisher} {Asian
  Committee for Future Accelerators},\ \bibinfo {year} {2010})\ pp.\ \bibinfo
  {pages} {492--494}\BibitemShut {NoStop}%
\bibitem [{\citenamefont {Martinez}\ \emph {et~al.}(2011)\citenamefont
  {Martinez}, \citenamefont {Zimmermann}, \citenamefont {Baer}, \citenamefont
  {Giovannozzi}, \citenamefont {Holzer}, \citenamefont {Nebot}, \citenamefont
  {Nordt}, \citenamefont {Sapinski},\ and\ \citenamefont {Yang}}]{Fuster2011}%
  \BibitemOpen
  \bibfield  {author} {\bibinfo {author} {\bibfnamefont {N.~F.}\ \bibnamefont
  {Martinez}}, \bibinfo {author} {\bibfnamefont {F.}~\bibnamefont
  {Zimmermann}}, \bibinfo {author} {\bibfnamefont {T.}~\bibnamefont {Baer}},
  \bibinfo {author} {\bibfnamefont {M.}~\bibnamefont {Giovannozzi}}, \bibinfo
  {author} {\bibfnamefont {E.~B.}\ \bibnamefont {Holzer}}, \bibinfo {author}
  {\bibfnamefont {E.}~\bibnamefont {Nebot}}, \bibinfo {author} {\bibfnamefont
  {A.}~\bibnamefont {Nordt}}, \bibinfo {author} {\bibfnamefont
  {M.}~\bibnamefont {Sapinski}},\ and\ \bibinfo {author} {\bibfnamefont
  {Z.}~\bibnamefont {Yang}},\ }\bibfield  {title} {\bibinfo {title}
  {{Simulation studies of macroparticles falling into the LHC proton beam}},\
  }in\ \href@noop {} {\emph {\bibinfo {booktitle} {Proceedings of the 2nd
  International Particle Accelerator Conference, MOPS017, San Sebastian,
  Spain}}}\ (\bibinfo  {publisher} {EPS-AG, Spain},\ \bibinfo {year} {2011})\
  pp.\ \bibinfo {pages} {634--636}\BibitemShut {NoStop}%
\bibitem [{\citenamefont {B\'elanger}\ \emph {et~al.}()\citenamefont
  {B\'elanger}, \citenamefont {Baartman}, \citenamefont {Iadarola},
  \citenamefont {Lechner}, \citenamefont {Lindstrom}, \citenamefont {Schmidt},\
  and\ \citenamefont {Wollmann}}]{Belanger2021}%
  \BibitemOpen
  \bibfield  {author} {\bibinfo {author} {\bibfnamefont {P.}~\bibnamefont
  {B\'elanger}}, \bibinfo {author} {\bibfnamefont {R.}~\bibnamefont
  {Baartman}}, \bibinfo {author} {\bibfnamefont {G.}~\bibnamefont {Iadarola}},
  \bibinfo {author} {\bibfnamefont {A.}~\bibnamefont {Lechner}}, \bibinfo
  {author} {\bibfnamefont {B.}~\bibnamefont {Lindstrom}}, \bibinfo {author}
  {\bibfnamefont {R.}~\bibnamefont {Schmidt}},\ and\ \bibinfo {author}
  {\bibfnamefont {D.}~\bibnamefont {Wollmann}},\ }\bibfield  {title} {\bibinfo
  {title} {Charging mechanisms and orbital dynamics of charged dust grains in
  the {LHC}},\ }\href@noop {} {\bibinfo  {journal} {Submitted to Phys. Rev.
  Accel. Beams}\ }\BibitemShut {NoStop}%
\bibitem [{\citenamefont {Abada}\ \emph {et~al.}(2019)\citenamefont {Abada}
  \emph {et~al.}}]{Abada2019}%
  \BibitemOpen
\bibfield  {journal} {  }\bibfield  {author} {\bibinfo {author} {\bibfnamefont
  {A.}~\bibnamefont {Abada}} \emph {et~al.},\ }\bibfield  {title} {\bibinfo
  {title} {{FCC-hh: The Hadron Collider}},\ }\href
  {https://doi.org/https://doi.org/10.1140/epjst/e2019-900087-0} {\bibfield
  {journal} {\bibinfo  {journal} {Eur. Phys. J. Spec. Top.}\ ,\ \bibinfo
  {pages} {755}} (\bibinfo {year} {2019})}\BibitemShut {NoStop}%
\bibitem [{\citenamefont {Böhlen}\ \emph {et~al.}(2014)\citenamefont
  {Böhlen}, \citenamefont {Cerutti}, \citenamefont {Chin}, \citenamefont
  {Fassò}, \citenamefont {Ferrari}, \citenamefont {Ortega}, \citenamefont
  {Mairani}, \citenamefont {Sala}, \citenamefont {Smirnov},\ and\ \citenamefont
  {Vlachoudis}}]{Bohlen2014}%
  \BibitemOpen
  \bibfield  {author} {\bibinfo {author} {\bibfnamefont {T.}~\bibnamefont
  {Böhlen}}, \bibinfo {author} {\bibfnamefont {F.}~\bibnamefont {Cerutti}},
  \bibinfo {author} {\bibfnamefont {M.}~\bibnamefont {Chin}}, \bibinfo {author}
  {\bibfnamefont {A.}~\bibnamefont {Fassò}}, \bibinfo {author} {\bibfnamefont
  {A.}~\bibnamefont {Ferrari}}, \bibinfo {author} {\bibfnamefont
  {P.}~\bibnamefont {Ortega}}, \bibinfo {author} {\bibfnamefont
  {A.}~\bibnamefont {Mairani}}, \bibinfo {author} {\bibfnamefont
  {P.}~\bibnamefont {Sala}}, \bibinfo {author} {\bibfnamefont {G.}~\bibnamefont
  {Smirnov}},\ and\ \bibinfo {author} {\bibfnamefont {V.}~\bibnamefont
  {Vlachoudis}},\ }\bibfield  {title} {\bibinfo {title} {The {FLUKA} code:
  Developments and challenges for high energy and medical applications},\
  }\href {https://doi.org/https://doi.org/10.1016/j.nds.2014.07.049} {\bibfield
   {journal} {\bibinfo  {journal} {Nuclear Data Sheets}\ }\textbf {\bibinfo
  {volume} {120}},\ \bibinfo {pages} {211} (\bibinfo {year}
  {2014})}\BibitemShut {NoStop}%
\bibitem [{\citenamefont {Battistoni}\ \emph {et~al.}(2015)\citenamefont
  {Battistoni}, \citenamefont {Boehlen}, \citenamefont {Cerutti}, \citenamefont
  {Chin}, \citenamefont {Esposito}, \citenamefont {Fassò}, \citenamefont
  {Ferrari}, \citenamefont {Lechner}, \citenamefont {Empl}, \citenamefont
  {Mairani}, \citenamefont {Mereghetti}, \citenamefont {Ortega}, \citenamefont
  {Ranft}, \citenamefont {Roesler}, \citenamefont {Sala}, \citenamefont
  {Vlachoudis},\ and\ \citenamefont {Smirnov}}]{Battistoni2015}%
  \BibitemOpen
  \bibfield  {author} {\bibinfo {author} {\bibfnamefont {G.}~\bibnamefont
  {Battistoni}}, \bibinfo {author} {\bibfnamefont {T.}~\bibnamefont {Boehlen}},
  \bibinfo {author} {\bibfnamefont {F.}~\bibnamefont {Cerutti}}, \bibinfo
  {author} {\bibfnamefont {P.~W.}\ \bibnamefont {Chin}}, \bibinfo {author}
  {\bibfnamefont {L.~S.}\ \bibnamefont {Esposito}}, \bibinfo {author}
  {\bibfnamefont {A.}~\bibnamefont {Fassò}}, \bibinfo {author} {\bibfnamefont
  {A.}~\bibnamefont {Ferrari}}, \bibinfo {author} {\bibfnamefont
  {A.}~\bibnamefont {Lechner}}, \bibinfo {author} {\bibfnamefont
  {A.}~\bibnamefont {Empl}}, \bibinfo {author} {\bibfnamefont {A.}~\bibnamefont
  {Mairani}}, \bibinfo {author} {\bibfnamefont {A.}~\bibnamefont {Mereghetti}},
  \bibinfo {author} {\bibfnamefont {P.~G.}\ \bibnamefont {Ortega}}, \bibinfo
  {author} {\bibfnamefont {J.}~\bibnamefont {Ranft}}, \bibinfo {author}
  {\bibfnamefont {S.}~\bibnamefont {Roesler}}, \bibinfo {author} {\bibfnamefont
  {P.~R.}\ \bibnamefont {Sala}}, \bibinfo {author} {\bibfnamefont
  {V.}~\bibnamefont {Vlachoudis}},\ and\ \bibinfo {author} {\bibfnamefont
  {G.}~\bibnamefont {Smirnov}},\ }\bibfield  {title} {\bibinfo {title}
  {Overview of the {FLUKA} code},\ }\href
  {https://doi.org/https://doi.org/10.1016/j.anucene.2014.11.007} {\bibfield
  {journal} {\bibinfo  {journal} {Annals of Nuclear Energy}\ }\textbf {\bibinfo
  {volume} {82}},\ \bibinfo {pages} {10} (\bibinfo {year} {2015})},\ \bibinfo
  {note} {{Joint} International Conference on Supercomputing in Nuclear
  Applications and Monte Carlo 2013, SNA + MC 2013.}\BibitemShut {Stop}%
\bibitem [{Flu()}]{FlukaWeb}%
  \BibitemOpen
  \href@noop {} {\bibinfo {title} {{FLUKA.CERN} website}},\ \bibinfo
  {howpublished} {\url{https://fluka.cern/}}\BibitemShut {NoStop}%
\bibitem [{\citenamefont {Lechner}\ \emph {et~al.}(2019)\citenamefont
  {Lechner}, \citenamefont {Auchmann}, \citenamefont {Baer}, \citenamefont
  {Bahamonde~Castro}, \citenamefont {Bruce}, \citenamefont {Cerutti},
  \citenamefont {Esposito}, \citenamefont {Ferrari}, \citenamefont {Jowett},
  \citenamefont {Mereghetti}, \citenamefont {Pietropaolo}, \citenamefont
  {Redaelli}, \citenamefont {Salvachua}, \citenamefont {Sapinski},
  \citenamefont {Schaumann}, \citenamefont {Shetty}, \citenamefont
  {Vlachoudis},\ and\ \citenamefont {Skordis}}]{Lechner2019}%
  \BibitemOpen
  \bibfield  {author} {\bibinfo {author} {\bibfnamefont {A.}~\bibnamefont
  {Lechner}}, \bibinfo {author} {\bibfnamefont {B.}~\bibnamefont {Auchmann}},
  \bibinfo {author} {\bibfnamefont {T.}~\bibnamefont {Baer}}, \bibinfo {author}
  {\bibfnamefont {C.}~\bibnamefont {Bahamonde~Castro}}, \bibinfo {author}
  {\bibfnamefont {R.}~\bibnamefont {Bruce}}, \bibinfo {author} {\bibfnamefont
  {F.}~\bibnamefont {Cerutti}}, \bibinfo {author} {\bibfnamefont {L.~S.}\
  \bibnamefont {Esposito}}, \bibinfo {author} {\bibfnamefont {A.}~\bibnamefont
  {Ferrari}}, \bibinfo {author} {\bibfnamefont {J.~M.}\ \bibnamefont {Jowett}},
  \bibinfo {author} {\bibfnamefont {A.}~\bibnamefont {Mereghetti}}, \bibinfo
  {author} {\bibfnamefont {F.}~\bibnamefont {Pietropaolo}}, \bibinfo {author}
  {\bibfnamefont {S.}~\bibnamefont {Redaelli}}, \bibinfo {author}
  {\bibfnamefont {B.}~\bibnamefont {Salvachua}}, \bibinfo {author}
  {\bibfnamefont {M.}~\bibnamefont {Sapinski}}, \bibinfo {author}
  {\bibfnamefont {M.}~\bibnamefont {Schaumann}}, \bibinfo {author}
  {\bibfnamefont {N.~V.}\ \bibnamefont {Shetty}}, \bibinfo {author}
  {\bibfnamefont {V.}~\bibnamefont {Vlachoudis}},\ and\ \bibinfo {author}
  {\bibfnamefont {E.}~\bibnamefont {Skordis}},\ }\bibfield  {title} {\bibinfo
  {title} {Validation of energy deposition simulations for proton and heavy ion
  losses in the cern large hadron collider},\ }\href
  {https://doi.org/https://doi.org/10.1103/PhysRevAccelBeams.22.071003}
  {\bibfield  {journal} {\bibinfo  {journal} {Phys. Rev. Accel. Beams}\
  }\textbf {\bibinfo {volume} {22}},\ \bibinfo {pages} {071003} (\bibinfo
  {year} {2019})}\BibitemShut {NoStop}%
\bibitem [{\citenamefont {Gröbner}(1995)}]{groebner1995}%
  \BibitemOpen
  \bibfield  {author} {\bibinfo {author} {\bibfnamefont {O.}~\bibnamefont
  {Gröbner}},\ }\bibfield  {title} {\bibinfo {title} {Vacuum system for
  {LHC}},\ }\href
  {https://doi.org/https://doi.org/10.1016/0042-207X(95)00042-9} {\bibfield
  {journal} {\bibinfo  {journal} {Vacuum}\ }\textbf {\bibinfo {volume} {46}},\
  \bibinfo {pages} {797} (\bibinfo {year} {1995})}\BibitemShut {NoStop}%
\bibitem [{\citenamefont {Veness}\ \emph {et~al.}(1999)\citenamefont {Veness},
  \citenamefont {Brunet}, \citenamefont {Gröbner}, \citenamefont {Lepeule},
  \citenamefont {Reymermier}, \citenamefont {Schneider}, \citenamefont
  {Skoczen}, \citenamefont {Kleimenok},\ and\ \citenamefont
  {Nikitin}}]{Veness1999}%
  \BibitemOpen
  \bibfield  {author} {\bibinfo {author} {\bibfnamefont {R.~J.~M.}\
  \bibnamefont {Veness}}, \bibinfo {author} {\bibfnamefont {J.~C.}\
  \bibnamefont {Brunet}}, \bibinfo {author} {\bibfnamefont {O.}~\bibnamefont
  {Gröbner}}, \bibinfo {author} {\bibfnamefont {P.}~\bibnamefont {Lepeule}},
  \bibinfo {author} {\bibfnamefont {C.}~\bibnamefont {Reymermier}}, \bibinfo
  {author} {\bibfnamefont {G.}~\bibnamefont {Schneider}}, \bibinfo {author}
  {\bibfnamefont {B.}~\bibnamefont {Skoczen}}, \bibinfo {author} {\bibfnamefont
  {V.}~\bibnamefont {Kleimenok}},\ and\ \bibinfo {author} {\bibfnamefont
  {I.~N.}\ \bibnamefont {Nikitin}},\ }\bibfield  {title} {\bibinfo {title}
  {{Beam Vacuum Interconnects for the LHC Cold Arcs}},\ }in\ \href@noop {}
  {\emph {\bibinfo {booktitle} {Proceedings of the 18th Biennial Particle
  Accelerator Conference, New York, NY, USA}}}\ (\bibinfo {year} {1999})\ pp.\
  \bibinfo {pages} {1399--1341}\BibitemShut {NoStop}%
\bibitem [{\citenamefont {Hempel}\ \emph {et~al.}(2013)\citenamefont {Hempel},
  \citenamefont {Baer}, \citenamefont {Pedersen}, \citenamefont {Dehning},
  \citenamefont {Effinger}, \citenamefont {Griesmayer}, \citenamefont
  {Lechner}, \citenamefont {Schmidt},\ and\ \citenamefont
  {Lohmann}}]{Hempel2012}%
  \BibitemOpen
  \bibfield  {author} {\bibinfo {author} {\bibfnamefont {M.}~\bibnamefont
  {Hempel}}, \bibinfo {author} {\bibfnamefont {T.}~\bibnamefont {Baer}},
  \bibinfo {author} {\bibfnamefont {S.~B.}\ \bibnamefont {Pedersen}}, \bibinfo
  {author} {\bibfnamefont {B.}~\bibnamefont {Dehning}}, \bibinfo {author}
  {\bibfnamefont {E.}~\bibnamefont {Effinger}}, \bibinfo {author}
  {\bibfnamefont {E.}~\bibnamefont {Griesmayer}}, \bibinfo {author}
  {\bibfnamefont {A.}~\bibnamefont {Lechner}}, \bibinfo {author} {\bibfnamefont
  {R.}~\bibnamefont {Schmidt}},\ and\ \bibinfo {author} {\bibfnamefont
  {W.}~\bibnamefont {Lohmann}},\ }\bibfield  {title} {\bibinfo {title}
  {Bunch-by-bunch beam loss diagnostics with diamond detectors at the {LHC}},\
  }in\ \href@noop {} {\emph {\bibinfo {booktitle} {Proceedings of the 52nd ICFA
  Advanced Beam Dynamics Workshop on High-Intensity and High-Brightness Hadron
  Beams (HB2012), MOP203, Beijing, China}}}\ (\bibinfo  {publisher} {JACoW,
  Geneva, Switzerland},\ \bibinfo {year} {2013})\ pp.\ \bibinfo {pages}
  {41--45}\BibitemShut {NoStop}%
\bibitem [{\citenamefont {Holzer}\ \emph {et~al.}(2005)\citenamefont {Holzer}
  \emph {et~al.}}]{Holzer2005}%
  \BibitemOpen
  \bibfield  {author} {\bibinfo {author} {\bibfnamefont {E.~B.}\ \bibnamefont
  {Holzer}} \emph {et~al.},\ }\bibfield  {title} {\bibinfo {title} {{Beam Loss
  Monitoring System for the LHC}},\ }in\ \href
  {https://doi.org/https://doi.org/10.1109/NSSMIC.2005.1596433} {\emph
  {\bibinfo {booktitle} {2005 IEEE Nuclear Science Symposium Conference Record,
  San Juan/Puerto Rico}}}\ (\bibinfo  {publisher} {IEEE, Piscataway, NJ},\
  \bibinfo {year} {2005})\ pp.\ \bibinfo {pages} {1052--1056}\BibitemShut
  {NoStop}%
\bibitem [{\citenamefont {Dehning}\ \emph {et~al.}(2007)\citenamefont {Dehning}
  \emph {et~al.}}]{Dehning2007}%
  \BibitemOpen
  \bibfield  {author} {\bibinfo {author} {\bibfnamefont {B.}~\bibnamefont
  {Dehning}} \emph {et~al.},\ }\bibfield  {title} {\bibinfo {title} {{The LHC
  beam loss measurement system}},\ }in\ \href
  {https://doi.org/https://doi.org/10.1109/PAC.2007.4439980} {\emph {\bibinfo
  {booktitle} {Proceedings of the 22nd Particle Accelerator Conference,
  FRPMN071, Albuquerque, New Mexico, USA}}}\ (\bibinfo  {publisher} {IEEE,
  Piscataway, NJ},\ \bibinfo {year} {2007})\ pp.\ \bibinfo {pages}
  {4192--4194}\BibitemShut {NoStop}%
\bibitem [{\citenamefont {Kalliokoski}\ \emph {et~al.}(2015)\citenamefont
  {Kalliokoski}, \citenamefont {Auchmann}, \citenamefont {Dehning},
  \citenamefont {Sousa}, \citenamefont {Effinger}, \citenamefont {Emery},
  \citenamefont {Grishin}, \citenamefont {Holzer}, \citenamefont {Jackson},
  \citenamefont {Kolad}, \citenamefont {Busto}, \citenamefont {Picha},
  \citenamefont {Roderick}, \citenamefont {Sapinski}, \citenamefont
  {Sobieszek},\ and\ \citenamefont {Zamantzas}}]{Kalliokoski2015}%
  \BibitemOpen
  \bibfield  {author} {\bibinfo {author} {\bibfnamefont {M.}~\bibnamefont
  {Kalliokoski}}, \bibinfo {author} {\bibfnamefont {B.}~\bibnamefont
  {Auchmann}}, \bibinfo {author} {\bibfnamefont {B.}~\bibnamefont {Dehning}},
  \bibinfo {author} {\bibfnamefont {F.~D.}\ \bibnamefont {Sousa}}, \bibinfo
  {author} {\bibfnamefont {E.}~\bibnamefont {Effinger}}, \bibinfo {author}
  {\bibfnamefont {J.}~\bibnamefont {Emery}}, \bibinfo {author} {\bibfnamefont
  {V.}~\bibnamefont {Grishin}}, \bibinfo {author} {\bibfnamefont
  {E.}~\bibnamefont {Holzer}}, \bibinfo {author} {\bibfnamefont
  {S.}~\bibnamefont {Jackson}}, \bibinfo {author} {\bibfnamefont
  {B.}~\bibnamefont {Kolad}}, \bibinfo {author} {\bibfnamefont {E.~N.~D.}\
  \bibnamefont {Busto}}, \bibinfo {author} {\bibfnamefont {O.}~\bibnamefont
  {Picha}}, \bibinfo {author} {\bibfnamefont {C.}~\bibnamefont {Roderick}},
  \bibinfo {author} {\bibfnamefont {M.}~\bibnamefont {Sapinski}}, \bibinfo
  {author} {\bibfnamefont {M.}~\bibnamefont {Sobieszek}},\ and\ \bibinfo
  {author} {\bibfnamefont {C.}~\bibnamefont {Zamantzas}},\ }\bibfield  {title}
  {\bibinfo {title} {{B}eam {L}oss {M}onitoring for {R}un 2 of the {LHC}},\
  }in\ \href {https://doi.org/https://doi.org/10.18429/JACoW-IPAC2015-MOPTY055}
  {\emph {\bibinfo {booktitle} {Proceedings of the 6th International Particle
  Accelerator Conference, MOPTY055, Richmond, VA, USA, May 3-8, 2015}}},\
  \bibinfo {series and number} {\bibinfo {number} {6}}\ (\bibinfo  {publisher}
  {JACoW, Geneva, Switzerland},\ \bibinfo {year} {2015})\ pp.\ \bibinfo {pages}
  {1057--1060}\BibitemShut {NoStop}%
\bibitem [{\citenamefont {Baer}\ \emph
  {et~al.}(2012{\natexlab{b}})\citenamefont {Baer}, \citenamefont {Barnes},
  \citenamefont {Carlier}, \citenamefont {Cerutti}, \citenamefont {Dehning},
  \citenamefont {Ducimetière}, \citenamefont {Ferrari}, \citenamefont
  {Garrel}, \citenamefont {Gérardin}, \citenamefont {Goddard}, \citenamefont
  {Holzer}, \citenamefont {Jackson}, \citenamefont {Jimenez}, \citenamefont
  {Kain}, \citenamefont {Lechner}, \citenamefont {Mertens}, \citenamefont
  {Misiowiec}, \citenamefont {Morón~Ballester}, \citenamefont {Nebot~del
  Busto}, \citenamefont {Norderhaug~Drosdal}, \citenamefont {Nordt},
  \citenamefont {Uythoven}, \citenamefont {Velghe}, \citenamefont {Vlachoudis},
  \citenamefont {Wenninger}, \citenamefont {Zamantzas}, \citenamefont
  {Zimmermann},\ and\ \citenamefont {Fuster~Martinez}}]{Baer2012a}%
  \BibitemOpen
  \bibfield  {author} {\bibinfo {author} {\bibfnamefont {T.}~\bibnamefont
  {Baer}}, \bibinfo {author} {\bibfnamefont {M.~J.}\ \bibnamefont {Barnes}},
  \bibinfo {author} {\bibfnamefont {E.}~\bibnamefont {Carlier}}, \bibinfo
  {author} {\bibfnamefont {F.}~\bibnamefont {Cerutti}}, \bibinfo {author}
  {\bibfnamefont {B.}~\bibnamefont {Dehning}}, \bibinfo {author} {\bibfnamefont
  {L.}~\bibnamefont {Ducimetière}}, \bibinfo {author} {\bibfnamefont
  {A.}~\bibnamefont {Ferrari}}, \bibinfo {author} {\bibfnamefont
  {N.}~\bibnamefont {Garrel}}, \bibinfo {author} {\bibfnamefont
  {A.}~\bibnamefont {Gérardin}}, \bibinfo {author} {\bibfnamefont
  {B.}~\bibnamefont {Goddard}}, \bibinfo {author} {\bibfnamefont {E.~B.}\
  \bibnamefont {Holzer}}, \bibinfo {author} {\bibfnamefont {S.}~\bibnamefont
  {Jackson}}, \bibinfo {author} {\bibfnamefont {J.~M.}\ \bibnamefont
  {Jimenez}}, \bibinfo {author} {\bibfnamefont {V.}~\bibnamefont {Kain}},
  \bibinfo {author} {\bibfnamefont {A.}~\bibnamefont {Lechner}}, \bibinfo
  {author} {\bibfnamefont {V.}~\bibnamefont {Mertens}}, \bibinfo {author}
  {\bibfnamefont {M.}~\bibnamefont {Misiowiec}}, \bibinfo {author}
  {\bibfnamefont {R.}~\bibnamefont {Morón~Ballester}}, \bibinfo {author}
  {\bibfnamefont {E.}~\bibnamefont {Nebot~del Busto}}, \bibinfo {author}
  {\bibfnamefont {L.}~\bibnamefont {Norderhaug~Drosdal}}, \bibinfo {author}
  {\bibfnamefont {A.}~\bibnamefont {Nordt}}, \bibinfo {author} {\bibfnamefont
  {J.}~\bibnamefont {Uythoven}}, \bibinfo {author} {\bibfnamefont
  {B.}~\bibnamefont {Velghe}}, \bibinfo {author} {\bibfnamefont
  {V.}~\bibnamefont {Vlachoudis}}, \bibinfo {author} {\bibfnamefont
  {J.}~\bibnamefont {Wenninger}}, \bibinfo {author} {\bibfnamefont
  {C.}~\bibnamefont {Zamantzas}}, \bibinfo {author} {\bibfnamefont
  {F.}~\bibnamefont {Zimmermann}},\ and\ \bibinfo {author} {\bibfnamefont
  {N.}~\bibnamefont {Fuster~Martinez}},\ }\bibfield  {title} {\bibinfo {title}
  {{UFOs in the LHC after LS1}},\ }in\ \href
  {https://doi.org/http://dx.doi.org/10.5170/CERN-2012-006.294} {\emph
  {\bibinfo {booktitle} {Proceedings of the Chamonix 2012 Workshop on LHC
  Performance}}}\ (\bibinfo {year} {2012})\ pp.\ \bibinfo {pages}
  {294--298}\BibitemShut {NoStop}%
\bibitem [{\citenamefont {Auchmann}\ \emph
  {et~al.}(2015{\natexlab{b}})\citenamefont {Auchmann}, \citenamefont {Baer},
  \citenamefont {Bednarek}, \citenamefont {Bellodi}, \citenamefont {Bracco},
  \citenamefont {Bruce}, \citenamefont {Cerutti}, \citenamefont {Chetvertkova},
  \citenamefont {Dehning}, \citenamefont {Granieri}, \citenamefont {Hofle},
  \citenamefont {Holzer}, \citenamefont {Lechner}, \citenamefont {Nebot
  Del~Busto}, \citenamefont {Priebe}, \citenamefont {Redaelli}, \citenamefont
  {Salvachua}, \citenamefont {Sapinski}, \citenamefont {Schmidt}, \citenamefont
  {Shetty}, \citenamefont {Skordis}, \citenamefont {Solfaroli}, \citenamefont
  {Steckert}, \citenamefont {Valuch}, \citenamefont {Verweij}, \citenamefont
  {Wenninger}, \citenamefont {Wollmann},\ and\ \citenamefont
  {Zerlauth}}]{Auchmann2015}%
  \BibitemOpen
  \bibfield  {author} {\bibinfo {author} {\bibfnamefont {B.}~\bibnamefont
  {Auchmann}}, \bibinfo {author} {\bibfnamefont {T.}~\bibnamefont {Baer}},
  \bibinfo {author} {\bibfnamefont {M.}~\bibnamefont {Bednarek}}, \bibinfo
  {author} {\bibfnamefont {G.}~\bibnamefont {Bellodi}}, \bibinfo {author}
  {\bibfnamefont {C.}~\bibnamefont {Bracco}}, \bibinfo {author} {\bibfnamefont
  {R.}~\bibnamefont {Bruce}}, \bibinfo {author} {\bibfnamefont
  {F.}~\bibnamefont {Cerutti}}, \bibinfo {author} {\bibfnamefont
  {V.}~\bibnamefont {Chetvertkova}}, \bibinfo {author} {\bibfnamefont
  {B.}~\bibnamefont {Dehning}}, \bibinfo {author} {\bibfnamefont {P.~P.}\
  \bibnamefont {Granieri}}, \bibinfo {author} {\bibfnamefont {W.}~\bibnamefont
  {Hofle}}, \bibinfo {author} {\bibfnamefont {E.~B.}\ \bibnamefont {Holzer}},
  \bibinfo {author} {\bibfnamefont {A.}~\bibnamefont {Lechner}}, \bibinfo
  {author} {\bibfnamefont {E.}~\bibnamefont {Nebot Del~Busto}}, \bibinfo
  {author} {\bibfnamefont {A.}~\bibnamefont {Priebe}}, \bibinfo {author}
  {\bibfnamefont {S.}~\bibnamefont {Redaelli}}, \bibinfo {author}
  {\bibfnamefont {B.}~\bibnamefont {Salvachua}}, \bibinfo {author}
  {\bibfnamefont {M.}~\bibnamefont {Sapinski}}, \bibinfo {author}
  {\bibfnamefont {R.}~\bibnamefont {Schmidt}}, \bibinfo {author} {\bibfnamefont
  {N.}~\bibnamefont {Shetty}}, \bibinfo {author} {\bibfnamefont
  {E.}~\bibnamefont {Skordis}}, \bibinfo {author} {\bibfnamefont
  {M.}~\bibnamefont {Solfaroli}}, \bibinfo {author} {\bibfnamefont
  {J.}~\bibnamefont {Steckert}}, \bibinfo {author} {\bibfnamefont
  {D.}~\bibnamefont {Valuch}}, \bibinfo {author} {\bibfnamefont
  {A.}~\bibnamefont {Verweij}}, \bibinfo {author} {\bibfnamefont
  {J.}~\bibnamefont {Wenninger}}, \bibinfo {author} {\bibfnamefont
  {D.}~\bibnamefont {Wollmann}},\ and\ \bibinfo {author} {\bibfnamefont
  {M.}~\bibnamefont {Zerlauth}},\ }\bibfield  {title} {\bibinfo {title}
  {Testing beam-induced quench levels of lhc superconducting magnets},\ }\href
  {https://doi.org/https://doi.org/10.1103/PhysRevSTAB.18.061002} {\bibfield
  {journal} {\bibinfo  {journal} {Phys. Rev. ST Accel. Beams}\ }\textbf
  {\bibinfo {volume} {18}},\ \bibinfo {pages} {061002} (\bibinfo {year}
  {2015}{\natexlab{b}})}\BibitemShut {NoStop}%
\bibitem [{\citenamefont {Holzer}\ \emph {et~al.}(2010)\citenamefont {Holzer},
  \citenamefont {Assmann}, \citenamefont {Bellodi}, \citenamefont {Bruce},
  \citenamefont {Dehning}, \citenamefont {Effinger}, \citenamefont {Emery},
  \citenamefont {Grishin}, \citenamefont {Hajdu}, \citenamefont {Jackson},
  \citenamefont {Jowett}, \citenamefont {Kurfuerst}, \citenamefont {Marsili},
  \citenamefont {Misiowiec}, \citenamefont {Nebot Del~Busto}, \citenamefont
  {Nordt}, \citenamefont {Redaelli}, \citenamefont {Roderick}, \citenamefont
  {Rossi}, \citenamefont {Sapinski}, \citenamefont {Wollmann},\ and\
  \citenamefont {Zamantzas}}]{Holzer2010}%
  \BibitemOpen
  \bibfield  {author} {\bibinfo {author} {\bibfnamefont {E.~B.}\ \bibnamefont
  {Holzer}}, \bibinfo {author} {\bibfnamefont {R.}~\bibnamefont {Assmann}},
  \bibinfo {author} {\bibfnamefont {G.}~\bibnamefont {Bellodi}}, \bibinfo
  {author} {\bibfnamefont {R.}~\bibnamefont {Bruce}}, \bibinfo {author}
  {\bibfnamefont {B.}~\bibnamefont {Dehning}}, \bibinfo {author} {\bibfnamefont
  {E.}~\bibnamefont {Effinger}}, \bibinfo {author} {\bibfnamefont
  {J.}~\bibnamefont {Emery}}, \bibinfo {author} {\bibfnamefont
  {V.}~\bibnamefont {Grishin}}, \bibinfo {author} {\bibfnamefont
  {C.}~\bibnamefont {Hajdu}}, \bibinfo {author} {\bibfnamefont
  {S.}~\bibnamefont {Jackson}}, \bibinfo {author} {\bibfnamefont
  {J.}~\bibnamefont {Jowett}}, \bibinfo {author} {\bibfnamefont
  {C.}~\bibnamefont {Kurfuerst}}, \bibinfo {author} {\bibfnamefont
  {A.}~\bibnamefont {Marsili}}, \bibinfo {author} {\bibfnamefont
  {M.}~\bibnamefont {Misiowiec}}, \bibinfo {author} {\bibfnamefont
  {E.}~\bibnamefont {Nebot Del~Busto}}, \bibinfo {author} {\bibfnamefont
  {A.}~\bibnamefont {Nordt}}, \bibinfo {author} {\bibfnamefont
  {S.}~\bibnamefont {Redaelli}}, \bibinfo {author} {\bibfnamefont
  {C.}~\bibnamefont {Roderick}}, \bibinfo {author} {\bibfnamefont
  {A.}~\bibnamefont {Rossi}}, \bibinfo {author} {\bibfnamefont
  {M.}~\bibnamefont {Sapinski}}, \bibinfo {author} {\bibfnamefont
  {D.}~\bibnamefont {Wollmann}},\ and\ \bibinfo {author} {\bibfnamefont
  {C.}~\bibnamefont {Zamantzas}},\ }\bibfield  {title} {\bibinfo {title}
  {{Losses away from collimators: statistics and extrapolation}},\ }in\
  \href@noop {} {\emph {\bibinfo {booktitle} {Proceedings of the 2nd Evian
  Workshop on LHC Beam Operation, Evian Les Bains, France}}}\ (\bibinfo {year}
  {2010})\ pp.\ \bibinfo {pages} {173--177}\BibitemShut {NoStop}%
\bibitem [{\citenamefont {Papadopoulou}\ \emph {et~al.}(2019)\citenamefont
  {Papadopoulou}, \citenamefont {Antoniou}, \citenamefont {Efthymiopoulos},
  \citenamefont {Hostettler}, \citenamefont {Iadarola}, \citenamefont
  {Karastathis}, \citenamefont {Kostoglou}, \citenamefont {Papaphilippou},\
  and\ \citenamefont {Trad}}]{Papadopoulou2019}%
  \BibitemOpen
  \bibfield  {author} {\bibinfo {author} {\bibfnamefont {S.}~\bibnamefont
  {Papadopoulou}}, \bibinfo {author} {\bibfnamefont {F.}~\bibnamefont
  {Antoniou}}, \bibinfo {author} {\bibfnamefont {I.}~\bibnamefont
  {Efthymiopoulos}}, \bibinfo {author} {\bibfnamefont {M.}~\bibnamefont
  {Hostettler}}, \bibinfo {author} {\bibfnamefont {G.}~\bibnamefont
  {Iadarola}}, \bibinfo {author} {\bibfnamefont {N.}~\bibnamefont
  {Karastathis}}, \bibinfo {author} {\bibfnamefont {S.}~\bibnamefont
  {Kostoglou}}, \bibinfo {author} {\bibfnamefont {Y.}~\bibnamefont
  {Papaphilippou}},\ and\ \bibinfo {author} {\bibfnamefont {G.}~\bibnamefont
  {Trad}},\ }\bibfield  {title} {\bibinfo {title} {{What do we understand on
  the Emittance Growth?}},\ }in\ \href {https://cds.cern.ch/record/2750416}
  {\emph {\bibinfo {booktitle} {Proceedings of the 9th Evian Workshop on LHC
  Beam Operation, Evian Les Bains, France}}}\ (\bibinfo  {publisher} {CERN,
  Geneva, Switzerland},\ \bibinfo {year} {2019})\ pp.\ \bibinfo {pages}
  {199--205}\BibitemShut {NoStop}%
\bibitem [{\citenamefont {Bottura}\ \emph {et~al.}(2019)\citenamefont
  {Bottura}, \citenamefont {Breschi}, \citenamefont {Felcini},\ and\
  \citenamefont {Lechner}}]{Bottura2019}%
  \BibitemOpen
  \bibfield  {author} {\bibinfo {author} {\bibfnamefont {L.}~\bibnamefont
  {Bottura}}, \bibinfo {author} {\bibfnamefont {M.}~\bibnamefont {Breschi}},
  \bibinfo {author} {\bibfnamefont {E.}~\bibnamefont {Felcini}},\ and\ \bibinfo
  {author} {\bibfnamefont {A.}~\bibnamefont {Lechner}},\ }\bibfield  {title}
  {\bibinfo {title} {Stability modeling of the {LHC Nb-Ti Rutherford} cables
  subjected to beam losses},\ }\href
  {https://doi.org/https://doi.org/10.1103/PhysRevAccelBeams.22.041002}
  {\bibfield  {journal} {\bibinfo  {journal} {Phys. Rev. Accel. Beams}\
  }\textbf {\bibinfo {volume} {22}},\ \bibinfo {pages} {041002} (\bibinfo
  {year} {2019})}\BibitemShut {NoStop}%
\end{thebibliography}%

\end{document}